\documentclass[
aps,
prl,
amsmath,
amssymb,
floatfix,
longbibliography,
letterpaper,
lengthcheck,
superscriptaddress
]{revtex4-2}

\usepackage{graphicx}
\usepackage{siunitx}
\DeclareSIUnit\angstrom{\protect \text {Å}}
\usepackage{floatflt,epsfig} 
\usepackage{mathtools}
\usepackage{color}
\usepackage{braket}
\usepackage{hyperref}
\usepackage[normalem]{ulem}
\usepackage[english]{babel}
\usepackage{blindtext}
\usepackage{lipsum}  
\usepackage{comment}
\usepackage{amsthm}
\usepackage{tabularx}
\usepackage{bm}
\usepackage{dsfont}
\usepackage{bbold}
\usepackage{ulem}
\usepackage{mwe,tikz}
\usepackage[percent]{overpic}
\usepackage{todonotes}
\usepackage{algorithm}
\usepackage{algpseudocode}

\DeclareUnicodeCharacter{0301}{\"{o}}

\newcommand {\diff} {\mathrm{d}}

\DeclareSIUnit{\atomicunit}{a.u.}

\begin{document}

\title{Collective Strong Coupling Modifies Aggregation and Solvation}

\author{Matteo Castagnola}
\affiliation{Department of Chemistry, Norwegian University of Science and Technology, 7491 Trondheim, Norway}

\author{Tor S. Haugland}
\affiliation{Department of Chemistry, Norwegian University of Science and Technology, 7491 Trondheim, Norway}

\author{Enrico Ronca}
\affiliation{Dipartimento di Chimica, Biologia e Biotecnologie, Università degli Studi di Perugia, Via Elce di Sotto, 8,06123, Perugia, Italy}

\author{Henrik Koch}
\email[Electronic address:\;]{henrik.koch@ntnu.no}
\affiliation{Department of Chemistry, Norwegian University of Science and Technology, 7491 Trondheim, Norway}
\affiliation{Scuola Normale Superiore, Piazza dei Cavalieri 7, 56126 Pisa, Italy}

\author{Christian Sch\"afer}
\email[Electronic address:\;]{christian.schaefer@chalmers.se}
\affiliation{
Department of Physics, Chalmers University of Technology, 412 96 G\"oteborg, Sweden}
\affiliation{
Department of Microtechnology and Nanoscience, MC2, Chalmers University of Technology, 412 96 G\"oteborg, Sweden}

\begin{abstract}
Intermolecular interactions are pivotal for aggregation, solvation, and crystallization.
We demonstrate that the collective strong coupling of several molecules to a single optical mode results in notable changes in the molecular excitations around an impurity, e.g., in the first aggregation or solvation shell. 
A competition between short-range Coulombic and long-range photonic correlation inverts the local transition density in a polaritonic state, suggesting notable changes in the polarizability of the solvation shell.
Our results provide an alternative perspective on recent work in polaritonic chemistry and pave the way for the rigorous treatment of cooperative effects in aggregation, solvation, and crystallization.
\end{abstract}

\maketitle
The coherent interaction of molecules with confined optical modes leads to hybrid light-matter states called polaritons \cite{bhuyan2023rise, hirai2020recent, hirai2023molecular}.
In the strong coupling regime, usually achieved by coupling several molecules to the optical cavity, experiments show significant modifications of chemical properties \cite{pang2020role, sau2021modifying, schwartz2013polariton, george2015ultra, wang2014quantum, zhong2017energy, hutchison2012modifying, mony2021photoisomerization, schwartz2011reversible, thomas2016ground, lather2019cavity, thomas2019tilting, munkhbat2018suppression, canaguier2013thermodynamics, hirai2020modulation, polak2020manipulating}. 
Due to their delocalized nature, polaritons in quantum optics are studied from a collective perspective where the molecules are modeled as few-level systems that indirectly interact solely through the photon field \cite{michetti2008simulation, michetti2005polariton, wellnitz2022disorder, du2018theory, garraway2011, hagenmuller2017cavity, feist2018polaritonic, ribeiro2018polariton}.
Yet, chemistry is governed by local interactions, and the molecular complexity requires refined chemical methods to analyze processes like reactions. 
The behavior of molecules is hence susceptible to their immediate surroundings.
For example, their spectral absorption and emission can be altered by their interplay with the solvent (solvatochromic) or by close proximity to identical molecules (concentration-dependent).
Experiments under strong coupling often use organic molecules that are sensitive to their surroundings, show intense excitations, or even form aggregates \cite{lidzey1998strong, hobson2002strong, coles2013imaging, lidzey1999room, salomon2013strong, sugawara2006strong, coles2014polariton, zhong2017energy, georgiou2021ultralong, tischler2007solid, schwartz2013polariton, ren2020efficient, cheng2022molecular, bigeon2017strong, wang2014quantum, george2015ultra, mony2021photoisomerization, timmer2023plasmon, vasa2013real}.
The chemical environment can then play an active role and exert significant influence, as emphasized by several vibrational strong coupling (VSC) experiments showing modification of assembly \cite{joseph2021supramolecular, hirai2021selective, sandeep2022manipulating} and reactivity \cite{lather2019cavity, singh2023solvent, piejko2023solvent}.
For a clear chemical understanding of polaritonic processes, we must then investigate the coexisting roles of the molecule, the solvent (chemical environment), and the cavity (optical environment). 
To this end, \textit{ab initio} quantum electrodynamics (QED) merges the knowledge of quantum optics and quantum chemistry to explore single-molecule effects retaining the chemical complexity \cite{fregoni2018manipulating, fregoni2019strong,sun2023modification,li2020cavity, pavovsevic2022wavefunction, doi:10.1021/jacs.1c13201,schafer2021shining,schafer2022emb, haugland2020coupled, haugland2020intermolecular, mordovina2020polaritonic, vidal2022polaritonic,schafer2023machine}. 
Even if collective states can show larger contributions from selected molecules \cite{sidler2020polaritonic,schafer2022emb}, local changes tend to reduce due to collective delocalization and an increasing number of quasi-dark states~\cite{yuen2019polariton}. 
While recent work indicates a larger relevance of dynamic electronic polarization~\cite{schafer2023machine,sidler2023unraveling,schnappinger2023cavity}, it remains puzzling how the polaritons' collective nature enters chemical processes.
The role of changes in the chemical environment (solvents, aggregates) 
under strong coupling has been widely disregarded up to this point. 

In this letter, we add another facet to the understanding of polaritonic chemistry by reintroducing the chemical environment.
Using QED coupled cluster (QED-CC) \cite{haugland2020coupled}, which is at present the most accurate approach for medium-sized molecular polaritons, we study the yet unexplored competition between photon and intermolecular interactions.
We extract the single-molecule response in collective ensembles and investigate the microscopic changes of aggregates or solvation systems in polaritonic chemistry.
We demonstrate local response modifications unambiguously distinguishable from collective delocalization and arising from the interplay between Coulomb and transverse fields.
Moreover, our results show a slower decrease than conventional local effects in polaritonics, i.e., not with the total number of coupled molecules but with the number of affected solvation shells.
As a result, the \textit{immediate} surroundings of the impurity, representing a solute, nucleation, or reaction center, undergo notable changes.
Structural changes in the chemical environment can then influence the dynamics of the impurity.
In the following, solvation, aggregation, and nucleation are considered synonymously due to their conceptual similarity.

\begin{figure*}[!ht]
    \centering
    \includegraphics[width=\textwidth]{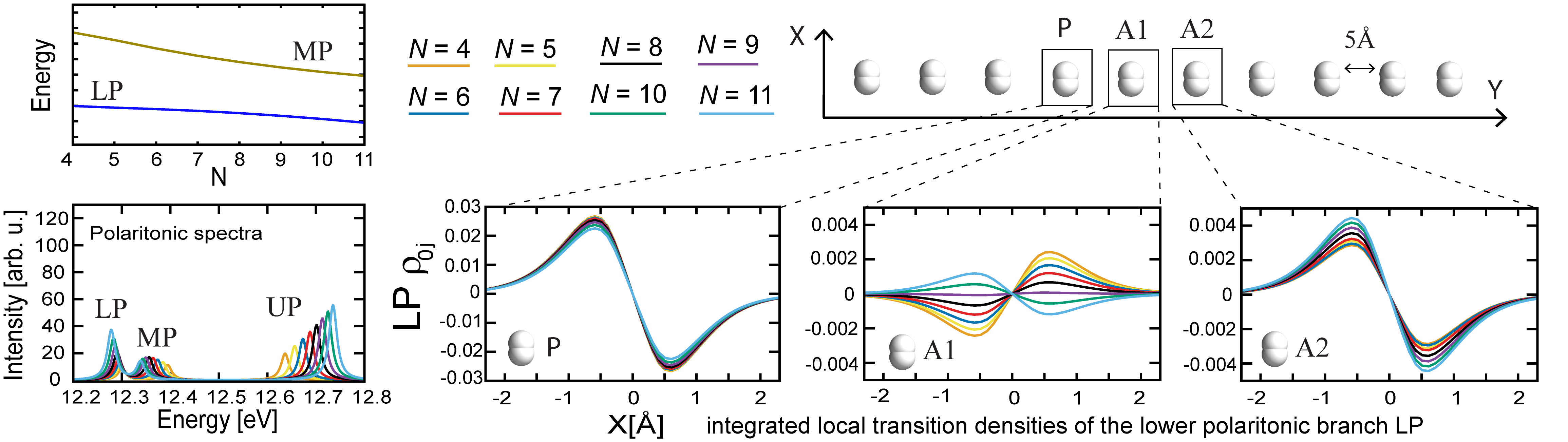}
    \caption{
    Polaritonic spectra for \qty[mode = text]{5}{\angstrom} intermolecular separation with coupling strength \qty[mode = text]{0.005}{\atomicunit}, polarization along the H$_2$ bonds, and photon frequency tuned to the undressed excitation of the isolated unperturbed dimers. 
    The energies of the lower and middle polaritons as a function of the number of dimers $N$ in the chain, depicted in the upper left panel, highlight an avoided crossing.
    The lower polariton LP integrated local transition densities for P, A1, and A2, are shown.
    The Coulomb forces compete with the transverse field (photon) to determine the alignment pattern of the transition moments, generating a sign flip in the transition density of A1.}
    \label{fig:image b}
\end{figure*}

\textit{Modelling Solvation and Collective Coupling - }
\begin{figure*}[!ht]
    \centering
    \includegraphics[width=\textwidth]{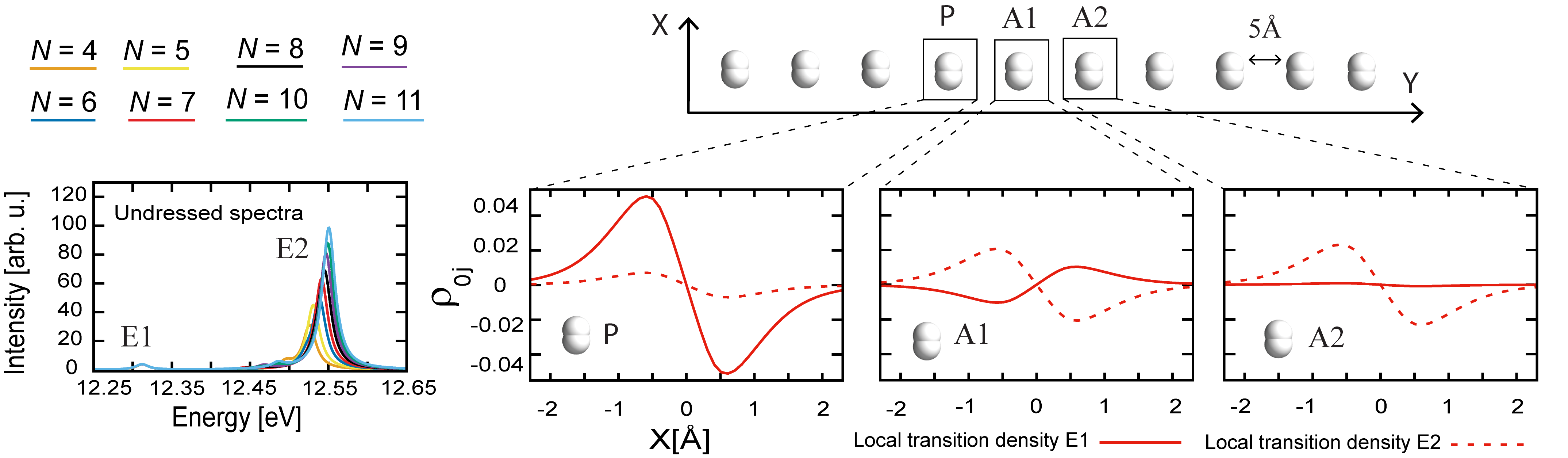}
    \caption{Structure of the $(\text{H}_2)_N$ aggregate and absorption spectra outside the cavity for \qty[mode = text]{5}{\angstrom} intermolecular separation.
    The local transition densities $\rho_{j0}(x)$ (i.e., the transition density $\varrho_{j0}(x,y,z)$ integrated perpendicularly to the dipole moment direction, that is, perpendicular to the H$_2$ bond $\rho^{I}_{j0}(x) = \int_{-\infty}^{+\infty}\diff z\int_{Y_I-2\text{\AA}}^{Y_I+2\text{\AA}}\diff y \;\varrho_{j0}(x,y,z)$ where $I=\text{P, A1, A2}$, see Sec. I in the SI) of the undressed excitations E1 (solid lines) and E2 (dotted lines) of $(\text{H}_2)_7$ are shown, showing specific transition moment alignment patterns caused by intermolecular interactions.
    Only the left densities are shown because the right densities display the same behavior (see Sec. I of the SI).
    \label{fig:image a}}
\end{figure*}
Molecular properties are significantly modified by chemical environments such as polymer matrices, solvents, or aggregates.
The interaction of a solute with its surrounding molecules leads to changes in the ground and excited electronic densities, which affect energy levels, molecular geometries, and even chemical reactivity.
In addition, the solvent also directly contributes to properties such as spectroscopic signals.
To study the impact of polaritons on the complex chemical response, we must simultaneously include the photon coupling and the Coulomb interactions in the solute-solvent system.
Therefore, we focus on a system with non-negligible intermolecular interactions as a prototype for the Coulomb forces in supramolecular structures. 
Our model system is a chain of $N$ hydrogen molecules that form an H-aggregate illustrated in \autoref{fig:image a}. 
An impurity (P), representing a solute or reactive molecule, is introduced by stretching the bond of the central hydrogen, and we investigate different intermolecular distances to alter the Coulomb forces. 
\autoref{fig:image a} shows the absorption spectrum of the undressed (out-of-cavity) chain for different $N$ and \qty[mode = text]{5}{\angstrom} intermolecular separation.
The global response of the system is governed by the transition density $\varrho_{j0}(x,y,z)$, which is for instance connected to the optical properties via the transition dipole moment $\bm{\mu}_{j0}=\int_{\mathbb{R}^3}\bm{r}\varrho_{j0}(x,y,z)\diff^3r$.
To characterize the local behavior of each dimer, we integrate $\varrho_{j0}(x,y,z)$ perpendicularly to the excitations transition dipole moments (that is, in the Y and Z directions) around P, its nearest-neighbor A1 (first solvation shell), and its next-nearest neighbor A2.
The integrated local transition densities $\rho^{I}_{j0}(x) = \int_{-\infty}^{+\infty}\diff z\int_{Y_I^{(1)}}^{Y_I^{(2)}}\diff y \;\varrho_{j0}(x,y,z)$ ($I=\text{P, A1, A2}$) for the excited states E1 and E2 are depicted in \autoref{fig:image a}.
The excitation E1 is mainly localized on P and A1 with anti-aligned local transition dipoles, while E2 is delocalized and shows an all-parallel transition moment alignment.
In \autoref{fig:image b}, we show the polaritonic absorption spectrum with the photon resonant to the (isolated) unperturbed H$_2$ excitation (with light-matter coupling strength $\lambda = \sqrt{{1}/(\varepsilon_0 V)}$ = \qty[mode = text]{0.005}{\atomicunit}).
Three polaritonic branches, the lower (LP), middle (MP), and upper polaritons (UP) emerge. 
The electron and electron-photon correlation is accurately described by the \textit{ab initio} QED-CC ansatz wave function, thus capturing any modification of inter- (dispersion) and intra-molecular forces \cite{haugland2020intermolecular, haugland2020coupled, haugland2023understanding, piejko2023solvent, cao2023cavity}.
Nevertheless, due to the low coupling strength, the molecular ground state is here essentially unaffected by the embedding in the optical environment, and polaritonic effects effectively arise from \textit{collective} coupling.
The dimers are indirectly coupled through the cavity, establishing an interplay between longitudinal (Coulomb) and transverse (photon) fields.
The simplicity and engineerability of our model allow us to separate intermolecular and cavity-induced effects, providing a satisfactory proof-of-concept model for investigating microscopic molecular responses in complex chemical and photonic environments.

\textit{Modification of Solvation - }
In \autoref{fig:image b}, we plot the LP local transition densities for P, A1, and A2.
As $N$ increases, the A1 local transition density changes sign, i.e., the transition dipole of A1 aligns with the other molecules. 
This effect requires that both the (undressed) excitations E1 and E2 contribute to the LP state, and hence occurs around the avoided crossing domain illustrated in \autoref{fig:image b}.
While the alignment clearly originates from \textit{collective} effects, its impact is evidently \textit{localized} in the first solvation shell and requires an explicit treatment of the solute-solvent interface.
The local change arises from competition between the short-range intermolecular forces, which pattern E1 and E2, and the collective interaction to the optical mode, which tends to align the molecular transition moments (see also Sec. III of the SI).
Indeed, for larger intermolecular separations, the LP local moments are aligned (see Sec. V of the SI), confirming that the sign change stems from the interplay between Coulomb and photon fields.
Which effect prevails depends on the setup, specifically on the excitation energies (solute-solvent system), the chain length $N$ (collective coupling), and the light-matter coupling strength $\lambda = \sqrt{{1}/(\varepsilon_0 V)}$ (optical device features), and the intermolecular forces (solute-solvent system and intermolecular separation).

\begin{figure}[!ht]
    \centering
    \includegraphics[width=.45\textwidth]{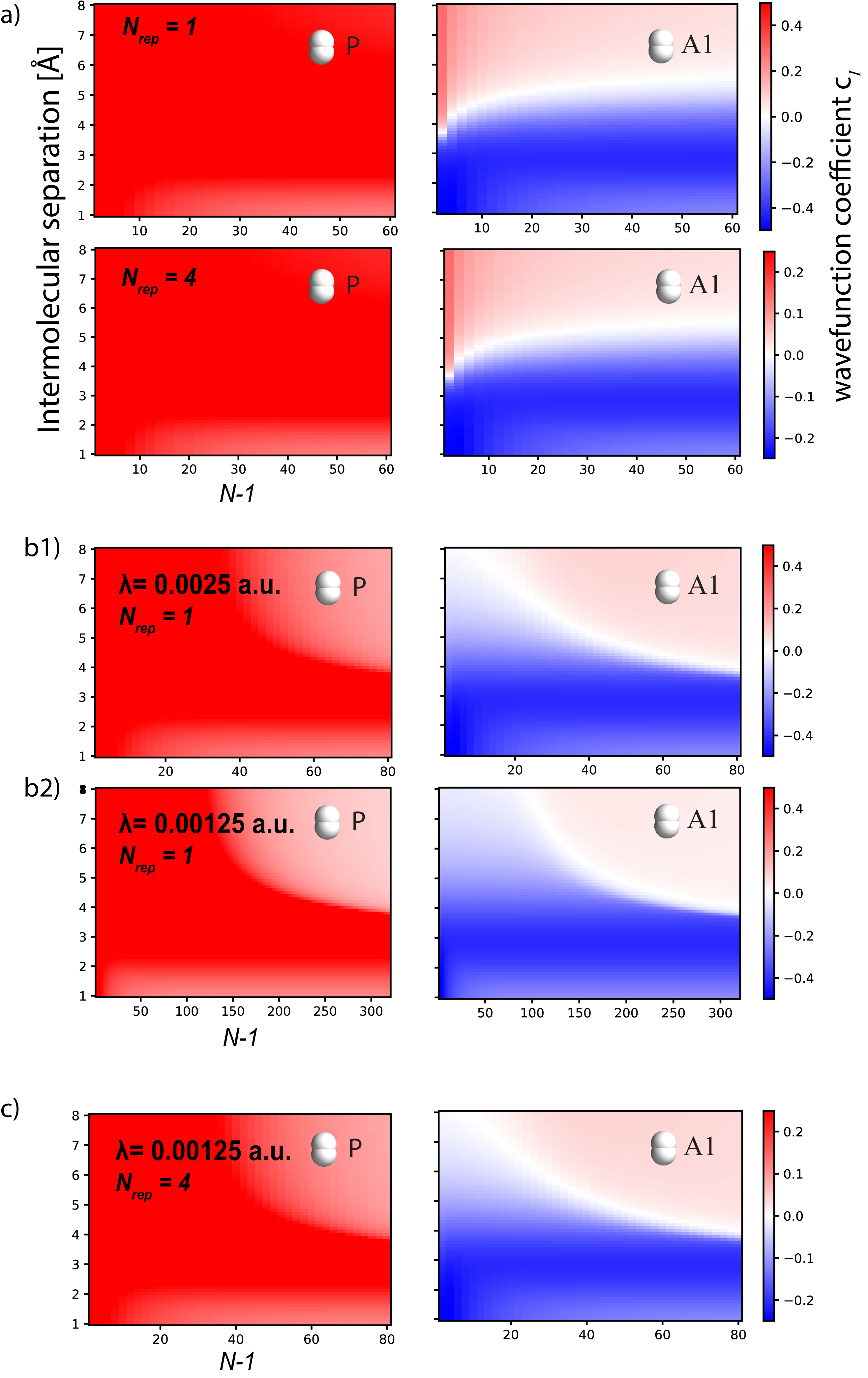}
    \caption{Wave function coefficients c$_I$ where $I=\text{P, A1}$ from the TC-Kasha model as a function of the environment molecules $N-1$ and the intermolecular separation. 
    a) The coupling strength is set to $\lambda=\frac{\text{\qty[mode=text]{0.015}{\atomicunit}}}{\sqrt{N_e}}$. We qualitatively recover the same sign flip for A1 as \autoref{fig:image b} for $N_{rep}$ = 1 (upper panel) and 4 (lower panel), but the coefficient values approximately scale as $1/\sqrt{N_{rep}}$. Notably, such a decay is much smaller than $1/\sqrt{N_{e}}$. 
    b) Wavefunction coefficients obtained by fixing $\lambda$ and $N_{rep}$ = 1 as $N$ varies. This is consistent with experimental setups where the same cavity (i.e., $\lambda$ fixed) is employed at different molecular concentrations (here, the solvent molecules $N-1$). 
    c) Same as (b) but for $N_{rep} =4$. Compared to (b1), the overall collective coupling is maintained, but due to the larger number of solute molecules, the coefficients are rescaled by $1/\sqrt{N_{rep}}$. The impurity concentration P then defines the decay of the wavefunction coefficients. Notice that P does not need to be in strong coupling.
    See Sec. V-A of the SI for more results.
    }
    \label{fig:model}
\end{figure}

The number of solvent molecules heavily outweighs the concentration of solutes. 
It is therefore instructive to verify how our observation scales in the thermodynamic limit $N_e, V \rightarrow \infty,~N_e/V  = \text{const.}$, i.e., with fixed $N_e/V$ but increasing particle number $N_e$ and quantization volume $V$ (controlling the fundamental coupling strength).
To this end, we resort to a simplified Tavis-Cummings (TC)-Kasha \cite{dicke1954coherence, tavis1968exact, kasha1963energy, hestand2018expanded} model (parameterized according to the electronic CC calculations), with a nearest-neighbor transition dipole-dipole coupling between the molecular excitations (see Sec. III-C and V-A of the SI). 
The total number of molecules $N_e = N_{rep} \times N$ is given by the number of molecules $N$ per aggregate (i.e., the same number used in \autoref{fig:image a} and \ref{fig:image b}), and how many times the system appears in the cavity $N_{rep}$. 
In \autoref{fig:model}(a), we report the eigenvector coefficients for a single P and A1 to quantify how the individual molecules contribute to the collective excitations of the TC-Kasha model. 
The ratio $N_e/V$ is fixed, i.e., the Rabi-splitting is kept constant while increasing $N$.
The simulations for $N_{rep}=$ \qtylist{1;4}{} show analogous results and thus prove that the described behavior is qualitatively resistant to the thermodynamic limit. 
Still, the single-molecule contribution (i.e., the TC-Kasha eigenvector coefficients of a single dimer) decreases by approximately $1/\sqrt{N_{rep}}$.
Similar conclusions are drawn from \autoref{fig:model}(b-c), where for each panel $\lambda$ and $N_{rep}$ are fixed while increasing $N$.
This allows us to investigate the effect of changing the number of coupled molecules and thus the collective effects.
Increasing the number $N-1$ of coupled solvent molecules at fixed $N_{rep}$ and $\lambda$ shortens the critical intermolecular distance at which the A1 coefficient shows a change. 
The panels \ref{fig:model}(b1) and (b2), computed at different couplings, show similar behavior and order of magnitude, but the $N$ for which the transition moment of A1 changes sign is shifted to longer chains.
We also confirm this trend by simulating two H$_2$ chains in our QED-CC calculations (see Sec. V-A of the SI).
Comparing \autoref{fig:model}(b1) and (c) shows the effect of increasing $ N_{rep} / N $ while rescaling $\lambda$ by $1/\sqrt{N_{rep}}$, which thus retains the overall collective coupling. 
The panels show the same qualitative behavior, with coefficients rescaled by $1/\sqrt{N_{rep}}$, as in \ref{fig:model}(a).
Therefore, \autoref{fig:model} demonstrates that the magnitude of the change in the environment response for fixed $N_{rep}/V$ decreases with the number of activated species P, which is significantly smaller than the number of coupled molecules $N_e$, and P does not need to be strongly coupled to the device.
In other words, if the solvent exists in vast excess as in experiments showing modifications of crystallization~\cite{joseph2021supramolecular, hirai2021selective, sandeep2022manipulating} and ionic conductivity~\cite{D3SC03364C}, each solvation shell will experience a considerably larger effect from the cavity than the bulk of the solvent molecules.
Moreover, keeping the solute-solvent ratio fixed (increasing $N_{rep}$) shortens the critical length $N$ as seen from \autoref{fig:model}(b) and (c).
Therefore, the intermolecular distance and chain length $N$ for which the transition moment of A1 changes sign depends solely on the solvent strong coupling defined by the ratio $N_{rep}\times (N-1)/V$.

\begin{figure}[!ht]
    \centering
    \includegraphics[width=.5\textwidth]{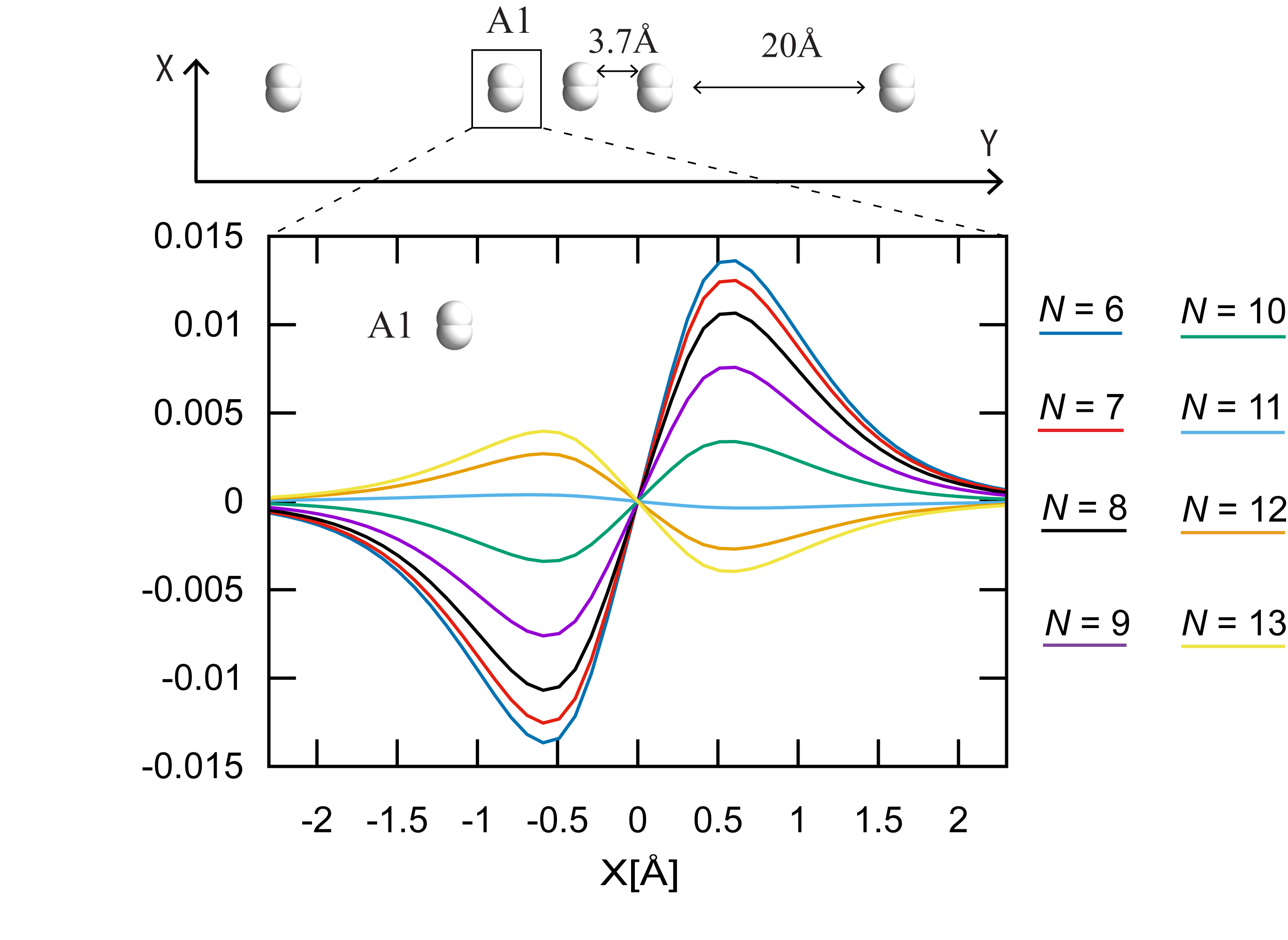}
    \caption{LP integrated local transition densities for A1 for a cluster $(\text{H}_2)_3$ with intermolecular separation of \qty[mode = text]{3.7}{\angstrom}, together with $N$-3 dimers intermolecular at a distance of \qty[mode = text]{20}{\angstrom}.
    Due to the larger intermolecular forces between A1 and P, the excitation E1 is strongly blueshifted. 
    Therefore, we use a slightly larger coupling and a larger number $N$ compared to \autoref{fig:image b} to reach the avoided crossing region between the LP and MP.
    The results show the same qualitative behavior as \autoref{fig:image b}.
    }
    \label{fig: cluster 3.7}
\end{figure}

So far, we considered all the molecules at the same distance.
However, analogous results are found for a small H$_2$ cluster, with the other hydrogens at larger distances.
This is illustrated in \autoref{fig: cluster 3.7} for $(\text{H}_2)_3$ with intermolecular distance \qty[mode = text]{3.7}{\angstrom}, and  slightly increased $\lambda$ = \qty[mode = text]{0.01}{\atomicunit} to compensate for the larger blueshift of E1. 
The remaining $N$-3 hydrogens are placed at \qty[mode = text]{20}{\angstrom} distance.
The results show the same behavior as in \autoref{fig:image b} and emphasize the locality induced by the short-range Coulomb interactions, clearly highlighting the interplay between the chemical surroundings and the collective optical dressing.

The TC-Kasha model successfully reproduces the modifications in the dynamic response of the first solvation shell.
Since the model is parametrized with bare (electronic) transitions, it stresses the importance of collective behavior, which alters the LP microscopic response near P once a critical cooperative coupling strength is reached.
Nevertheless, the TC-Kasha model has apparent limitations in that it will perform poorly in modelling the electronic dynamic of more complex systems and it cannot account for non-linear effects included in the QED-CC calculations, i.e., a self-consistent and more realistic treatment is likely to give similar or more favorable scaling. 
It should also be noted that the strength and structure of solvation vary widely with the type of intermolecular interaction, such that the onset of this effect can change dramatically with the specific choice of solvent and solute.
Quantum chemistry then provides a flexible tool-set to explore such solvent effects beyond the heavily simplified dipole-dipole picture~\cite{senn2009qm, jensen2003discrete, tomasi2005quantum, curutchet2009electronic, steindal2011excitation, giovannini2023continuum}.

\begin{figure}[!ht]
    \centering
    \includegraphics[width=.5\textwidth]{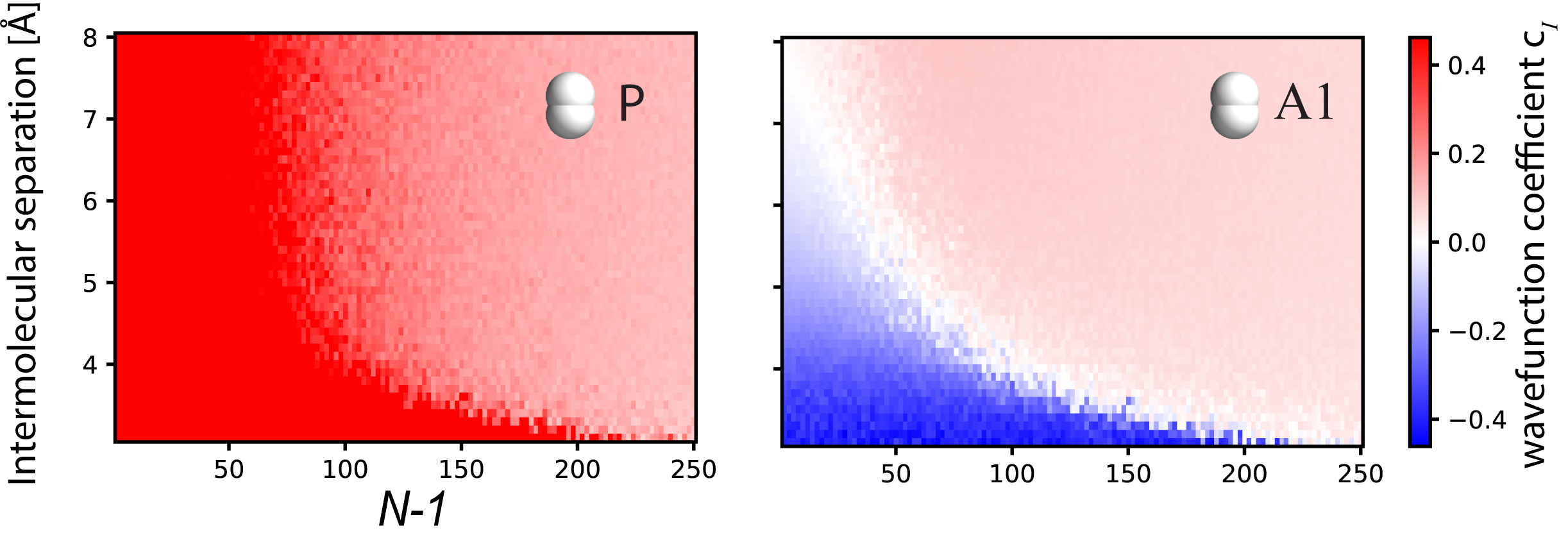}
    \caption{Wavefunction coefficients obtained from a disordered TC-Kasha model where the first and second solvation shells around P are fixed while orientational disorder and inhomogeneous broadening are applied to the other dimers. The coupling strength is set to \qty[mode = text]{0.005}{\atomicunit}. We still recover the same qualitative behavior and coefficient magnitude of \autoref{fig:image b} and \ref{fig:model}(b), although for larger $N$ because of the smaller collective coupling to the device due to disorder.
    Analogous results are obtained for different coupling strengths, see Sec. V-A of the SI for more results.
    }
    \label{fig:model disorder}
\end{figure}

\textit{Disorder and Resonance dependence - }
While aggregation, solvation, and crystallization imply a somewhat structured environment, disorder (in the form of orientation or inhomogeneous broadening) is inevitable in chemistry under standard ambient conditions.
Inhomogeneous broadening and the molecular orientation relative to the cavity polarization, for all solvent molecules besides A1 (\textit{long-range} disorder), is included in our TC-Kasha model by sampling the excitation energies from a Gaussian distribution and assigning random H$_2$ orientations (see SI Sec.~V-A for details).
As demonstrated in \autoref{fig:model disorder} and Sec.~V-A in the SI, the long-range disorder merely increases the required number of molecules $N$, but the overall solute-solvent effect remains unaffected.
At the same time, \textit{local} disorder severely impacts the intermolecular interactions since the Coulomb forces are strongly directional, i.e., other forms of aggregation will impact the solute differently.
In the SI, we report the local transition densities for a J-aggregate (head-tail) $(\text{H}_2)_N$ configuration and show a sign flip for the local transition dipole of A1 \textit{in the MP} instead of the LP. 
The modification of the solvent-solute response is conceptually identical and merely moves to a different polaritonic state.
This stems from the different transition moment patterns in the undressed excitations of the J-aggregate. 
Hence, we can expect that a realistic solvation system will exhibit the discussed effect partially in the LP and the MP.

Lastly, the solvent-solute polarization features a clear resonant behavior.
When the cavity is detuned from the E2 excitation (entirely off-resonant or tuned to E1) the local changes are either negligible or indistinguishable from collective delocalization effects.
If the field is tuned to E2 (instead of the excitation of an isolated unperturbed H$_2$), the results are in line with the foregoing discussion, although require a slightly higher $N$ to achieve the sign flip in A1.
These results are consistent with the position of the avoided crossing between the LP and MP, which depends on the photon frequency and determines the mixing of E1 and E2 (and thus the local impact of the strong coupling regime).

\textit{Conclusion - }
Using state-of-the-art \textit{ab initio} QED-CC and quantum-optics models, we illustrate how the intermolecular forces interplay with collective coupling in cooperative systems and induce considerable changes in the dynamic polarizability of the first solvation shell.
Such changes arise from a competition between the collective interaction with the optical mode and the local Coulomb interactions, which favors aligned (J-aggregate) or anti-aligned (H-aggregate) patterns in the transition moments.
Which effect prevails depends on the local chemical environment around the solute and the \textit{collective} coupling strength.
Cooperative coupling to a solvent thus induces strong modifications in the dynamic polarization of the solute's first solvation shell. 
A change in the dynamic polarization of the first solvation shell is then likely to affect the solute dynamics, possibly leading to changes in solvent rearrangement, chemical reactivity, nucleation, aggregation, and ionic conductivity.
Such a mechanism could be experimentally investigated using ultrafast UV electronic spectroscopy \cite{timmer2023plasmon, vasa2013real, wang2014interplay, vasa2010ultrafast}.
Moreover, our results suggest that these effects feature a clear resonance dependence and will be relevant even in the thermodynamic limit and for disordered systems.
Hence, our work provides an alternative perspective on cooperative strong coupling and the underlying mechanism that modifies chemistry via changes in the interaction within solvents and aggregates. 
Several VSC experiments have shown that the resonant coupling can induce modifications in solvent effects \cite{lather2019cavity, singh2023solvent, piejko2023solvent} and assembly \cite{joseph2021supramolecular, hirai2021selective, sandeep2022manipulating}, a feature proposed to emerge from collective coupling and structural reorganization rather than significant individual molecular changes~\cite{ebbesen2023direct}.
Longitudinal interactions are ubiquitous and fundamental in understanding chemical effects, and the concepts developed in this letter can help and inspire further discussions on both electronic and VSC. 
While we fixed the nuclear positions in this work, motivated by a separation of time scales, it is apparent that vibrational strong coupling will require the inclusion of nuclear motion. 
We can therefore expect that changes in the first solvation shell might result in the reorganization of the solvation structure.
These results also encourage the development and refinement of multiscale approaches 
\cite{luk2017multiscale, schafer2022emb} to extend our study to experimentally investigated systems.

\begin{acknowledgments}
We thank G\"oran Johansson for insightful discussions. 
C.S. acknowledges funding from the Horizon Europe research and innovation program of the European Union under the Marie Sk{\l}odowska-Curie grant agreement no.\ 101065117 and the Swedish Research Council (VR) through Grant No. 2016-06059.
M.C. and H.K. acknowledge funding from the European Research Council (ERC) under the European Union’s Horizon 2020 Research and Innovation Programme (grant agreement No. 101020016).
E.R. acknowledges funding from the European Research Council (ERC) under the European Union’s Horizon Europe Research and Innovation Programme (Grant n. ERC-StG-2021-101040197 - QED-SPIN).
Partially funded by the European Union. 
Views and opinions expressed are, however, those of the author(s) only and do not necessarily reflect those of the European Union or REA. 
Neither the European Union nor the granting authority can be held responsible.
\end{acknowledgments}

\bibliography{lit}

\end{document}


\title{Supplemental Information: Collective Strong Coupling Modifies Aggregation and Solvation}

\author{Matteo Castagnola}
\affiliation{Department of Chemistry, Norwegian University of Science and Technology, 7491 Trondheim, Norway}

\author{Tor S. Haugland}
\affiliation{Department of Chemistry, Norwegian University of Science and Technology, 7491 Trondheim, Norway}

\author{Enrico Ronca}
\affiliation{Dipartimento di Chimica, Biologia e Biotecnologie, Università degli Studi di Perugia, Via Elce di Sotto, 8,06123, Perugia, Italy}

\author{Henrik Koch}
\email[Electronic address:\;]{henrik.koch@ntnu.no}
\affiliation{Department of Chemistry, Norwegian University of Science and Technology, 7491 Trondheim, Norway}
\affiliation{Scuola Normale Superiore, Piazza dei Cavalieri 7, 56126 Pisa, Italy}

\author{Christian Sch\"afer}
\email[Electronic address:\;]{christian.schaefer.physics@gmail.com}
\affiliation{
Department of Physics, Chalmers University of Technology, 412 96 G\"oteborg, Sweden}
\affiliation{
Department of Microtechnology and Nanoscience, MC2, Chalmers University of Technology, 412 96 G\"oteborg, Sweden}
\maketitle

\section{Computational Details}\label{Sec:comput details}
\begin{figure}
    \centering
    \includegraphics[width=.45\textwidth]{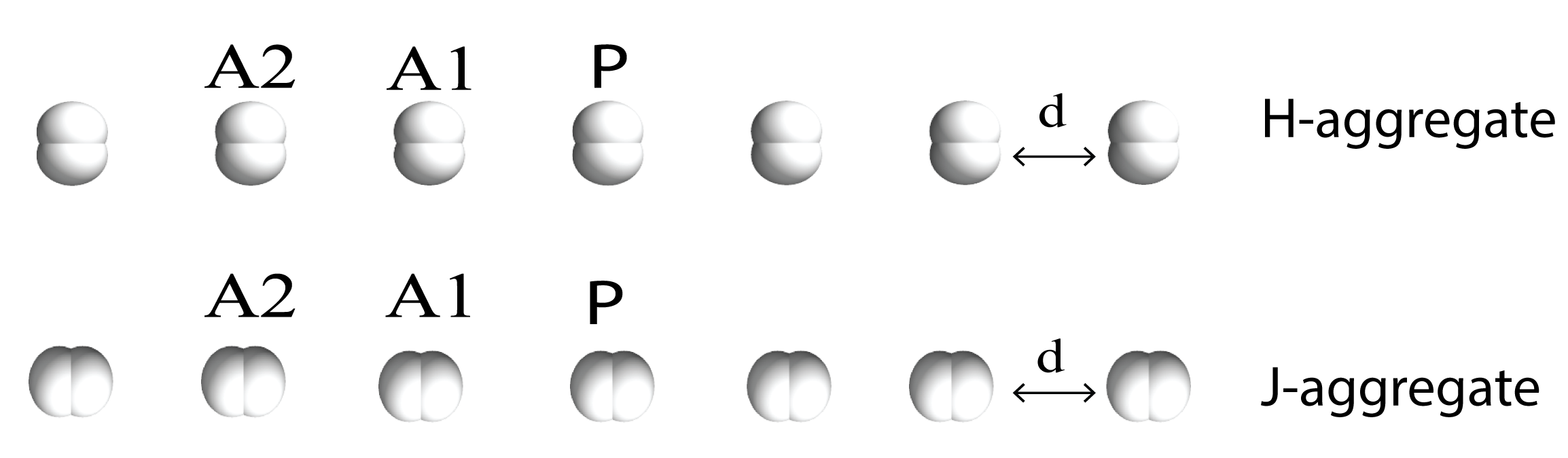}
    \caption{Configurations of the hydrogen aggregates.}
    \label{fig:Si structures}
\end{figure}

The model system we focus on in this letter is a chain of $N$ parallel equidistant molecules.
The dimers are arranged either in an H-aggregate parallel-bond or a J-aggregate head-tail configuration as depicted in  \autoref{fig:Si structures}, with constant intermolecular separation d.
The central dimer, labeled P, is slightly stretched with a bond length of 0.78\AA, while the other chain dimers have a bond length of 0.76\AA.
The chain lies in the XY plane, with the hydrogens aligned on X.
The first and second adjacent dimers of P are labeled A1 and A2.
The system is modeled at CCSD and QED-CCSD-2 level with the aug-cc-pVDZ basis set (see \ref{Sec:ccsd} for a brief description of the methods), and the light-matter coupling constant for the QED calculations is set to \qty{0.005}{\atomicunit} with polarization along X.
The stretched dimer has an (isolated and undressed) excitation energy of \qty{12.337}{\eV}, slightly lower than the unperturbed excitation of \qty{12.498}{\eV}.
The transition dipole moment is along the bonds (X-axis).
For each excitation, the transition densities $\varrho_{j0}(x,y,z)$ are computed and saved in a cube file with a grid of {0.1\AA} spacing.
The local transition densities along X $\rho^{I}_{j0}(x)$ (direction of the transition moment) are computed by numerically integrating the cube on Z and Y around the hydrogen $I=$ P, A1, A2 (within \qty{2.0}{\angstrom} from each atom)
\begin{equation}
    \rho^{I}_{j0}(x) = \int_{-\infty}^{+\infty}\diff z\int_{Y_I-\qty{2.0}{\angstrom}}^{Y_I+\qty{2.0}{\angstrom}}\diff y \;\varrho_{j0}(x,y,z)\quad I=\text{P, A1, A2}.
\end{equation}
All the calculations are computed on a local branch of the $e^\mathcal{T}$ program \cite{folkestad2020t}.
The Hamiltonian in the dipole approximation and second quantization is
\begin{align}
    H&=\sum_{pq}h_{pq}E_{pq}+\frac{1}{2}\sum_{pqrs}g_{pqrs}e_{pqrs}\nonumber\\
    &+\sqrt{\frac{\omega}{2}}(\bm{\lambda}\cdot \bm{d})_{pq}E_{pq}(b^\dagger+b)\\
    &+\frac{1}{2}\sum_{pqrs}(\bm{\lambda}\cdot \bm{d})_{pq}(\bm{\lambda}\cdot \bm{d})_{rs}E_{pq}E_{rs}\nonumber\\
    &+\omega\left({b}^\dagger{b}+\frac{1}{2}\right)
\end{align}
where $b^\dagger$ ($b$) is the creation (annihilation) operator for the photon mode of frequency $\omega$, $p$ and $q$ are indices for the one-electron orbitals, $E_{pq}$ and $e_{pqrs}$ are the standard one and two electron singlet excitation operators \cite{helgaker2014molecular}, $\bm{\lambda}$ is the light-matter coupling vector along the photon polarization, $h_{pq}$ and $g_{pqrs}$ are the one and two electron integrals, and $\bm d$ is the dipole operator.

\section{QED-CC}\label{Sec:ccsd}
The QED-CC ansatz is
\begin{align}
    \ket{\text{QED-CC}}&=\exp(T)\ket{\text{HF},0},
\end{align}
where $T$ is the cluster operator
\begin{align}
    T&=T_e+T_p+T_{int},
\end{align}
and $\ket{\text{HF},0}$ is the direct product of the QED-HF Slater determinant and the electromagnetic vacuum \cite{haugland2020coupled}.
The electronic cluster $T_e$ is an electronic excitation operator \cite{helgaker2014molecular}
\begin{gather}
    T_e=\sum_\mu t_\mu\tau_\mu\label{T_e_CC}\qquad
    \tau_\mu\ket{\text{HF}}=\ket{\mu},
\end{gather}
where $\ket{\mu}$ is an excited electronic determinant.
The cluster operator $T_p$ includes pure photonic excitations
\begin{gather}\label{Tp}
    T_p=\sum_{{n}=1}^{\bar{n}}\Gamma_{{n}}=\sum_{{n}=1}^{\bar{n}}\gamma_{{n}}(b^\dagger)^n,
\end{gather}
where we considered a single cavity mode. 
Notice that the bosonic space must be truncated up to a finite number ${\bar{n}}$ of photon occupations.
Finally, the operator $T_{int}$ contains excitations of both matter and photons
\begin{gather}\label{T_ep_QEDCC}
    T_{int}=\sum_{{n}=1}^{\bar{n}}(S_1^{{n}}+S_2^{{n}}+\dots+S^{{n}}_{N_e}),
\end{gather}
where for instance
\begin{align}
    S_1^{{n}}&=\sum_{ai}s_{ai}^{{n}}E_{ai}(b^\dagger)^{n}\\
    S_2^{{n}}&=\frac{1}{2}\sum_{aibj}s_{aibj}^{{n}}E_{ai}E_{bj}(b^\dagger)^{n}.
\end{align}
Notice that QED-CC is defined for the Hamiltonian transformed with the QED-HF coherent-state transformation $U_{\text{QED-HF}}$\cite{haugland2020coupled}
\begin{align}\label{QEDHF-transf-Ham}
    U^\dagger_{\text{QED-HF}}H&\,U_{\text{QED-HF}}=\nonumber\\
    &\sum_{pq}h_{pq}E_{pq}+\frac{1}{2}\sum_{pqrs}g_{pqrs}e_{pqrs}+h_{nuc}\nonumber\\
    +&\sum_\alpha\sqrt{\frac{\omega_\alpha}{2}}(\bm{\lambda}_\alpha\cdot (\bm{d}-\braket{\bm{d}}_{\text{QED-HF}}))(b^\dagger_\alpha+b_\alpha)\nonumber\\
   + &\frac{1}{2}\sum_\alpha(\bm{\lambda}_\alpha\cdot (\bm{d}-\braket{\bm{d}}_{\text{QED-HF}}))^2\nonumber\\
    +&\sum_{\alpha}\omega_{\alpha}{b}^\dagger_{\alpha}{b}_{\alpha}
\end{align}
\begin{align}
    U_{\text{QED-HF}} &= \prod_\alpha \text{exp}\bigg(-\frac{\bm{\lambda}_\alpha\cdot \braket{\bm{d}}_{\text{QED-HF}}}{\sqrt{2\omega_\alpha}}(b^\dagger_\alpha - b_\alpha)\bigg)\label{coherent-transf},
\end{align}
where $\braket{\bm{d}}_{\text{QED-HF}}$ is the mean QED-HF dipole moment.
The QED-HF orbitals are obtained by minimizing the energy of the transformed Hamiltonian in \autoref{QEDHF-transf-Ham}, with respect to the orbital coefficients.
This leads to a modified Fock matrix such that Brillouin's theorem $F_{ia}=0$ determines the coefficients
\begin{align}
    &F_{pq}=F_{pq}^e+\nonumber\\
    &\frac{1}{2}\bigg(\sum_a(\bm{\lambda}\cdot\bm{d}_{pa})(\bm{\lambda}\cdot\bm{d}_{aq})-\sum_i (\bm{\lambda}\cdot\bm{d}_{pi})(\bm{\lambda}\cdot\bm{d}_{iq})\bigg),
\end{align}
where $F^e_{pq}$ is the standard HF matrix \cite{helgaker2014molecular}, and $a$, $i$ refer to virtual and occupied orbitals.
The QED-CC state is then obtained by projection of \autoref{QEDHF-transf-Ham} onto the space spanned by the electron-photon excitations $\ket{\mu,{{n}}}$ \cite{haugland2020coupled}
\begin{align}
    &\braket{\mu,{n}|e^{-T}He^{T}|\text{HF},0}=0\\
    &\braket{\text{HF},0|H|\text{QED-CC}}=E_{\text{QED-CC}}.
\end{align}
The QED-CC dual state is defined as 
\begin{equation}
    \bra{\Lambda}=\bra{\text{HF},0}+\sum\bar{t}_{\mu {\bm{n}}}\bra{\mu,{{n}}}e^{-T},
\end{equation}
where $\bar{t}_{\mu {n}}$ are the Lagrangian multipliers \cite{haugland2020coupled, helgaker2014molecular} obtained by the equation 
\begin{equation}
    \braket{\Lambda|[H,\tau_\mu (b^\dagger)^n]|\text{QED-CC}} = 0.
\end{equation}
QED-CC is based on a hierarchy of approximations defined by a truncation of the cluster operators, such that the equations are projected only onto a subspace of the full Fock space \cite{helgaker2014molecular, haugland2020coupled}.
In this letter, we make use of the QED-CCSD-2 method, in which the electronic cluster includes only single and double excitations
\begin{equation}
    T_{e}= \sum_{ai}t_{ai}E_{ai} + \frac{1}{2}\sum_{aibj}t_{aibj}E_{ai}E_{bj},
\end{equation}
and the photonic operator includes single and double photonic creation operators
\begin{equation}
    T_p = \gamma_1 b^\dagger + \gamma_2 (b^\dagger)^2.
\end{equation}
The interaction cluster $T_{int}$ includes all the products between the operators in $T_p$ and $T_e$.
The excited states are defined in the equation of motion (EOM) framework via a CI-like parametrization, which includes the same states as in the cluster operator
\begin{align}
    \ket{\mathcal{R}}&=e^{T}\sum_{\mu,n}c_{\mu,n}\ket{\mu,n}\\
    \bra{\mathcal{L}}&=e^{-T}\sum_{\mu,n}\tilde{c}_{\mu,n}\bra{\mu,n}.
\end{align}
The left and right excited state coefficients are therefore obtained by diagonalization of the non-hermitian CC-transformed Hamiltonian \cite{helgaker2014molecular, haugland2020coupled}
\begin{equation}\label{eq: CC transf hamiltonian}
    \bar H = e^{-T}He^T = \begin{pmatrix}
        E_{\text{QED-CC}} & \eta_{\mu,n}\\
        0&A_{\mu {{n}}, \nu {m}}
    \end{pmatrix}
\end{equation}
where
\begin{align}
    \eta_{\mu{{n}}}&=\braket{\text{HF},0|e^{-T}{H}e^{T}|\mu,{{n}}}\\
    A_{\mu {{n}}, \nu {m}}&= \braket{\mu,{{n}}|[e^{-T}{H}e^{T},\tau_\nu(b^\dagger)^{m}]|\text{HF},0}.
\end{align}
From the left $\bra{\mathcal{L}_k}$ and right $\ket{\mathcal{R}_k}$ eigenvectors of \autoref{eq: CC transf hamiltonian} we define the transition and state CC densities
\begin{align}
    D^k_{pq} &= \braket{\mathcal{L}_k|E_{pq}|\mathcal{R}_k}\\
    D^{0k}_{pq} &= \braket{\Lambda|E_{pq}|\mathcal{R}_k}\\
    D^{k0}_{pq} &= \braket{\mathcal{L}_k|E_{pq}|\text{QED-CC}},
\end{align}
from which we can compute excited and transition properties such as the oscillator strengths.
Notice that since the CC theory is not hermitian, the left and right transition densities are not adjoints.

\section{Simple models}\label{Sec:TC}
\subsection{Jaynes-Cummings model}
The coupling of an optical mode with a single excitation is often described by the Jaynes-Cummings Hamiltonian $H_{JC}$, where the molecule is described as a two-level system coupled to the photon by an electric dipole interaction in the rotating wave approximation \cite{jaynes1963}
\begin{equation}\label{Eq: JC}
    H_{JC} = \omega_{mol}\ket{e}\bra{e} + \tilde g_0 \,\big(\ket{g}\bra{e}\, b^\dagger + \ket{e}\bra{g}\, b\big) + \omega_{ph}b^\dagger b.
\end{equation}
In \autoref{Eq: JC}, $b^\dagger$ ($b$) is the creation (annihilation) operator for the photons, $\ket{g}$ and $\ket{e}$ are the ground and excited molecular states, with energy difference $\omega_{mol}$ (the ground state energy is arbitrarily set to zero), and $\tilde g_0$ is the coupling strength
\begin{equation}
    \tilde g_0 = \sqrt{\frac{\omega_{ph}}{2}} \bm d_{ge}\cdot \bm{\lambda},
\end{equation}
where $\bm{\lambda}$ is the fundamental light-matter coupling strength directed along the polarization of the field and $\bm d_{ge}$ is the transition dipole moment.
For the Hamiltonian in \autoref{Eq: JC}, because of the rotating wave approximation, the excitation number operator $\hat \eta$
\begin{equation}
    \hat \eta = \ket{e}\bra{e} + \sum_n \ket{n}\bra{n},
\end{equation}
where $\ket{n}$ is the electromagnetic n-photon number state, commutes with $H_{JC}$ and therefore the Hamiltonian is block diagonal in the subspaces $\{\ket{e}\ket{n-1},\ket{g}\ket{n}\}$.
We then focus on the "vacuum Rabi splitting," with the subspace containing at most one photon excitation (excitation number $\eta=1$).
Then, in the resonant case $\omega_{mol}=\omega_{ph}$, the eigenstates of $H_{JC}$ are the cavity polaritons arising from the symmetric and antisymmetric combinations of the photon and the molecular excitation
\begin{equation}
    \ket{\pm} = \frac{1}{\sqrt{2}}\big(\ket{e}\ket{0} \pm \ket{g}\ket{1} \big),
\end{equation}
where $\ket{1}$ and $\ket{0}$ are respectively the one-photon and vacuum electromagnetic states.
The upper and lower polaritons are separated by the vacuum Rabi splitting $\Omega$
\begin{equation}
    \Omega = 2 \tilde g_0,
\end{equation}
while the ground state is the unchanged molecular ground state in the electromagnetic vacuum $\ket{g}\ket{0}$.
In the general subspace of excitation number $\eta=\tilde \eta$, the Rabi splitting scales as $\sqrt{\Tilde{\eta}}$.

\subsection{Tavis-Cummings model}
In the case of $N$ identical molecules coupled to the same photonic mode, the TC or Dicke Hamiltonian reads\cite{tavis1968exact, dicke1954coherence}
\begin{align}
    H_{TC} =& \omega_{mol}\sum_{k=1}^{N}\ket{e_k}\bra{e_k}+ \omega_{ph}b^\dagger b\nonumber \\
    +& \tilde g_0 \sum_{k=1}^{N}\big(\ket{g_k}\bra{e_k}\, b^\dagger + \ket{e_k}\bra{g_k}\, b\big) ,
\end{align}
where the molecules are assumed to have the same excitation energy $\omega_{mol}$, the same coupling strength $\tilde g_0$, and the index $k$ runs over the different replicas.
We now assume the system to be resonant $\omega_{ph}=\omega_{mol}=\omega$, and the Hamiltonian in the first excited manifold is
\begin{equation}\label{Eq: H_TC matrix}
    \begin{pmatrix}
        \omega & \tilde g_0 & \tilde g_0 &\tilde g_0 & \cdots\\
        \tilde g_0 & \omega & 0 & 0 & \cdots\\
        \tilde g_0 & 0 & \omega& 0 & \cdots\\
        \vdots & \vdots & \ddots& \ddots & \ddots
    \end{pmatrix}.
\end{equation}
The system is more easily diagonalized by using the collective state basis for matter
\begin{align}
    \ket{G} &= \ket{g_1, g_2, g_3, \dots }\\
    \ket{E} & = \frac{1}{\sqrt{N}}\sum_{k=1}^N\ket{e_k, g_{j\neq k}}\\
    \ket{D_j} & =\frac{1}{\sqrt{N}}\sum_{k=1}^Ne^{i \frac{2\pi}{N}j}\ket{e_k, g_{j\neq k}}\quad j =1,\dots,  N-1.
\end{align}
The states $\ket{D_j}\ket{0}$ are eigenvectors of $H_{TC}$ with eigenvalues $\omega$ and are referred to as \textit{dark states} since they have no contributions from the excited photon field.
The upper and lower polaritons $\ket{\pm}$ are
\begin{equation}
    \ket{\pm}= \frac{1}{\sqrt{2}}\bigg(\ket{G}\ket{1}\pm\ket{E}\ket{0}\bigg)
\end{equation}
and have contributions from both the photon field and the ${N}$ molecules, and are separated by a Rabi splitting $\Omega_{TC}$
\begin{equation}
    \Omega_{TC} = \sqrt{{N}} \Omega,
\end{equation}
that is, the Rabi splitting scales as the squared root of the number of coupled replicas.
Notice that the sign of the transition dipole moment is arbitrary.
If we flip the sign for one of the replicas, which we can assume without loss of generality to be the first one, the Hamiltonian reads
\begin{equation}\label{Eq: H_TC matrix flip}
    \begin{pmatrix}
        \omega & -\tilde g_0 & \tilde g_0 &\tilde g_0 & \cdots\\
        -\tilde g_0 & \omega & 0 & 0 & \cdots\\
        \tilde g_0 & 0 & \omega& 0 & \cdots\\
        \vdots & \vdots & \ddots& \ddots & \ddots
    \end{pmatrix}.
\end{equation}
The matrices in \autoref{Eq: H_TC matrix} and \autoref{Eq: H_TC matrix flip} are similarly equivalent, connected by the transformation
\begin{equation}
    \begin{pmatrix}
        1&0&0&0&\dots\\
        0&-1&0&0&\dots\\
        0&0&1&0&\dots\\
        \vdots&\ddots&\ddots&\ddots&\ddots
    \end{pmatrix}
\end{equation}
Therefore, the new polaritonic states have the coefficients of $\ket{e_1, g_2, g_3, \dots}$ with a flipped sign compared to the other representation.
This ensures that all of the transition dipoles constructively sum and not cancel with each other.
Therefore, in the polaritons, all of the transition moments must be aligned to maximize the electric-dipole interaction with the cavity field.
On the other hand, in the dark states, the phases of the molecular excitations are such that the transition moment sum vanishes.

\begin{figure*}[!ht]
    \centering
    \includegraphics[width=\textwidth]{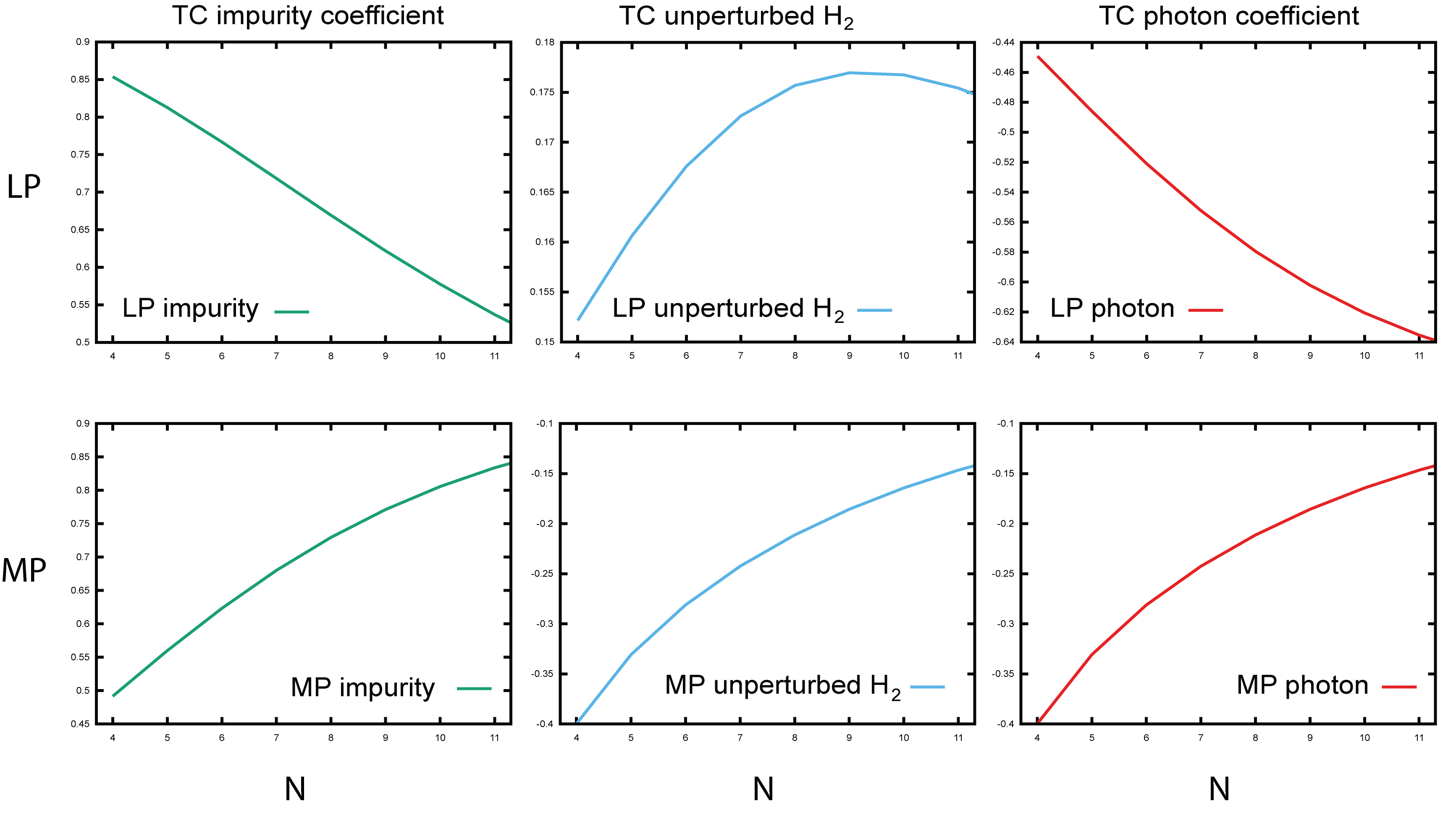}
    \caption{In the LP (upper panels), the perturbed and unperturbed H$_2$ excitations are in phase (same sign of the CI coefficients) while in antiphase with the photon contribution.
    The coefficient of the impurity decreases with ${N}$, and the photon character becomes more and more relevant, while the unperturbed dimer contribution reaches a maximum around $N=9$ and then decreases again.
    In the MP (lower panels), the perturbed dimers are in antiphase with the photon and the unperturbed H$_2$ molecules.
    The polariton becomes more localized on the impurity while the photon and the impurity contribution approach zero.}
    \label{fig:SI JC coeff}
\end{figure*}
\subsection{Disordered TC}
The molecular disorder affects both the molecular energies (inhomogeneous broadening) and the coupling strengths.
Following Houdr\'e et al.\cite{houdre1996vacuum}, the disordered TC Hamiltonian is
\begin{align}
    H_{DTC} &= \sum_{k=1}^N\omega_{k}\ket{e_k}\bra{e_k} \nonumber\\
    &+  \sum_{k=1}^N\tilde g_k\big(\ket{g_k}\bra{e_k}\, b^\dagger + \ket{e_k}\bra{g_k}\, b\big) + \omega_{ph}b^\dagger b,
\end{align}
and the polaritonic energies $E$ are defined by the equation
\begin{equation}
   \sum_{k=1}^N\frac{\tilde g_k^2}{(\omega_k-E)}-(\omega_{ph}-E)=0.
\end{equation}
Each molecule is then endowed with a different excitation energy (inhomogeneous broadening) and coupling strength (orientational and positional disorder).
If we disregard the inhomogeneous broadening of the excitations and set $\omega_k=\omega_{ph}=\omega$, the orientational and positional disorder only affects the coupling strengths $\tilde g_k$ via the scalar product $\bm\lambda\cdot\bm d_{ge}$ and the system have again $N-1$ degenerate states of energy $\omega$ and two polaritonic excitations
\begin{equation}\label{eq: disordered polaritons}
    \ket{\pm}= \frac{1}{\sqrt{2}}\bigg(\ket{G}\ket{0}\pm\sum_{k=1}^N\frac{\tilde g_k}{\sqrt{\sum_{i=1}^N\tilde g_i^2}}\ket{e_kg_{j\neq k}}\ket{1}\bigg)
\end{equation}
with a Rabi splitting
\begin{equation}
    \Omega_{DTC}=2\sqrt{\sum_{k=1}^N\tilde g_k^2}.
\end{equation}
Therefore, the Rabi splitting is determined by the \textit{quadratic average} of the different single-molecule couplings, and the contribution of each excited molecule to the polaritons is weighted by its coupling \cite{houdre1996vacuum}.
Notice that from \autoref{eq: disordered polaritons} we see that if all the molecules have the same module coupling $|\tilde g_k| \equiv \tilde g$ each contribution is weighted by $\tilde g_k / \tilde g = sgn(\tilde g_k)$, that is, it is weighted by the sign of the coupling strength so that all of the contributions from the transition moments add up.
The Rabi splitting is smaller than in the absence of disorder, but the overall polariton effects are the same.\\
If the inhomogeneous broadening of the molecular excitation is introduced such that
\begin{equation}
    \omega_k = \omega + \epsilon_k \qquad \epsilon_k \ll \omega,
\end{equation}
we can resort to a perturbative solution for the polaritonic energies as discussed in Ref. \cite{houdre1996vacuum}
\begin{equation}
    \omega_\pm = \omega \pm \sqrt{\sum_k\tilde g_k^2}-\frac{1}{2}\frac{{\sum_k\tilde \epsilon_k g_k^2}}{{\sum_k\tilde g_k^2}}+O(\epsilon^2).
\end{equation}
The remaining $N-1$ states are such that
\begin{equation}
    E_k \in [\omega_k,\omega_{k+1}]\quad \omega_k<\omega_{k+1},\quad k=1,\dots,N-1.
\end{equation}
Therefore, for the photon coupling to an inhomogeneously broadened set of oscillators, the TC predicts that the Rabi splitting is unchanged and, to first-order, the eigenstates are the same as in \autoref{eq: disordered polaritons}\cite{houdre1996vacuum}.

\subsection{TC analysis of the hydrogen chain}
For the $(\text{H}_2)_N$ system under study, the TC Hamiltonian reads
\begin{align}
    H'_{TC} =& \omega\sum_{k=1}^{N-1}\ket{e_k}\bra{e_k}+ \omega b^\dagger b + \nonumber \omega'\ket{e'}\bra{e'}\\
    +& \tilde g_0 \sum_{k=1}^{N-1}\big(\ket{g_k}\bra{e_k}\, b^\dagger + \ket{e_k}\bra{g_k}\, b\big) \nonumber\\
    +& \tilde {g_0}' \big(\ket{g'}\bra{e'}\, b^\dagger + \ket{e'}\bra{g'}\, b\big),
\end{align}
where the primed quantities refer to the stretched dimer and the photon is resonant with the unperturbed H$_2$ molecules.
The eigenstates of $H'_{TC}$ are obtained by diagonalizing the matrix
\begin{equation}\label{Eq: H'_TC matrix}
    \begin{pmatrix}
        \omega & \tilde {g_0}' & \tilde g_0 &\tilde g_0 & \cdots\\
        \tilde {g_0}' & \omega' & 0 & 0 & \cdots\\
        \tilde g_0 & 0 & \omega& 0 & \cdots\\
        \vdots & \vdots & \ddots& \ddots & \ddots
    \end{pmatrix}.
\end{equation}
In  \autoref{fig:SI JC coeff}, we plot the coefficient of the photon and H$_2$ excited states that contribute to the LP and MP from equation \autoref{Eq: H'_TC matrix} for $4\le N \le 11$.
Notice that different scales are applied to each panel.
In the LP, the perturbed and unperturbed H$_2$ excitations are in phase (same sign of the CI coefficients), and are in antiphase with the photon contribution.
The coefficient of the impurity decreases with N and the photon character becomes more and more relevant, while the unperturbed dimer contribution reaches a maximum around $N=9$ and then decreases again.
Notice that, since in the TC the intermolecular forces are completely disregarded, each unperturbed H$_2$ will give the same contribution to the polaritonic states.
The differences observed in the dimers A1 and A2 shown in the letter are due to the intermolecular forces.
In the MP, the perturbed dimers are in antiphase with the photon and the unperturbed H$_2$ molecules.
The polariton becomes more localized on the impurity while the photon and the impurity contribution approach zero.

\clearpage
\section{Hydrogen aggregates electronic structure properties (no QED)}
\subsection{H-aggregates}
\begin{figure}[!ht]
    \centering
    \includegraphics[width=.5\textwidth]{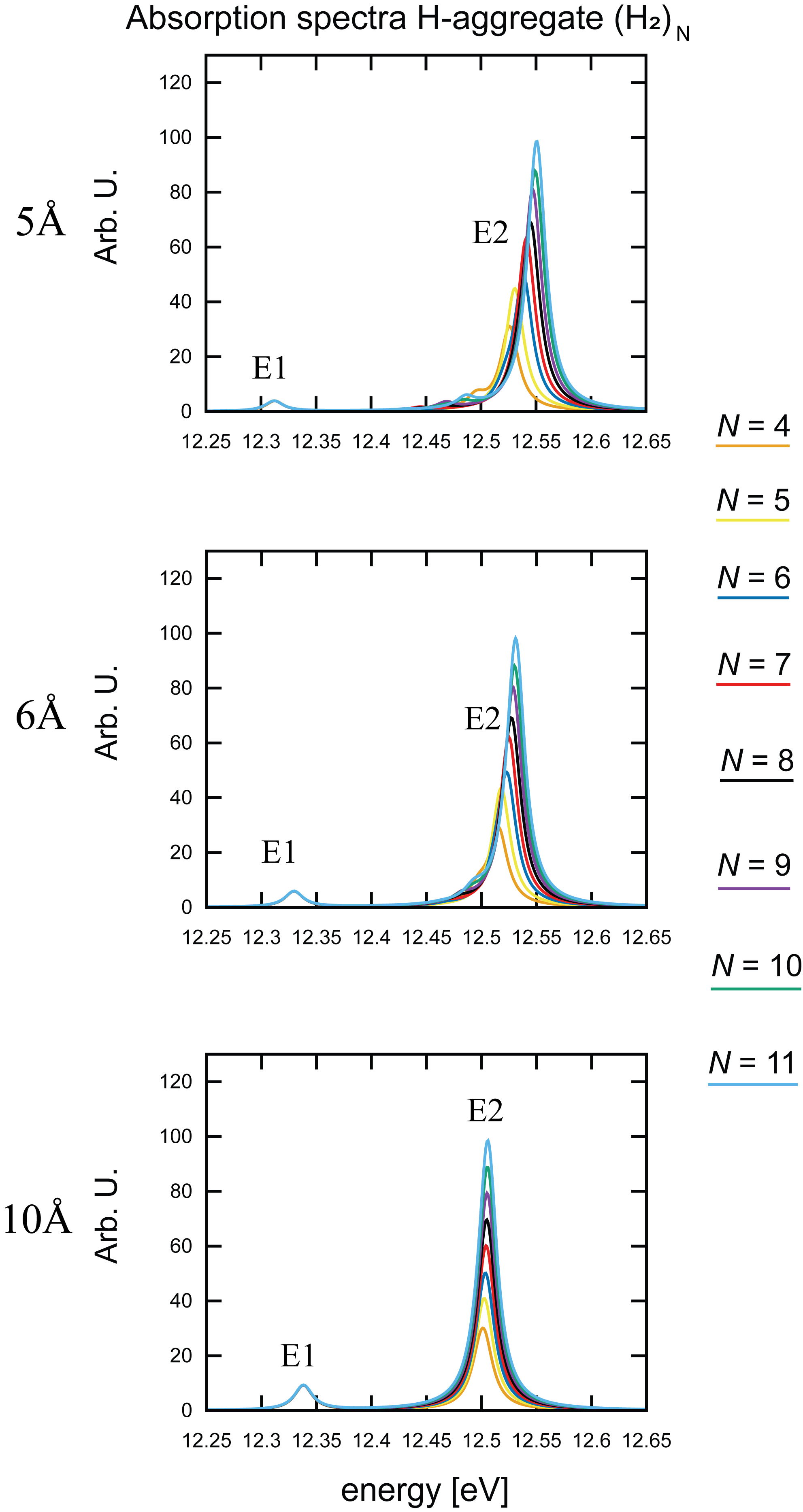}
    \caption{Absorption spectra for an H-aggregate $(\text{H}_2)_{N\ge4}$ for \qty{5}{\angstrom} (upper panel), \qty{6}{\angstrom} (central panel), and \qty{10}{\angstrom} (lower panel) intermolecular separation.
    The excitation E2 depends on the number of dimers in the chain, and it blueshifts with stronger intermolecular forces and larger $N$.
    The excitation E1 is mainly localized on P and its neighbor dimers and is almost independent of $N$, while it redshifts with the longitudinal forces.
    $N$-2 quasi-dark states gain oscillator strength due to the intermolecular couplings, showing smaller shoulder peaks near E2.
    }
    \label{fig:nocaviity spectra SI}
\end{figure}
In  \autoref{fig:nocaviity spectra SI}, we show the electronic (undressed) absorption spectra of the $(\text{H}_2)_{N\ge4}$ H-aggregate (see  \autoref{fig:Si structures}) for different intermolecular separations.
As the system enlarges, the upper excitation E2 is blueshifted due to the intermolecular forces; therefore, the shift is less significant for large intermolecular distances.
Notice that E2 gains oscillator strength with $N$, suggesting that all the unperturbed dimers contribute to this excitation.
The excitation E1 is almost independent of the chain length, and it redshifts and loses intensity for larger longitudinal couplings (shorter intermolecular separations).
Therefore, this suggests that E1 is mainly associated with the perturbed dimers P, and the nearest dimers produce an effective screening of its transition dipole moment.
The system also possesses $N-2$ quasi-dark states, which gain some oscillator strength due to the intermolecular forces.
The contribution of each dimer to E1 and E2 can be extracted from the transition densities integrated around each H$_2$ as explained in \autoref{Sec:comput details} and \autoref{Sec:ccsd}.

\begin{figure*}[!ht]
    \centering
    \includegraphics[width=\textwidth]{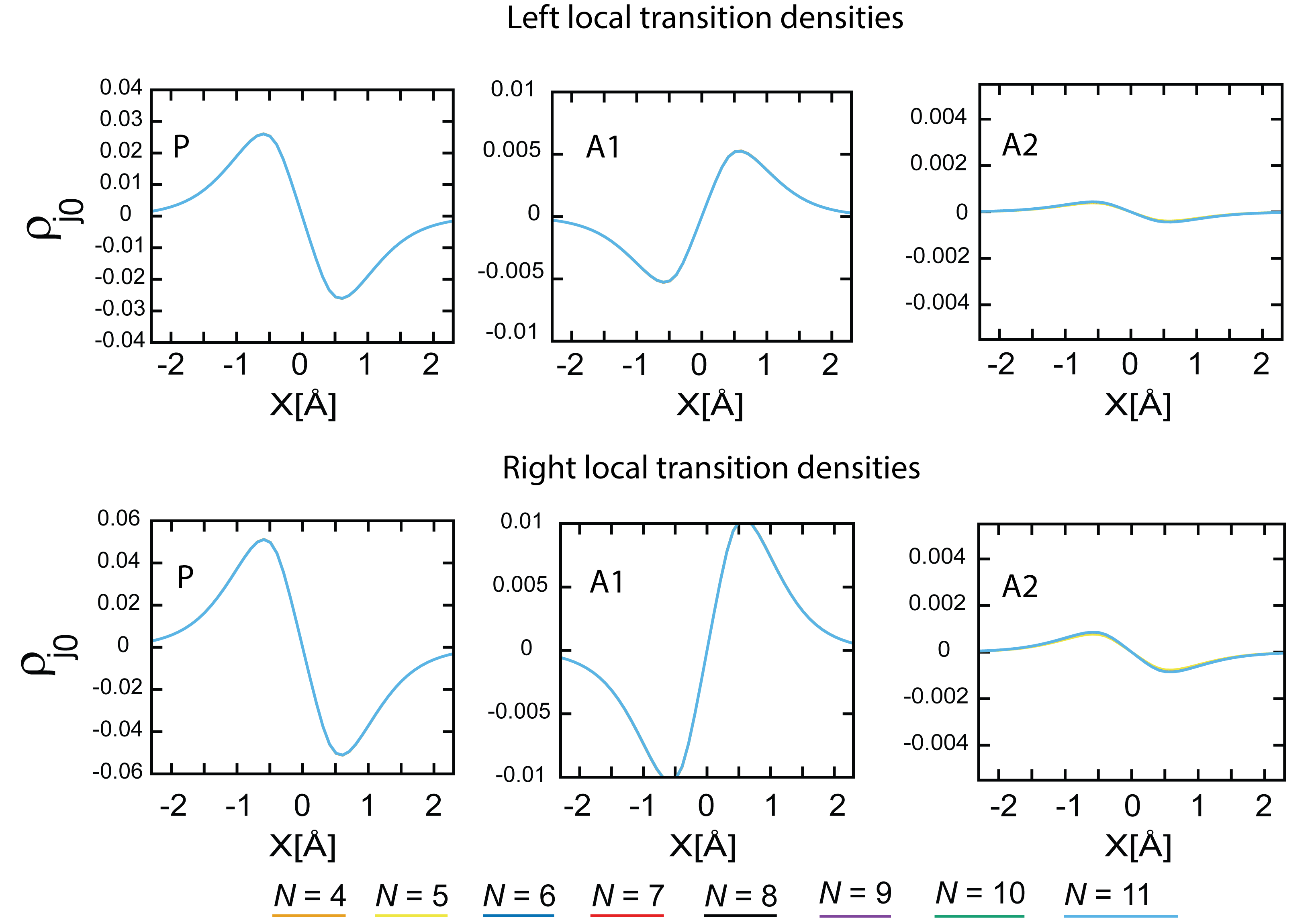}
    \caption{Local left and right transition densities for E1 for an intermolecular separation of \qty{5}{\angstrom} in an H-aggregate configuration (parallel-bond arrangement, see  \autoref{fig:Si structures}).
    The excitation is independent of the chain length $N$ and mainly localized on P, A1, and A2 with decreasing contributions.
    The local transition dipole moments are anti-aligned, producing an effective screening of the excitation compared to the isolated P.
    }
    \label{fig: SI nocavity E1 5}
\end{figure*}

\begin{figure*}[!ht]
    \centering
    \includegraphics[width=\textwidth]{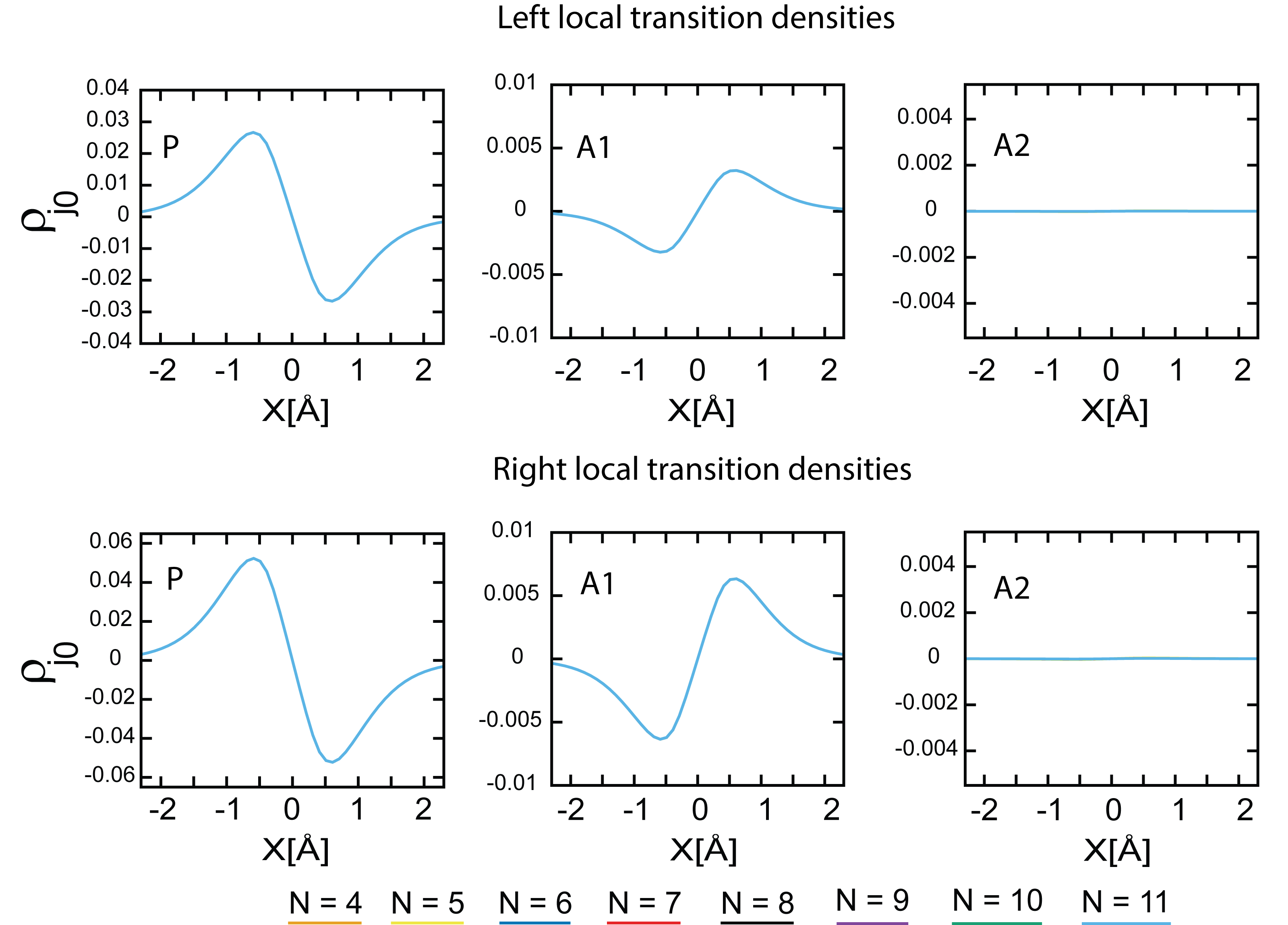}
    \caption{Local left and right transition densities for E1 for an intermolecular separation of {\qty{6}{\angstrom}} in an H-aggregate configuration (parallel-bond arrangement, see  \autoref{fig:Si structures}).
    The excitation is almost independent of $N$ and mainly localized on P, A1, and A2 with decreasing contributions.
    Compared to  \autoref{fig: SI nocavity E1 5}, the intermolecular forces are less relevant, and therefore, the excitation is more localized on P and less on A1 and A2. Still, the overall anti-alignment pattern of the molecular transition dipoles is maintained.
    }
    \label{fig: SI nocavity E1 6}
\end{figure*}

In  \autoref{fig: SI nocavity E1 5} and \autoref{fig: SI nocavity E1 6}, we show the left and right local transition densities of E1 for P, A1, and A2 for \qtylist{5;6}{\angstrom} intermolecular separation, respectively.
As for the absorption spectra of  \autoref{fig:nocaviity spectra SI}, the local densities for E1 are independent of $N$ and are mainly localized on P.
Due to the intermolecular forces, an excitation of P induces a response on its neighbor dimers, with a less important effect for more distant dimers.
As the intermolecular distances become larger, the excitation becomes more localized on P and less on the other dimers, as expected when decreasing the intermolecular couplings.
Adjacent molecular transition dipoles are anti-aligned, partially screening the excitation, resulting in smaller oscillator strength than the isolated P.
Notice that for A2, we only plot the local densities for $N\ge 6$ since for $N=4,5$, its "first solvation shell" is incomplete. Therefore, its properties have significant variations from $N=4$ to $6$ due to local intermolecular forces.

\begin{figure*}[!ht]
    \centering
    \includegraphics[width=\textwidth]{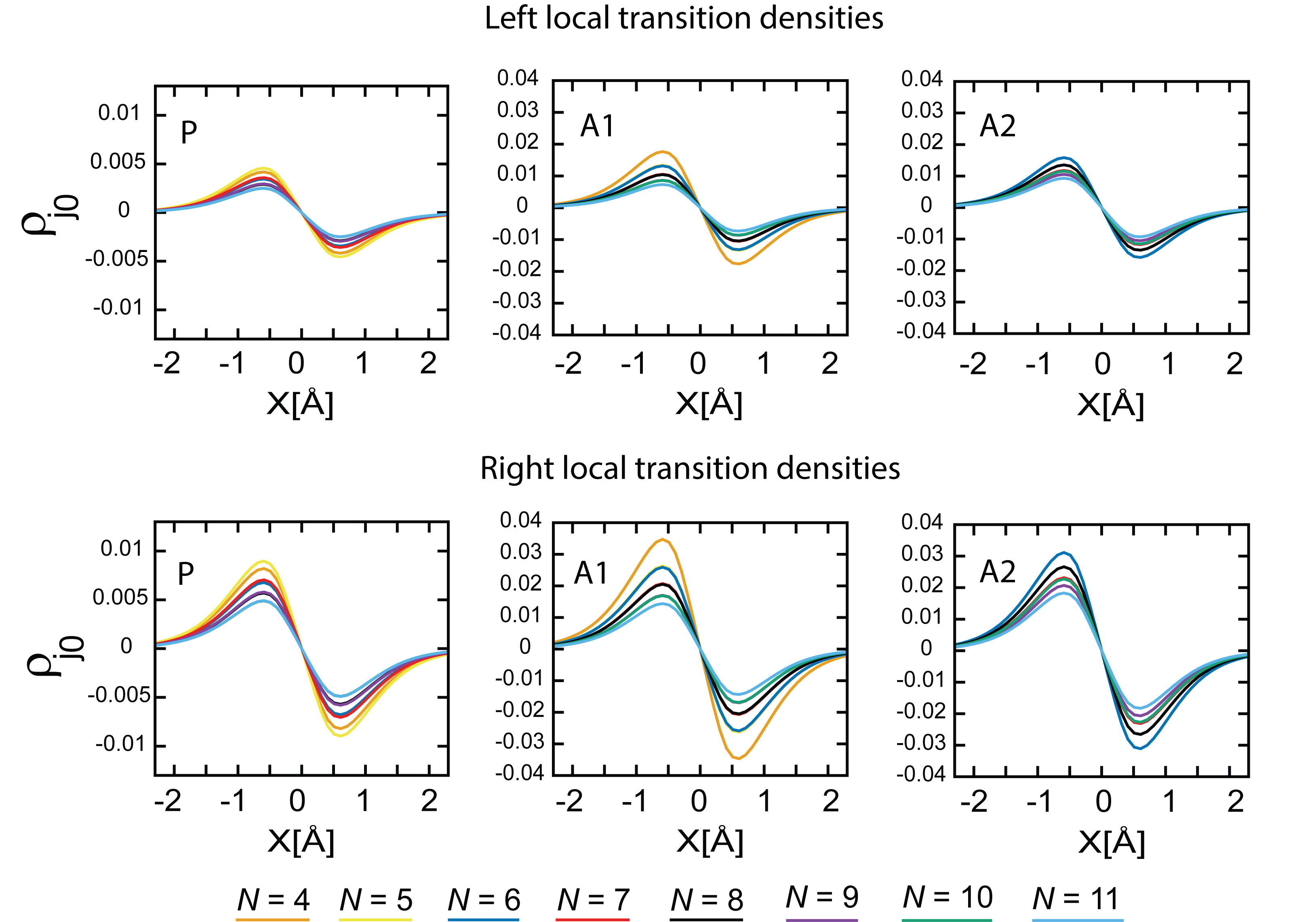}
    \caption{Local left and right transition densities for E2 for an intermolecular separation of \qty{5}{\angstrom} in an H-aggregate configuration (parallel-bond arrangement, see  \autoref{fig:Si structures}).
    The excitation blueshifts and gains oscillator strength with the chain length $N$ and is delocalized over the whole aggregate with a larger contribution from the unperturbed molecules.
    The local transition dipole moments are all aligned.
    }
    \label{fig: SI nocavity E2 5}
\end{figure*}

\begin{figure*}[!ht]
    \centering
    \includegraphics[width=\textwidth]{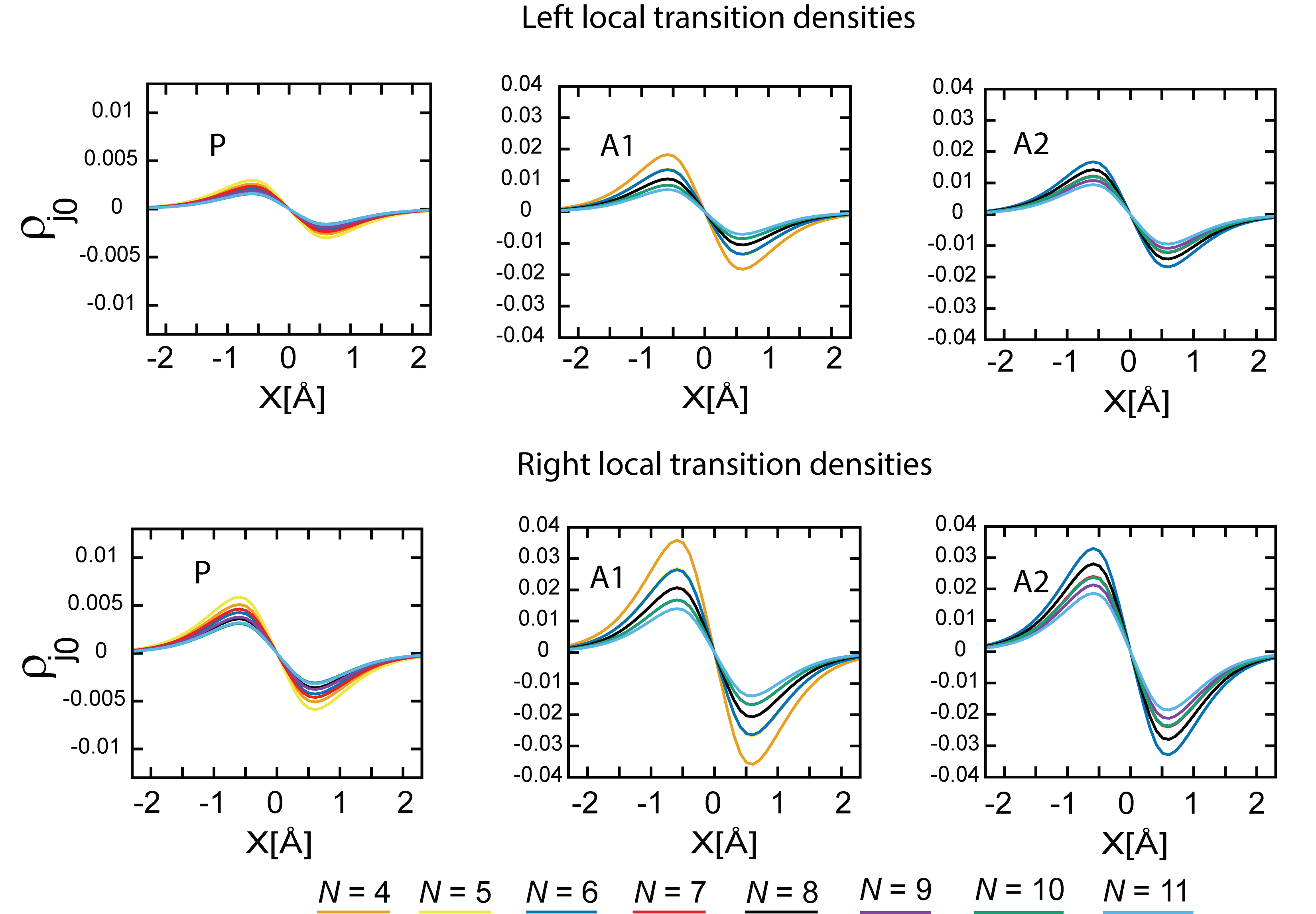}
    \caption{Local left and right transition densities for E2 for an intermolecular separation of \qtylist{6}{\angstrom} in an H-aggregate configuration (parallel-bond arrangement, see  \autoref{fig:Si structures}).
    The excitation blueshifts and gains oscillator strength with the chain length $N$. It is delocalized over the whole aggregate, with a more significant contribution from the unperturbed molecules.
    The local transition dipole moments are all aligned, and compared to  \autoref{fig: SI nocavity E2 5}, there is a smaller contribution from P and a similar contribution from the rest of the chain.
    For larger distances, that is, smaller intermolecular forces, the blueshift is less pronounced.
    }
    \label{fig: SI nocavity E2 6}
\end{figure*}

In  \autoref{fig: SI nocavity E2 5} and \autoref{fig: SI nocavity E2 6}, we show the left and right local transition densities of E2 for P, A1, and A2 for \qtylist{5;6}{\angstrom} intermolecular separation, respectively.
In this case, the local transition dipoles are all aligned, and all the dimers in the chain give a similar contribution.
Therefore, the properties of E2 strongly depend on the length of the whole chain and the intermolecular forces acting on each dimer. 
With larger intermolecular distances, the contribution of P to E2 decreases, while the other dimers maintain a similar local transition density.

\clearpage

\subsection{J-aggregates}
\begin{figure*}[!ht]
    \centering
    \includegraphics[width=\textwidth]{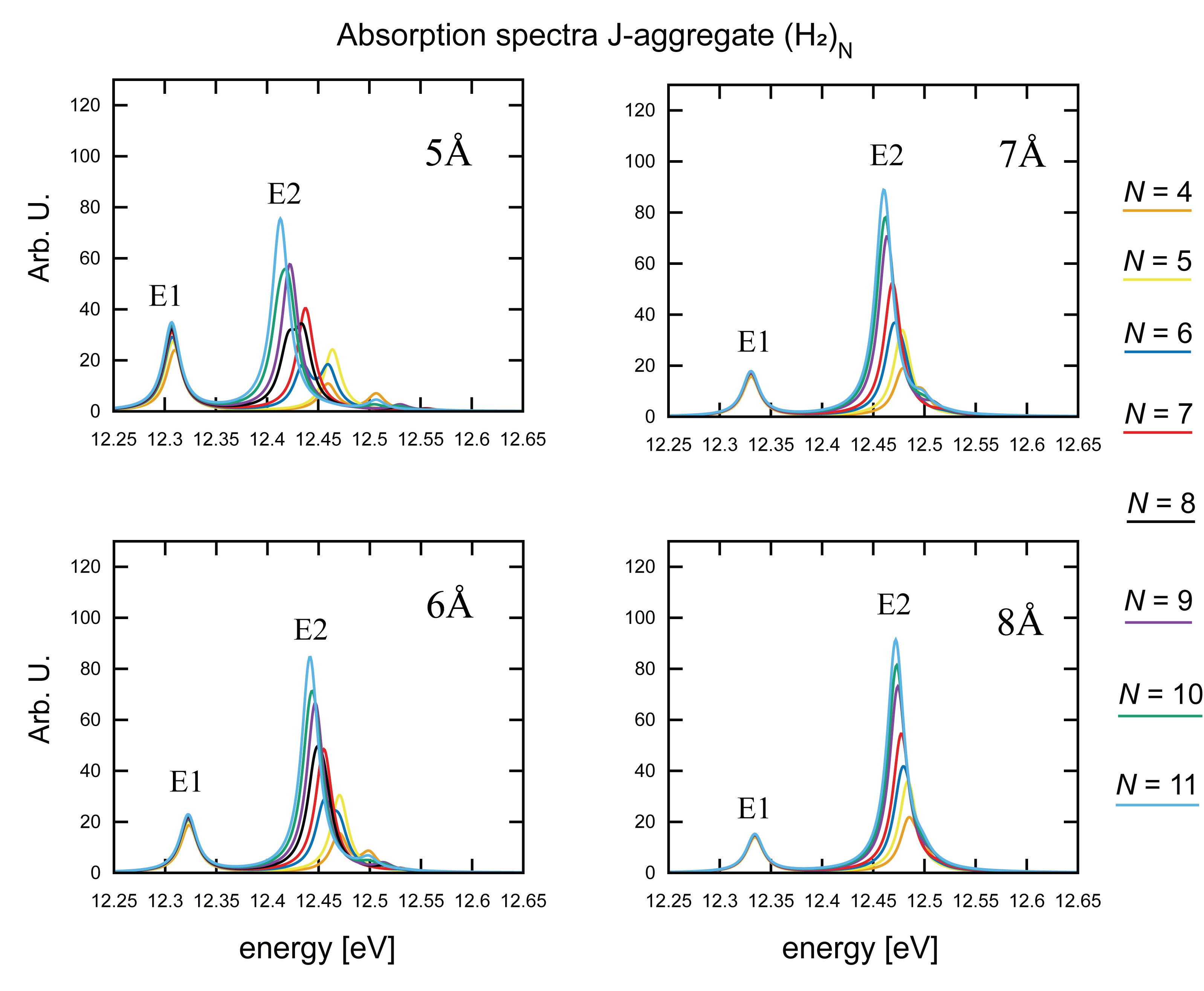}
    \caption{Absorption spectra for a head-tail aggregate (J-aggregate, see  \autoref{fig:Si structures}) $(H_2)_{N\ge4}$ for \qtylist{5}{\angstrom} (upper left panel), \qtylist{6}{\angstrom} (lower left panel), \qtylist{7}{\angstrom} (upper right panel), and \qtylist{8}{\angstrom} (lower right panel) intermolecular separation.
    The excitation E2 depends on the number of dimers and redshifts with stronger intermolecular forces and larger $N$.
    The excitation E1 is mainly localized on P and its neighbor dimers with little dependence $N$ and redshifts with the longitudinal forces.
    $N$-2 quasi-dark states gain oscillator strength due to the intermolecular couplings, showing peaks near E2.
    }
    \label{fig:nocaviity spectra Jaggregate SI}
\end{figure*}
\begin{figure*}[!ht]
    \centering
    \includegraphics[width=\textwidth]{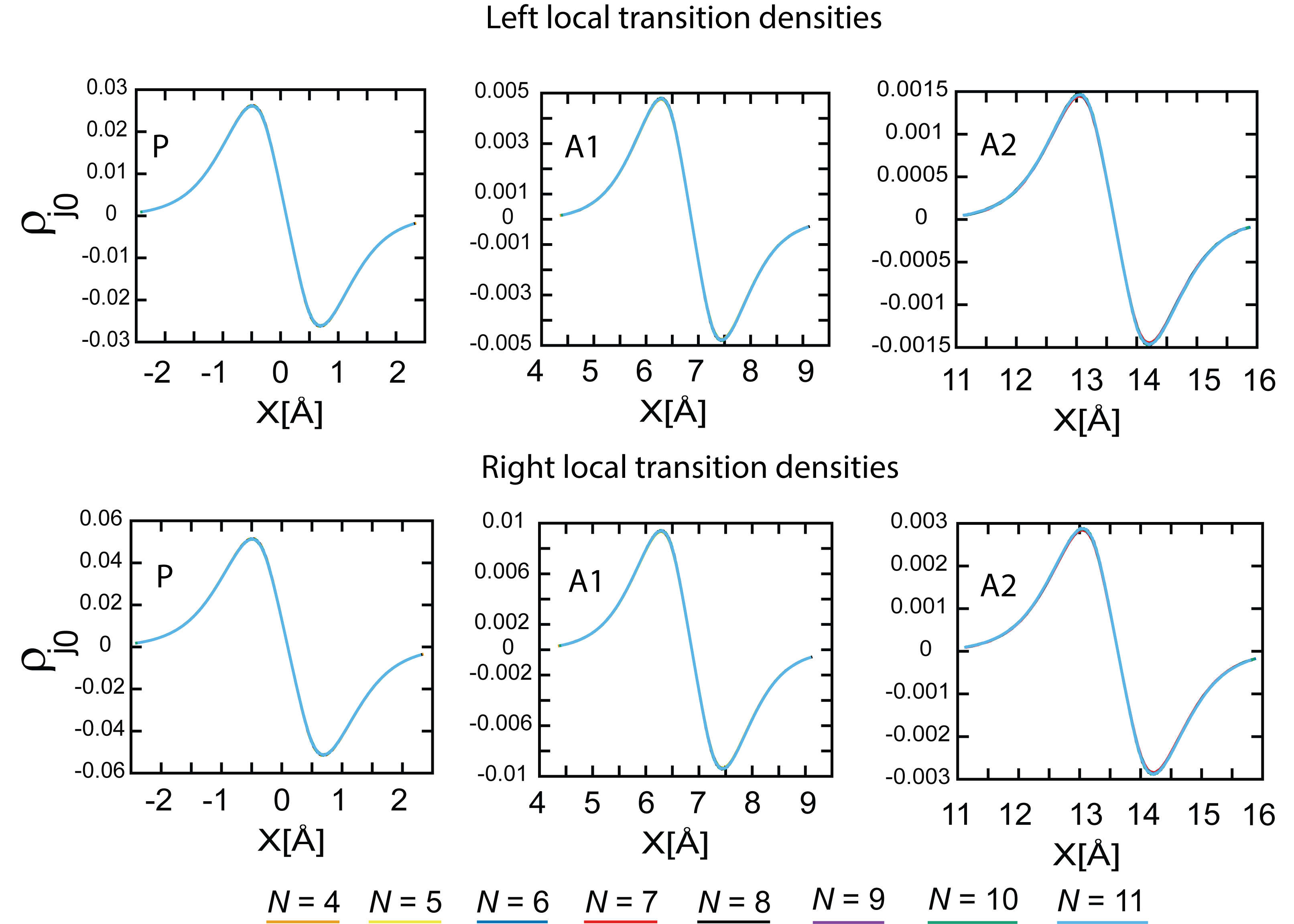}
    \caption{Local left and right transition densities for E1 for an intermolecular separation of \qtylist{6}{\angstrom} in a J-aggregate configuration (head-tail arrangement, see  \autoref{fig:Si structures}).
    The excitation is independent of the chain length $N$ and mainly localized on P, A1, and A2 with decreasing contributions.
    The local transition dipole moments are aligned, increasing the oscillator strength compared to the isolated P.
    }
    \label{fig: SI nocavity E1 6 J-aggregate}
\end{figure*}
\begin{figure*}[!ht]
    \centering
    \includegraphics[width=\textwidth]{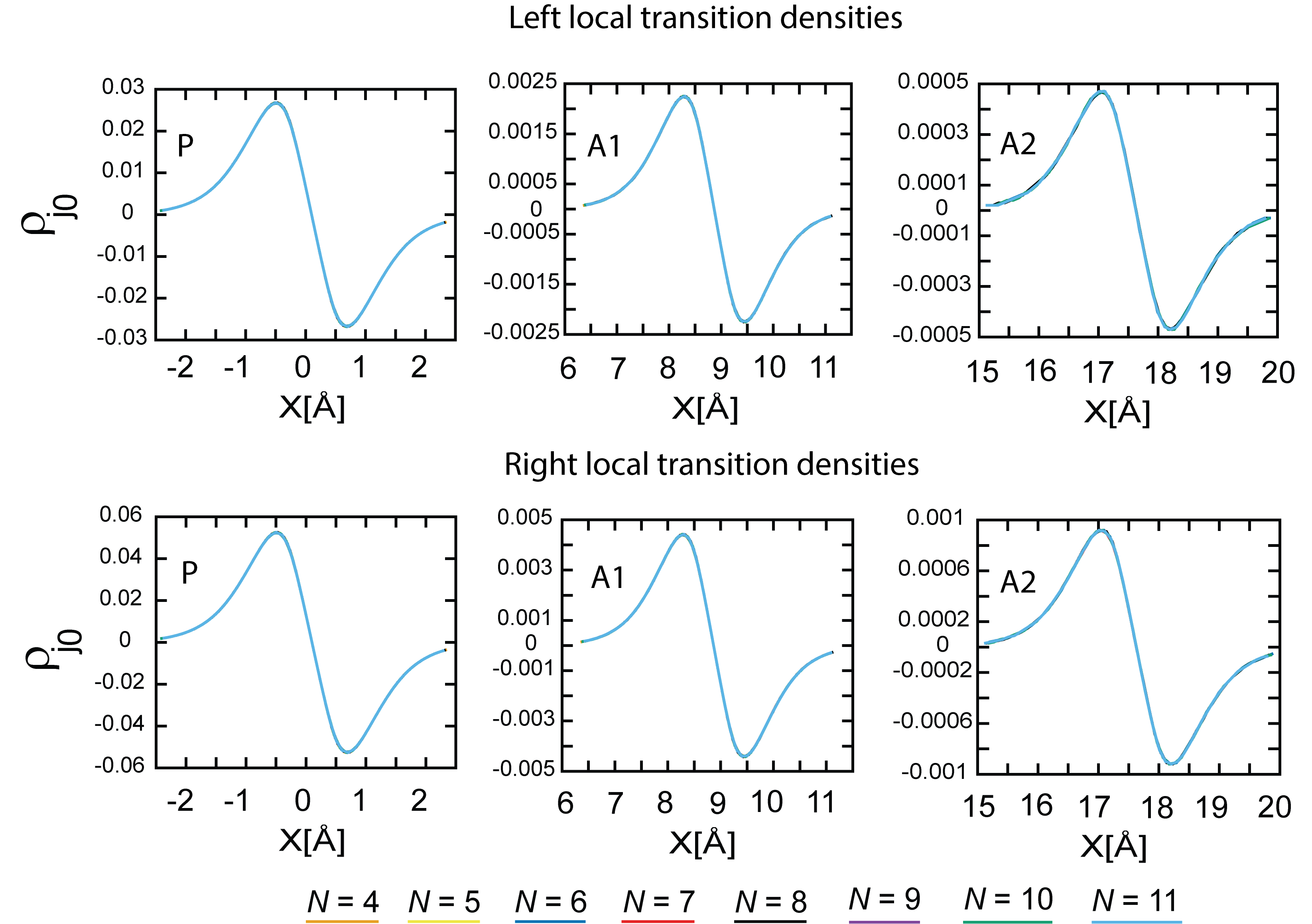}
    \caption{Local left and right transition densities for E1 for an intermolecular separation of \qtylist{8}{\angstrom} in a J-aggregate configuration (head-tail arrangement, see  \autoref{fig:Si structures}).
    The excitation is almost independent of $N$ and mainly localized on P, A1, and A2 with decreasing contributions.
    Compared to  \autoref{fig: SI nocavity E1 6 J-aggregate}, the intermolecular forces are less relevant, and therefore, the excitation is more localized on P and less on A1 and A2, but the overall alignment pattern of the molecular transition dipoles is maintained.
    }
    \label{fig: SI nocavity E1 8 J-aggregate}
\end{figure*}

Since the Coulomb forces are strongly directional, modifying the local arrangement severely affects the system's properties.
In  \autoref{fig:nocaviity spectra Jaggregate SI}, we show the absorption spectra for a head-tail configuration (J-aggregate, see  \autoref{fig:Si structures}) for different intermolecular distances.
Compared to the H-aggregate of  \autoref{fig:nocaviity spectra SI}, the effects of the intermolecular forces are more pronounced, as seen from the more complicated structure of the spectra for \qtylist{5;6}{\angstrom}.
For d = \qtylist{5;6}{\angstrom}, the $N$-2 quasi-dark state gained relevant oscillator strength, and the appreciable changes of E1 from $N$ = 4 to $N$ = 6 suggest that the dimer A2 and even father away give a significant contribution to E1.
For larger intermolecular separations, E1 is almost independent of $N$ as in  \autoref{fig:nocaviity spectra SI}.
The excitations E1 and E2 redshift for increasing chain length and decreasing intermolecular separation, while E2 blueshifts for the H-aggregate.
Unlike the H-aggregate case, E1 gains oscillator strength from embedding in the H$_2$ chain, while E2 loses intensity with increasing intermolecular forces.
Since the excitation energy E1 shows little dependence on the chain length, we deduce that it is still mainly localized on P, A1, and A2.
E2 becomes more intense with $N$, suggesting that all dimers contribute to this excitation.
The differences between the H- and J- aggregates result from the different effects of the local Coulomb (longitudinal) forces on the dimers.
In  \autoref{fig: SI nocavity E1 6 J-aggregate} and \autoref{fig: SI nocavity E1 8 J-aggregate}, we report the left and right local transition densities of E1 for P, A1, and A2 for \qtylist{6;8}{\angstrom} intermolecular separation, respectively.
Contrary to the H-aggregate arrangement, in this configuration, the local transition moments are all aligned, increasing the excitation's intensity.
The densities are almost independent of the chain length $N$, and the contribution of the neighbor dimers A1 and A2 increases with decreasing intermolecular separation.

\begin{figure*}[!ht]
    \centering
    \includegraphics[width=\textwidth]{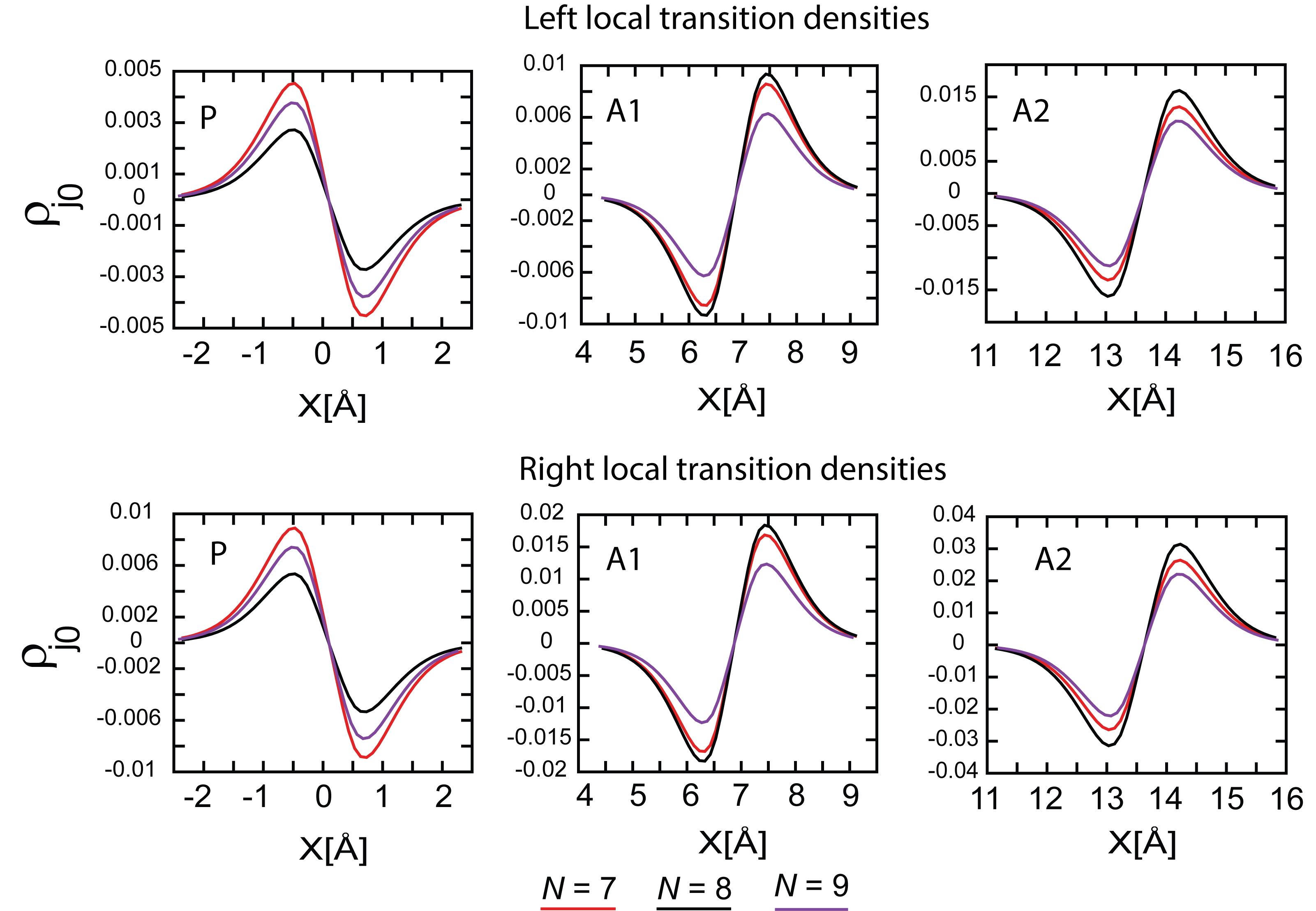}
    \caption{Local left and right transition densities for E2 for an intermolecular separation of \qtylist{6}{\angstrom} in a J-aggregate configuration (head-tail arrangement, see  \autoref{fig:Si structures}).
    The excitation delocalized over the whole chain with a larger contribution from the unperturbed dimers.
    The local transition dipole moments are anti-aligned, contrary to E2 for the H-aggregate configuration.
    }
    \label{fig: SI nocavity E2 5 J-aggregate 6}
\end{figure*}

\begin{figure*}[!ht]
    \centering
    \includegraphics[width=\textwidth]{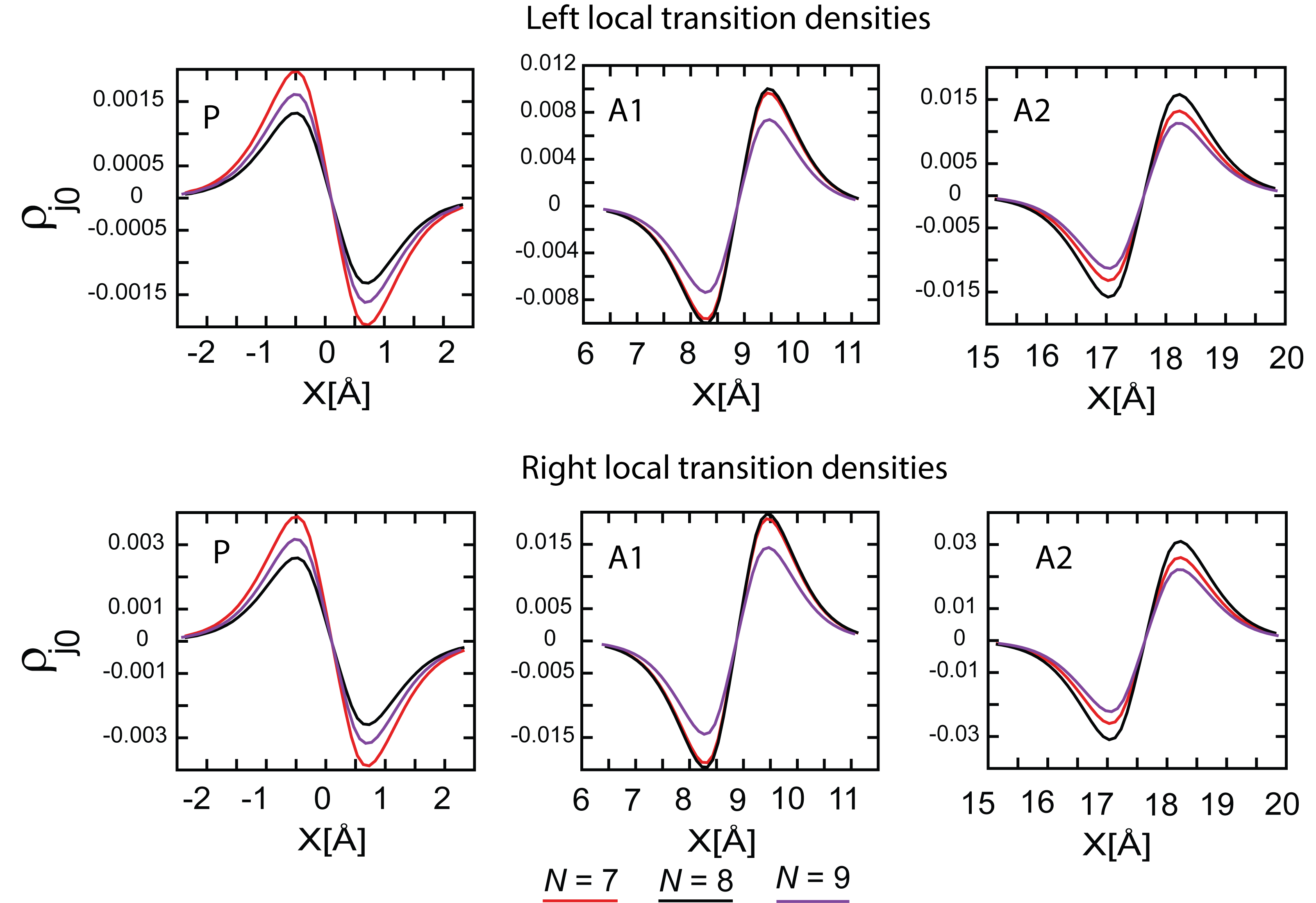}
    \caption{Local left and right transition densities for E2 for an intermolecular separation of \qtylist{8}{\angstrom} in a J-aggregate configuration (head-tail arrangement, see  \autoref{fig:Si structures}).
    The excitation delocalized over the whole chain with a larger contribution from the unperturbed dimers.
    The local transition dipole moments are anti-aligned, contrary to E2 for the H-aggregate configuration, and compared to  \autoref{fig: SI nocavity E2 5 J-aggregate 6} there is a smaller contribution from P and a similar contribution from the rest of the chain.
    }
    \label{fig: SI nocavity E2 5 J-aggregate 8}
\end{figure*}

In  \autoref{fig: SI nocavity E2 5 J-aggregate 6} and \autoref{fig: SI nocavity E2 5 J-aggregate 8}, we show the left and right local transition densities of E2 for P, A1, and A2 for \qtylist{6;8}{\angstrom} intermolecular separation, respectively.
In this configuration, the transition dipole moment of P is anti-aligned with the rest of the chain.
As for the H-aggregates, E2 is delocalized over the whole chain with a more relevant contribution from the unperturbed dimers, and the intermolecular forces tend to increase the contribution of P.
However, P's local transition dipole moment is anti-aligned with the rest of the hydrogens.
Therefore, the alignment pattern of P and A1 in E1 and E2 is swapped compared to the H-aggregate arrangement.
This reflects the severe consequences that intermolecular forces can have on the local properties of the system.

\clearpage

\subsection{Kasha-Frenkel model for H-and J-aggregates}
The spectral properties of the H- and J-aggregates can be qualitatively understood using the simple Kasha-Fernkel model \cite{kasha1965exciton, kasha1963energy, mcrae1958enhancement, kasha1950characterization,hestand2018expanded}.
The interplay between the excitation of different molecules is modeled by a dipole-dipole interaction $V_{AB}$ between their transition dipoles 
\begin{equation}\label{Eq: dipole-dipole coupling}
    V_{AB}=\frac{(\bm{\mu}_A\cdot\bm{\mu}_B)R_{AB}^2-3(\bm{\mu}_A\cdot\bm{R}_{AB})(\bm{\mu}_B\cdot\bm{R}_{AB})}{\epsilon R^5_{AB}},
\end{equation}
where $\epsilon$ is the medium dielectric constant, $A$ and $B$ label the two interacting molecules with isolated transition dipole moments $\bm{\mu}_A$ and $\bm{\mu}_A$ (usually associated with the $S_0\to S_1$ excitation), and $\bm{R}_{AB}$ is the vector, of length $R_{AB}$, connecting their center of mass.
The simple Frenkel exciton Hamiltonian, which describes the first-excitation manifold, is
\begin{equation}\label{Eq: Kasha hamiltonian}
    H_{ex}= E_{S_0\to S_1} + \sum_{m,n}V_{mn}\ket{m}\bra{n}
\end{equation}
where the states $\ket{n}$ correspond to the n-th molecule excited while the others are in the ground state, and $E_{S_0\to S_1}$ is the excitation energy of the molecular transition in the supramolecular aggregate, which includes an energy shift from the gas-phase excitation due to the molecular environment \cite{hestand2018expanded}.
Diagonalizing the Hamiltonian $H_{ex}$ for a fixed number N of molecules (or imposing periodic boundary conditions), we can reproduce the blueshift (redshift) of the excitation energy of the H-aggregates (J-aggregates).
The difference in the energy shift is given by the sign of the intermolecular coupling $V_{AB}$ in the different configurations, which is negative for J-aggregates and positive for H-aggregates.
Such a simple model qualitatively accounts for the delocalized nature of the excitation and for the photoluminescence properties of these aggregates \cite{mcrae1958enhancement, kasha1950characterization,hestand2018expanded}, and can be easily generalized including a different molecule (impurity) in the excitation manifold.
\begin{figure*}[!ht]
    \centering
    \includegraphics[width=.95\textwidth]{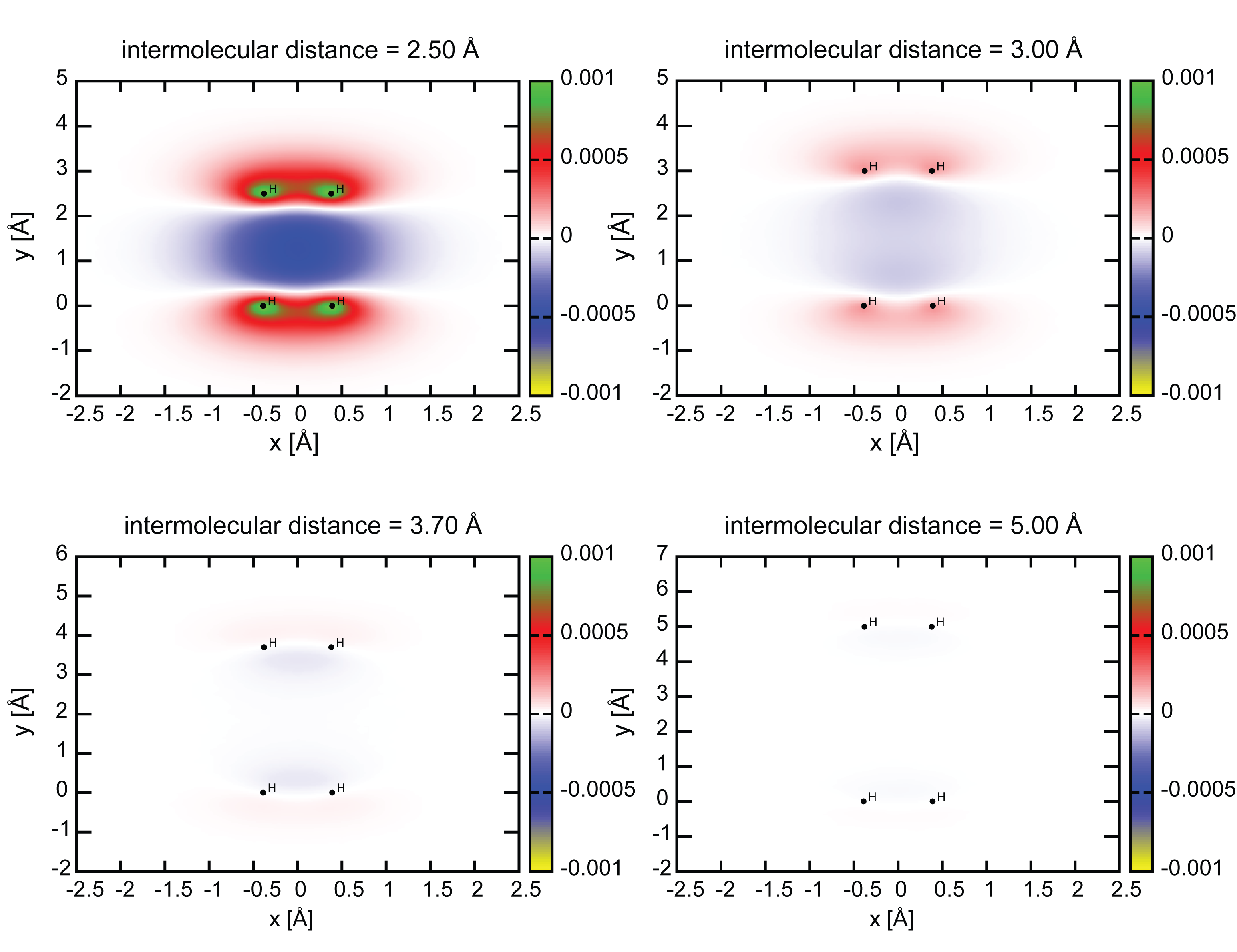}
    \caption{Variations in the ground-state densities of two parallel H$_2$ dimers (with bond length \qtylist{0.76;0.78}{\angstrom} placed on the XY plane, 
    the perturbed dimer is placed at $y$ = 0) with respect to the isolated dimers for different intermolecular distances.
    While the long-range interaction via the London dispersion force decays fast, there is a significant charge redistribution at small intermolecular distances.
    At small distances, the modifications are due to both Coulomb coupling and wavefunction overlaps, and therefore, their computation requires a proper quantum description.
    Similar modifications occur for the excited state densities.
    Density modifications also occur for larger distances for systems carrying permanent dipole moments.
    Moreover, specific interactions such as hydrogen bonds can significantly alter molecular densities.
    }
    \label{fig: SI GS densities}
\end{figure*}\\
\begin{figure*}[!ht]
    \centering
    \includegraphics[width=.95\textwidth]{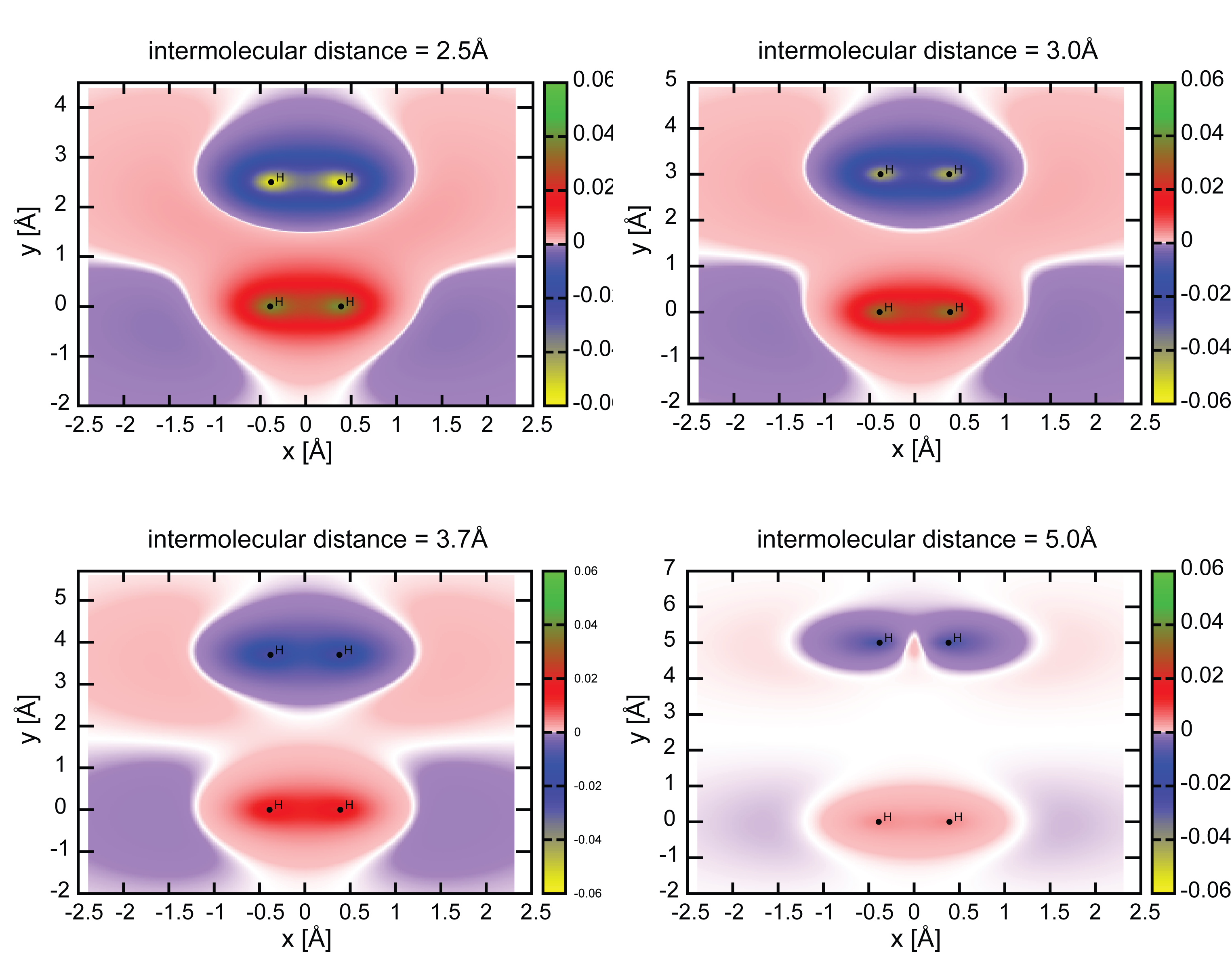}
    \caption{Variations in the first excited state (mainly associated with the impurity) density of two parallel H$_2$ dimers (with bond length \qtylist{0.76;0.78}{\angstrom} placed on the XY plane, 
    the perturbed dimer is placed at $y$ = 0) compared to the isolated ground state of the unperturbed H$_2$ and the excited state of the impurity, for different intermolecular distances.
    The differences are due to polarization and collective delocalization arising from the intermolecular coupling.
    }
    \label{fig: SI ES1GS densities}
\end{figure*}\\
\begin{figure*}[!ht]
    \centering
    \includegraphics[width=.95\textwidth]{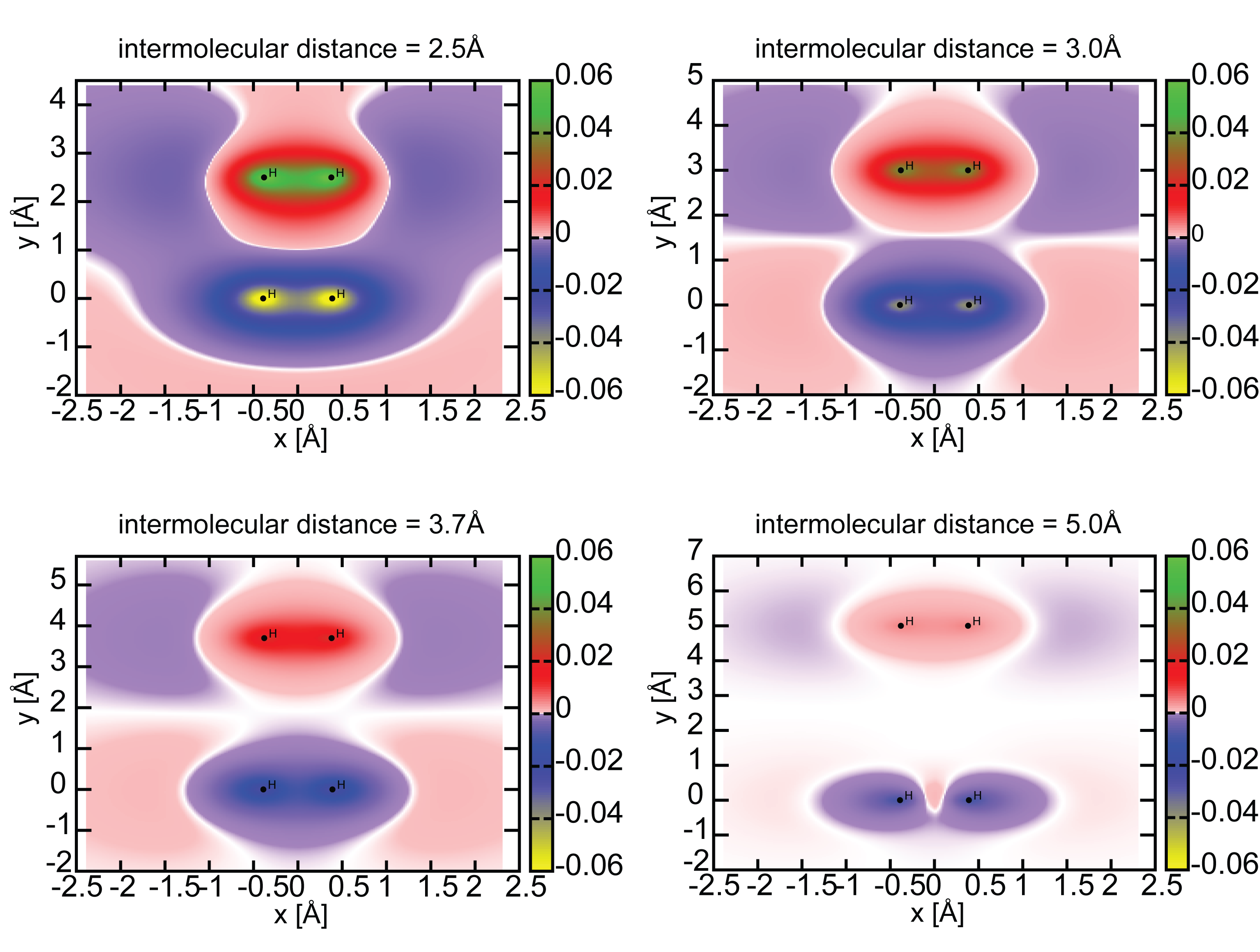}
    \caption{Variations in the second excited state (mainly associated with the unperturbed dimer) density of two parallel H$_2$ dimers (with bond length \qtylist{0.76;0.78}{\angstrom} placed on the XY plane, 
    the perturbed dimer is placed at $y$ = 0) compared to the isolated ground state of the impurity and the excited state of the unperturbed H$_2$, for different intermolecular distances.
    The differences are due to polarization and collective delocalization arising from the intermolecular coupling.
    }
    \label{fig: SI ES2GS densities}
\end{figure*}\\

While this simple model is straightforward and qualitatively correct, several relevant chemical aspects are disregarded.
First, we approximate the excitation manifold, including only the first excited states, while more excited states should be included to obtain a more realistic description.
In addition, an \textit{a priori} knowledge of the "environment shifted" excitation energy $E_{S_0\to S_1}$ might not be known and could require an \textit{ab initio} computation, for instance, via polarizable continuum models (PCM) \cite{tomasi1999medium, amovilli1998recent, tomasi2005quantum}.
The effect of molecular vibrations is also fundamental for an appropriate description of the photoluminescence properties of aggregates \cite{hestand2018expanded}.
Moreover, the point-dipole coupling is a crude approximation and significant improvement is achieved by using extended charge distributions \cite{czikklely1970extended} or approaches based on quantum chemistry \cite{gierschner2013luminescent, beljonne2000interchain, wong2009ideal, krueger1998calculation, madjet2009deciphering} which can capture the full complexity of the short-range interaction.
Indeed, in  \autoref{fig: SI GS densities}, we show the changes in the ground state density in the XY plane for an $(\text{H}_2)_2$ dimer, compared to the isolated molecules.
As in the previous calculations, the two dimers have a bond length of \qtylist{0.76;0.78}{\angstrom} and form an H-aggregate.
While at large distances, the intermolecular interaction is described by London forces and therefore decays fast with the distance $R^{-6}$, at smaller intermolecular distances, the intermolecular coupling has a complex form and depends on the wavefunction overlap.
Similar modifications also occur in the excited state densities.
In  \autoref{fig: SI ES1GS densities}, we show the difference between the first excited density of the H$_2$H'$_2$ dimer (mainly associated with the excitation of the perturbed dimer) and the sum of the \textit{isolated} ground state density of the unperturbed H$_2$ and the excited state density of the impurity.
As can be seen, the differences are more relevant than the ground state changes, and due to intermolecular forces, the excitation of one dimer polarized the other H$_2$.
Analogous results are seen in  \autoref{fig: SI ES2GS densities}, where we plot the second excited state density of H$_2$H$'
_2$ dimer, mainly associated with the excitation of the unperturbed dimer, and subtract the excited density of the unperturbed dimer and the ground state density of the impurity.
Such density modifications include rescaling due to collective delocalization and polarization due to intermolecular forces.
While for the simple H$_2$ system the intermolecular forces are small, for different molecules with permanent dipole moments or carrying specific interaction sites such as hydrogen bonds or $\pi-\pi$ interactions, the density modifications can be more relevant and even influence chemical processes.
These considerations apply to any supramolecular structure, including solute-solvent systems in which the solvation can significantly alter the molecular properties via hydrogen bonds, dipole-dipole, van der Waals, or other specific interactions, in addition to the effects of a different environment dielectric constant $\epsilon$.
Indeed, it is well known that solvation can modify chemical reactivity, and a suitable description of both the reactive molecule and the molecular surroundings is necessary to obtain qualitative and quantitative correct predictions.
Therefore, a detailed expression of the interactions with the chemical environment is generally necessary in experimental setups such as solid matrices or solutions.

\clearpage

\section{H-aggregate polaritonic excitations}
\begin{figure}[!ht]
    \centering
    \includegraphics[width=.45\textwidth]{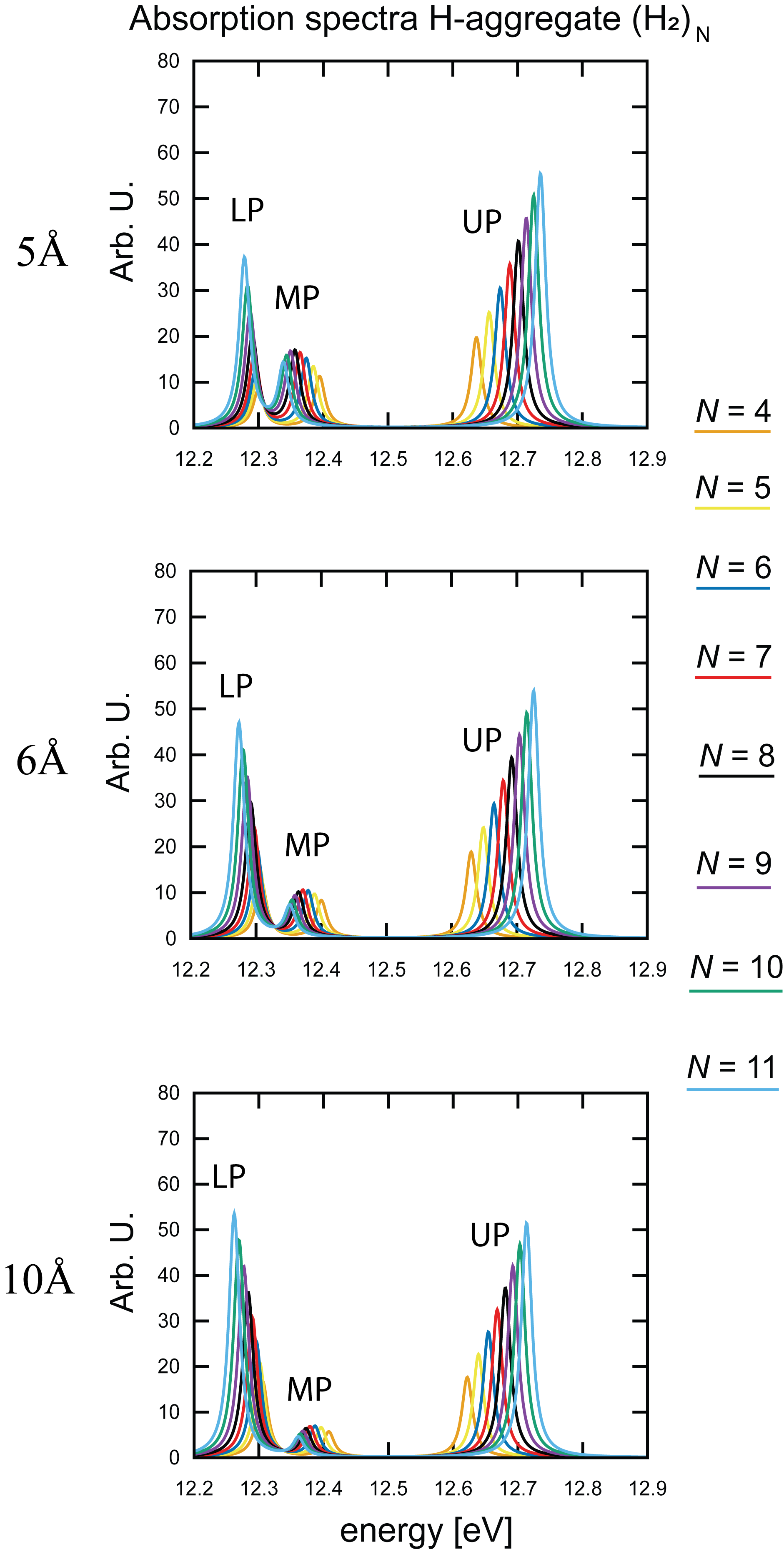}
    \caption{Polaritonic absorption spectra for an H-aggregate $(\text{H}_2)_{N\ge4}$ for \qty{5}{\angstrom} (upper panel), \qty{6}{\angstrom} (central panel), and \qty{10}{\angstrom} (lower panel) intermolecular separation.
    From the interaction between the photon field and the H-aggregate electronic excitations E1 and E2, three polaritonic branches, the lower, middle, and upper polariton, emerge.
    }
    \label{fig:cavity spectra SI}
\end{figure}
\begin{figure}[!ht]
    \centering
    \includegraphics[width=.45\textwidth]{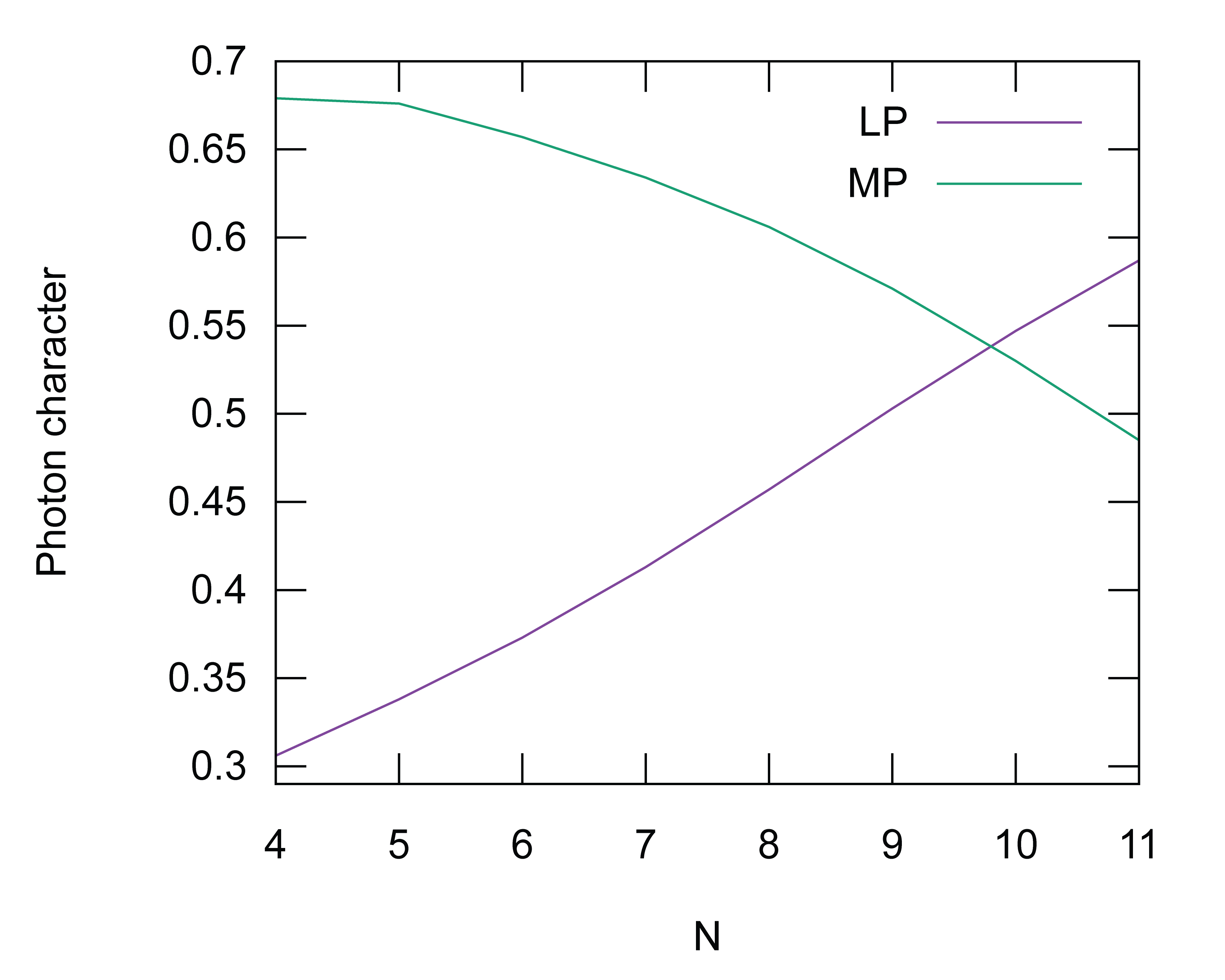}
    \caption{LP and MP photon character as a function of the chain length $N$ for \qty{5}{\angstrom} intermolecular distance.
    }
    \label{fig: LP MP photon character with N}
\end{figure}
\begin{figure}[!ht]
    \centering
    \includegraphics[width=.45\textwidth]{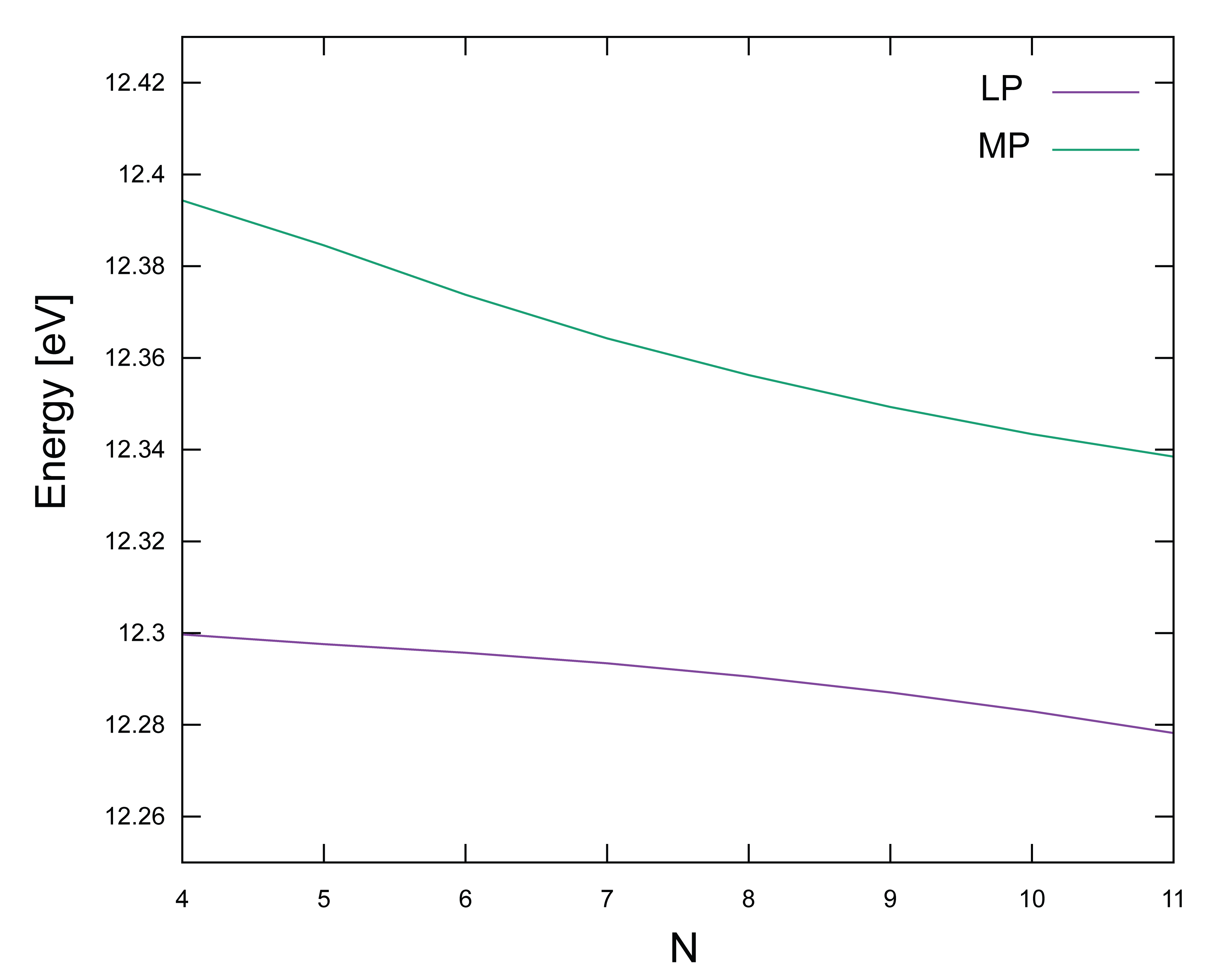}
    \caption{LP and MP energy as a function of the chain length $N$ for 5{\AA} intermolecular distance.
    }
    \label{fig: LP MP energy with N}
\end{figure}

In  \autoref{fig:cavity spectra SI}, we show the polaritonic (field-dressed) absorption spectra of the $(\text{H}_2)_{N\ge4}$ H-aggregate (see  \autoref{fig:Si structures}) for different intermolecular separations.
The cavity is tuned to the undressed excitation of the (isolated) unperturbed dimers, and from the interaction between the photon field and the H-aggregate electronic excitations E1 and E2, three polaritonic branches, the lower, middle, and upper polariton emerge.
As the intermolecular forces become more relevant (shorter distances), the MP gains oscillator strength while the LP loses it.
While the upper and lower polaritonic branches gain intensity with the chain length $N$, the MP shows a more involved trend.
In  \autoref{fig: LP MP photon character with N}, we show the LP and MP photon characters computed from the transition densities for an intermolecular separation of \qty{5}{\angstrom}.
The photon character of the MP steadily decreases with the chain length $N$, while the opposite happens for the LP.
The trends are consistent with the TC analysis of  \autoref{fig:SI JC coeff}, and analogous trends are found for other intermolecular distances.
Therefore, the matter excitations E1 and E2 gain weight in the MP properties as $N$ increases.
In  \autoref{fig: LP MP energy with N}, we report the energy of the LP and MP as a function of the number of dimers $N$ in the chain for \qty{5}{\angstrom} intermolecular distance.
The avoided crossing between the lower and upper polaritonic branches suggests relevant mixing between the E1 and E2 excitations.
Analogous avoided crossings are obtained in the TC model and for different H$_2$ separations, but they occur at different chain lengths since the intermolecular forces modify the undressed excitations.
We can then analyze the local effects of such mixing by extracting the integrated local transition densities, as explained in \autoref{Sec:comput details} and \autoref{Sec:ccsd}.

\begin{figure*}[!ht]
    \centering
    \includegraphics[width=\textwidth]{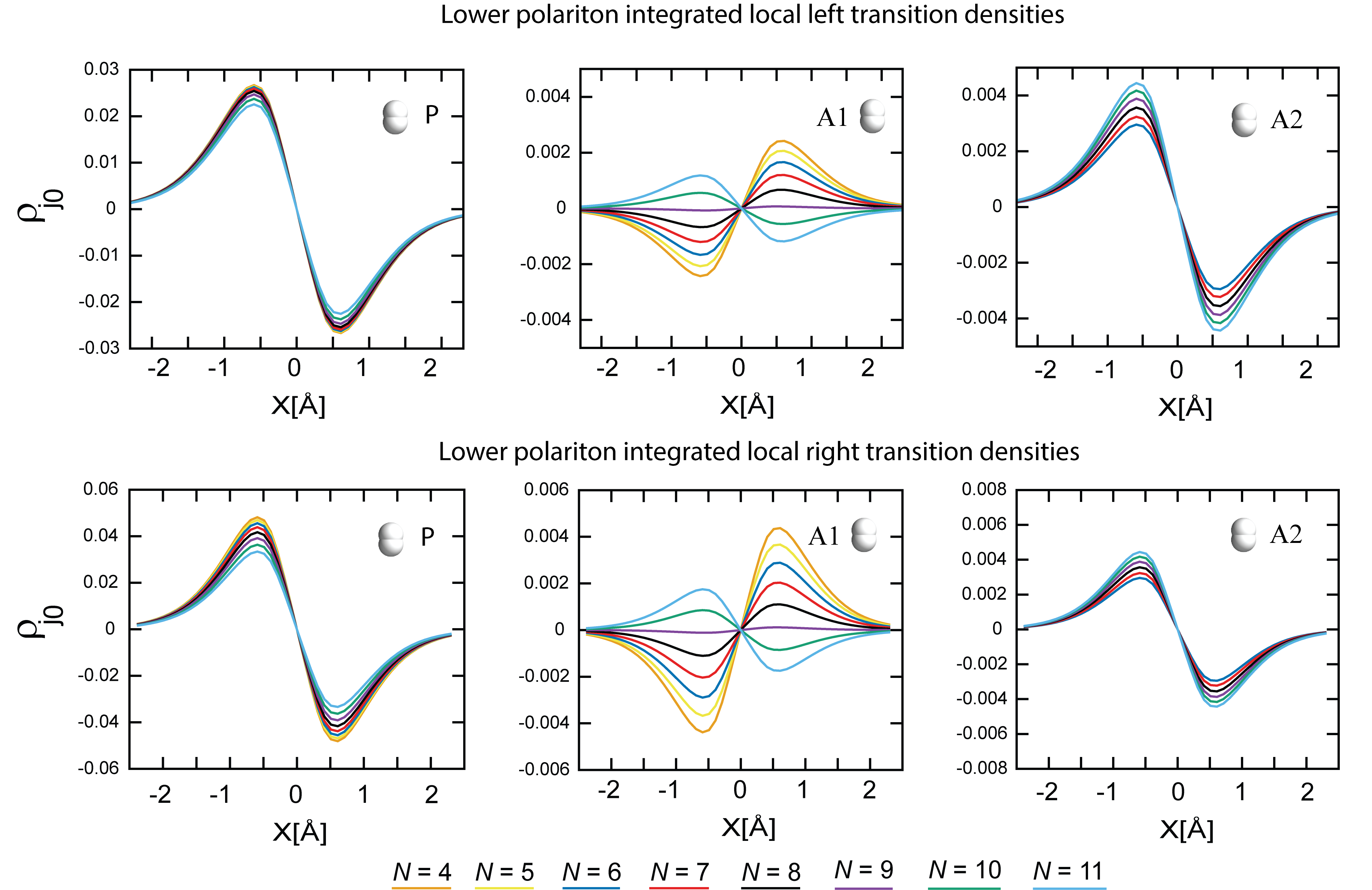}
    \caption{Local left and right transition densities for the lower polariton for an intermolecular separation of \qty{5}{\angstrom} in an H-aggregate configuration (parallel-bond arrangement, see  \autoref{fig:Si structures}).
    As we increase the number of molecules coupled to the photon field, the local transition dipole moment of A1 changes sign.
    This change in the alignment patterns of the transition moments is a consequence of the competition between the longitudinal intermolecular forces and the collective coupling to the optical mode.
    This points out the local role of the concentration in polaritonic chemistry and hints at a modification of the solvation response under cooperative strong coupling.
    }
    \label{fig: SI LP 5}
\end{figure*}

\begin{figure*}[!ht]
    \centering
    \includegraphics[width=\textwidth]{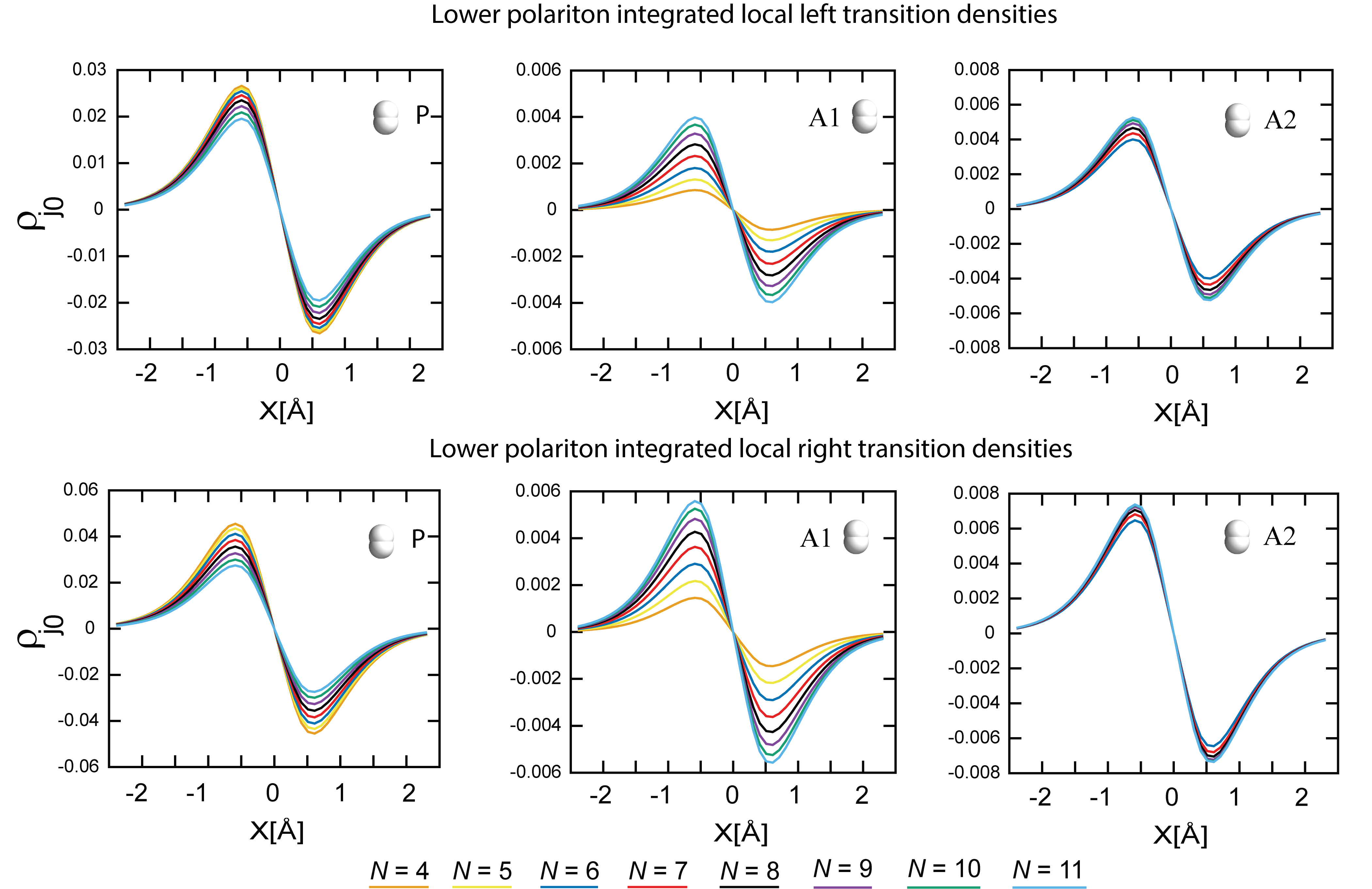}
    \caption{Local left and right transition densities for the lower polariton for an intermolecular separation of \qty{6}{\angstrom} in an H-aggregate configuration (parallel-bond arrangement, see  \autoref{fig:Si structures}).
    Compared to  \autoref{fig: SI LP 5}, the local transition dipole moments are already aligned even for short chains, so there is no sign-flip in the transition moment of A1. 
    Nevertheless, we see that as $N$ increases, the response of A1 becomes more and more relevant and more similar to A2.
    This behavior is still a consequence of the intermolecular forces between the dimers, but the collective photon effects here prevail since, at larger distances, the longitudinal coupling is smaller.
    }
    \label{fig: SI LP 6}
\end{figure*}

In  \autoref{fig: SI LP 5} and \autoref{fig: SI LP 6}, we plot the LP left and right integrated local transition densities for the H-aggregate configuration with an intermolecular distance of \qtylist{5;6}{\angstrom}, respectively.
The upper panels of  \autoref{fig: SI LP 5} are the same results as  2 in the main text.
We see that the responses of A1 and A2 are different because of the intermolecular forces.
In the undressed excitation E1, the local transition dipole moments are anti-aligned, while for E2, they are aligned.
In the polaritonic branches, the coupling to the photon mixes E1 and E2 and tends to align the local transition moments as discussed in \autoref{Sec:TC} for the TC model.
Therefore, a competition between longitudinal and transverse fields results in a change in the alignment pattern of the transition dipole moments in  \autoref{fig: SI LP 5}, for \qty{5}{\angstrom} intermolecular distance.
For smaller intermolecular forces, at \qty{6}{\angstrom} distance, the photon effects prevail, and the transition moments are always aligned in the LP.
This confirms that the pattern change results from the competition between intermolecular forces and collective strong coupling.
Nevertheless, for \qty{6}{\angstrom} separation, we see significant changes in the response of A1. 

\begin{figure*}[!ht]
    \centering
    \includegraphics[width=\textwidth]{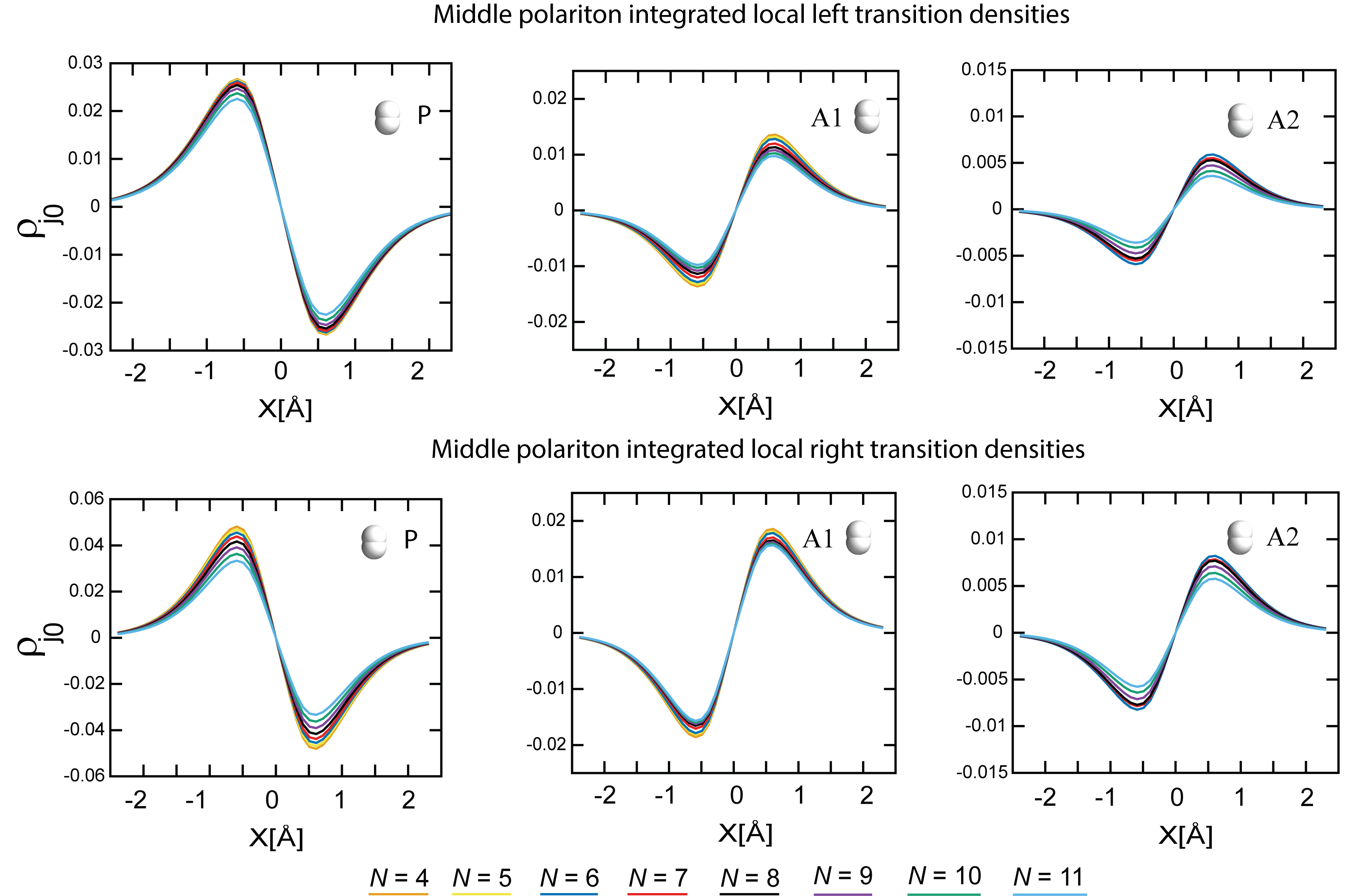}
    \caption{Local left and right transition densities for the middle polariton for an intermolecular separation of \qty{5}{\angstrom} in an H-aggregate configuration (parallel-bond arrangement, see  \autoref{fig:Si structures}).
    The P and A1 local transition dipole moments are antialigned as in E1, but, on the opposite, A2 is aligned to A1.
    The alignment A1-A2 is due to photon field effects from mixing with E2 and the favored alignment of the dimers due to the photon field.
    Since the contribution of A2 to E1 is small, it is easily overcome by the transverse field effects.
    The MP is mainly localized on P but also shows significant contributions from A1 and A2.
    }
    \label{fig: SI MP 5}
\end{figure*}

\begin{figure*}[!ht]
    \centering
    \includegraphics[width=\textwidth]{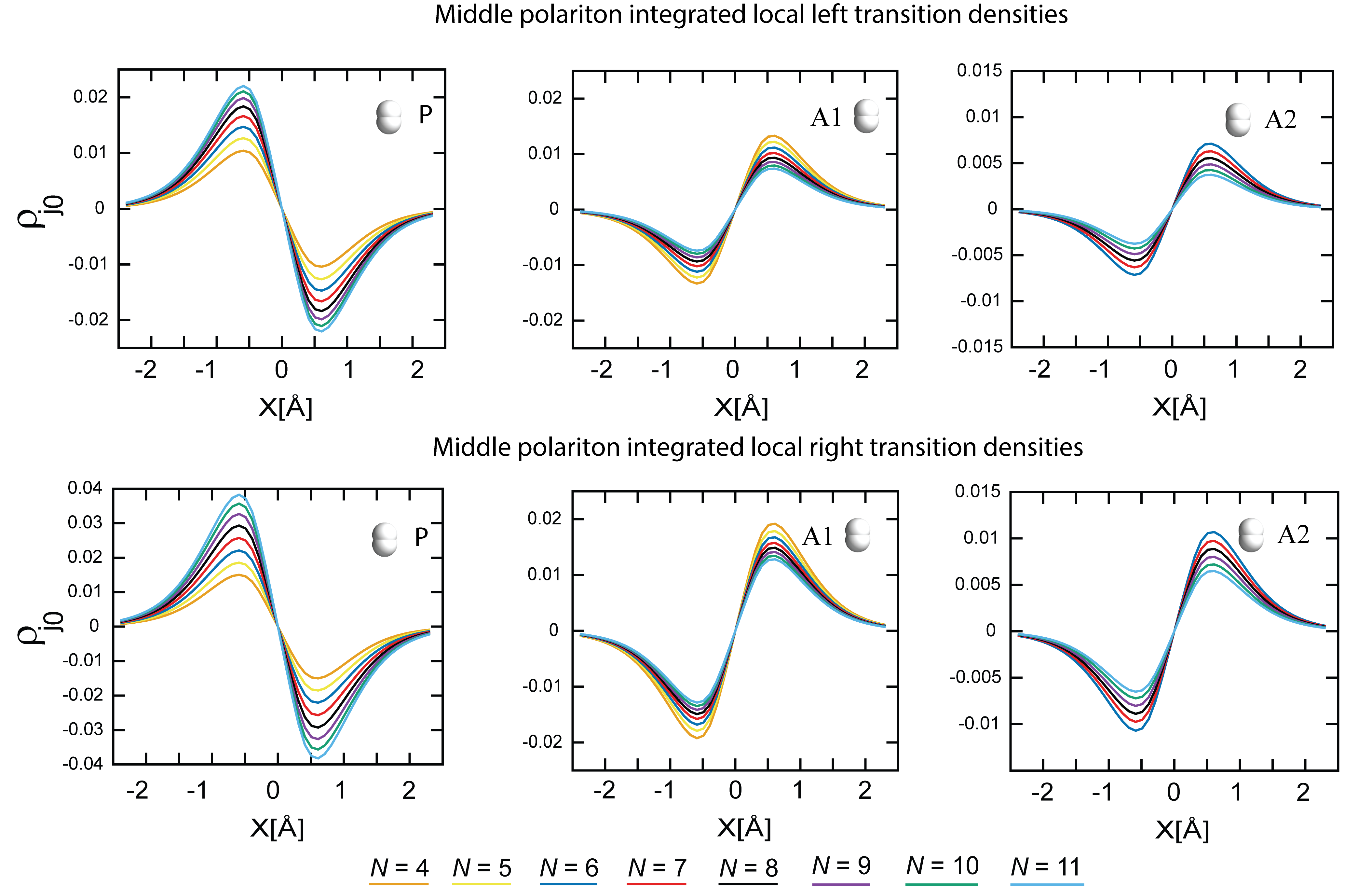}
    \caption{Local left and right transition densities for the middle polariton for an intermolecular separation of \qty{6}{\angstrom} in an H-aggregate configuration (parallel-bond arrangement, see  \autoref{fig:Si structures}).
    The P and A1 local transition dipole moments are antialigned as in E1, but, on the opposite, A2 is aligned to A1.
    The alignment A1-A2 is due to photon field effects from mixing with E2 and the favored alignment of the dimers due to the photon field.
    Since the contribution of A2 to E1 is small, it is easily overcome by the transverse field effects.
    The pattern is the same as  \autoref{fig: SI MP 5}, but the MP is more delocalized over different dimers, with more similar contributions from A1 and P.
    }
    \label{fig: SI MP 6}
\end{figure*}

In  \autoref{fig: SI MP 5} and \autoref{fig: SI MP 6}, we plot the MP left and right integrated local transition densities for the H-aggregate configuration with an intermolecular distance of \qtylist{5;6}{\angstrom}, respectively.
Compared to the LP branch, the perturbed dimer P transition moment is anti-aligned with the rest of the chain for both distances.
Notice that the MP is mainly localized on P, and the relation between the transition moments of A1 and P resembles E1.
Nevertheless, the dimer A2 is now aligned to A1, contrary to E1.
This is likely to be due to the mixing of E1 and E2 since, in E2, all of the dimers have a similar contribution in an aligned pattern. 
In contrast, the contribution of A2 in E1 is tiny and, therefore, easily overcome by E2 and the transverse field effects.
Compared to  \autoref{fig: SI MP 5}, for \qty{6}{\angstrom} intermolecular separation, the MP is more delocalized with similar contributions from P and A1, while for \qty{5}{\angstrom}, it is more localized on P.
Notice also that for \qty{6}{\angstrom}, the contribution of P increases with $N$, while it decreases for A1 and A2 ( \autoref{fig: SI MP 6}).
Instead, for \qty{5}{\angstrom} intermolecular separation ( \autoref{fig: SI MP 5}), the changes with $N$ are less evident, and the contribution of P decreases with $N$, behaving similarly to A1 and A2.\\

\begin{figure*}[!ht]
    \centering
    \includegraphics[width=\textwidth]{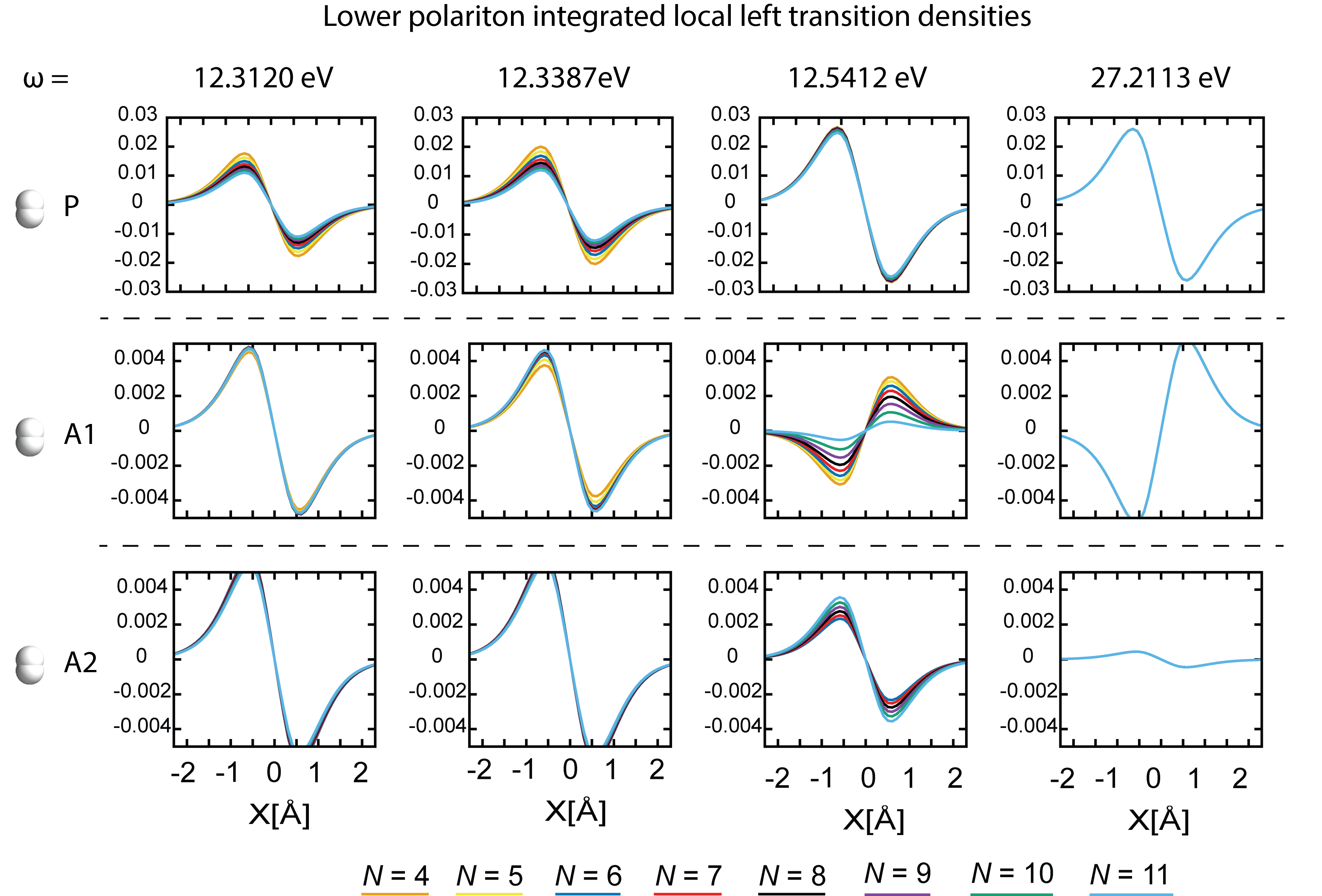}
    \caption{LP integrated local transition densities for P, A1 and A2 for different photon frequencies.\\
    When the photon is resonant with the impurity (either E1 or the undressed P excitation, first two columns), there are some modifications in the P transition density and smaller changes in A1 and A2.
    The excitations resemble E2, and no change in the alignment pattern of the transition moments is predicted. \\
    When the excitation is resonant with E2 (third column), we observe a trend similar to the cavity tuned to the undressed excitation of A1. 
    Still, the collective strong coupling is insufficient to achieve a flip in the A1 transition dipole, and longer chains are necessary.\\
    Finally, when the photon is completely detuned (rightmost column), there is negligible mixing between E1, E2, and the electromagnetic states, and therefore the excitations are analogous to the undressed ones.
    Therefore, we recover the resonant condition as a fundamental knob to observe local modifications of collective strong coupling since it determines the efficiency in the mixing of the undressed states.
    }
    \label{fig: SI LP frequency scan}
\end{figure*}

\begin{figure*}[!ht]
    \centering
    \includegraphics[width=\textwidth]{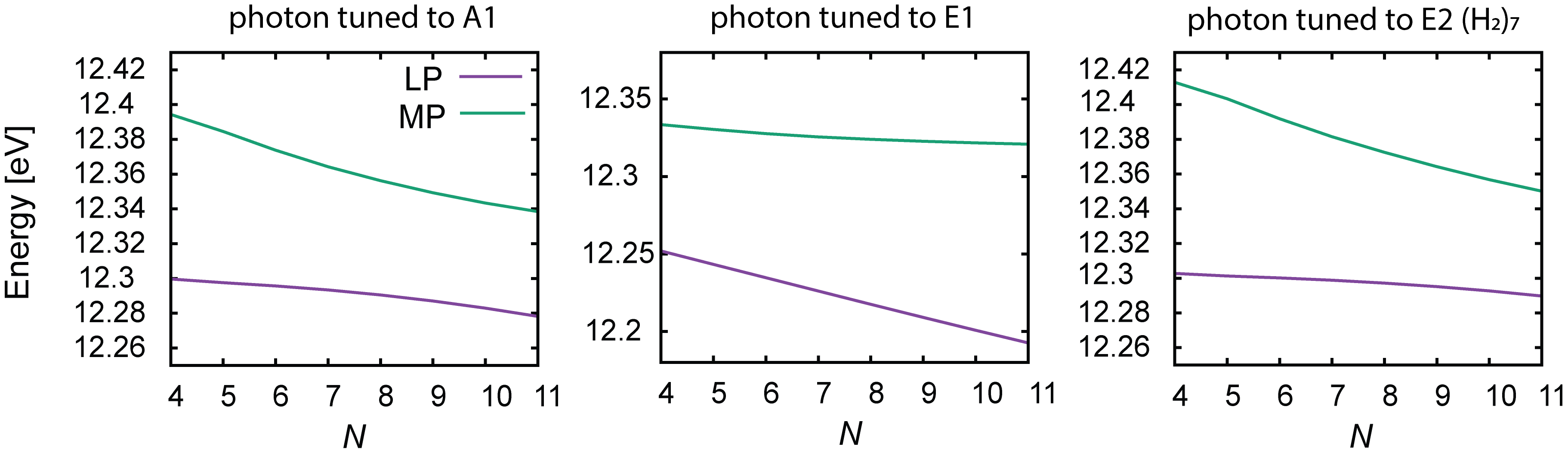}
    \caption{LP and MP energies for three different photon frequencies as a function of the chain length.
    The avoided crossings between the polaritonic branches indicate a relevant mixing between the molecular states.
    In resonance with the perturbed excitation (central panel), the avoided crossing is less pronounced and already overcome for small N, suggesting a less efficient interaction.
    This result is consistent with the less pronounced local transition density modifications.}
    \label{fig: SI avoided crossings}
\end{figure*}

The cavity tuning is a fundamental experimental knob, and empirical observations show that detuning the photon disrupts the polaritonic modifications.
Therefore, we investigate the effect of the photon frequency on our system.
Notice that the definition of the resonance condition is nontrivial since the excitation E2 blueshifts with increasing $N$.
We then consider four cavity frequencies for \qty{5}{\angstrom} intermolecular separation: i) tuned to the (undressed) excitation of the isolated perturbed dimer P, ii) tuned to the excitation E1 (basically independent of $N$), iii) tuned to E2 for the aggregate $(\text{H}_2)_7$, and iv) completely detuned at higher energies.
Resonance with E1 or P provides a model for the photon tuned with the impurity excitation.
For the resonance with E2, we expect effects similar to the above results.
Finally, when the photon is completely detuned, we expect no modifications compared to the undressed case.
In  \autoref{fig: SI LP frequency scan}, we show the LP local left transition densities of P, A1, and A2  (H-aggregate, intermolecular separation of \qty{5}{\angstrom}).
When the photon is tuned to the impurity (either resonant with the undressed excitation of the isolated P molecule or resonant to the excitation E1, first two columns), we observe small local changes that are more pronounced for P.
The transition moments are aligned as in E2, and the changes in A1 and A2 are modest.
When the cavity is tuned to E2 of $(\text{H}_2)_7$ (third column), we see significant changes in the local transition moment of A1, which resemble the results for the photon tuned to the undressed excitation of E1.
Finally, when the photon is completely detuned to the excitation (rightmost panels of  \autoref{fig: SI LP frequency scan}), there is no local change in the transition densities, and the LP shows the same pattern as E1.
Since the cavity has a much larger energy than the matter excitation, the hybridization is almost negligible, and the states are effectively the same as the undressed excitations.
However, the collective strong coupling for such $N$ is not enough to flip the sign of the A1 transition densities, and longer chains would be necessary.
We then recover the resonance condition as a fundamental knob to observe local modifications of molecular properties under cooperative strong coupling.
In our model, the cavity frequency determines the efficiency in the hybridization of E1, E2, and the cavity photon, therefore determining the changes in the alignment pattern of the transition dipoles.
In  \autoref{fig: SI avoided crossings}, we show the LP and MP energies as a function of $N$ for the cavity resonant to A1 (left), E1 (center), and E2 in (H$_2$)$_7$(right).
In all three cases, the energies display an avoided crossing between the polaritonic branches.
Still, when the photon is tuned to the undressed excitation of the isolated A1 (or to E2, first and last columns), the avoided crossing is more pronounced compared to the tuning to E1 (central column).
This suggests a more efficient mixing of the undressed excitation and the one-photon state, resulting in the observed local modifications.
On the other hand, when the cavity is tuned to E1, the avoided crossing is less pronounced and already overcome for small chain lengths, suggesting a less efficient mixing in the polaritonic branches.
This is consistent with the changes in the patterns of the local transition moments.

\begin{figure*}[!ht]
    \centering
    \includegraphics[width=\textwidth]{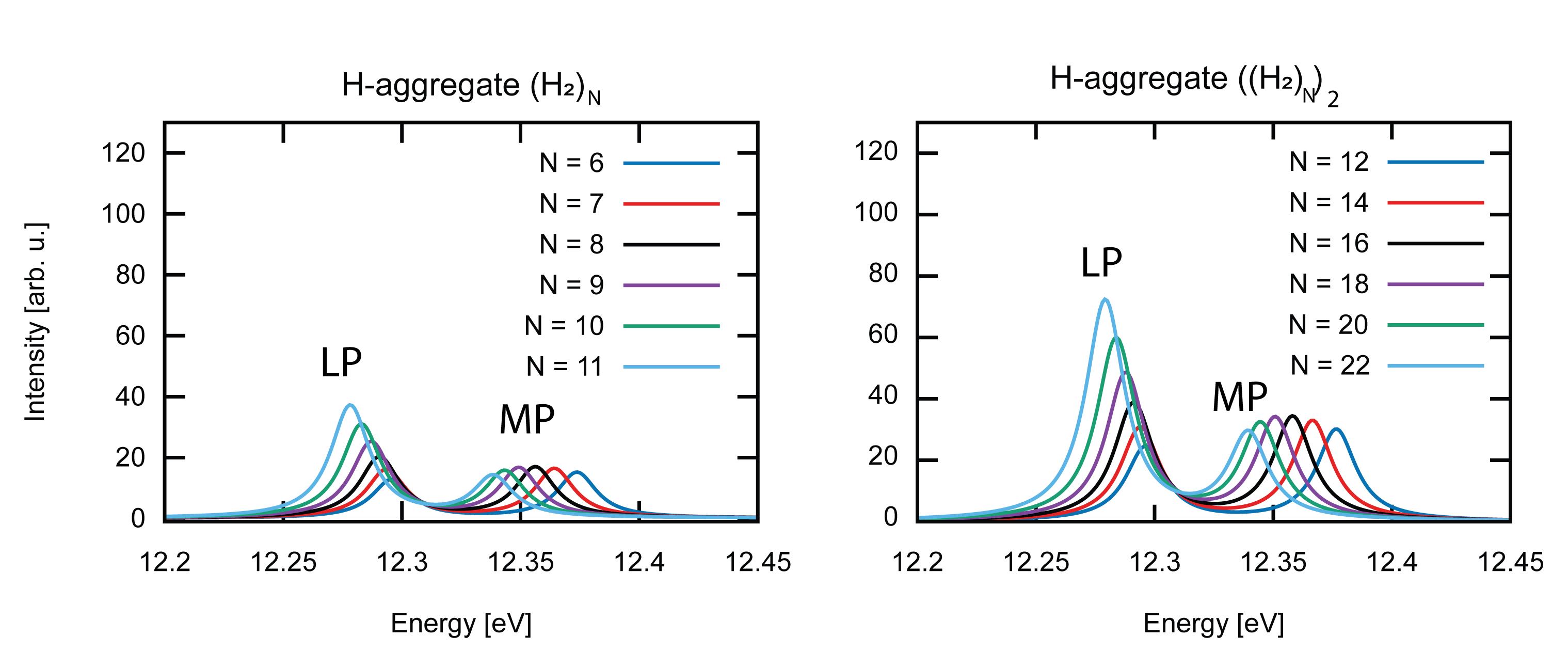}
    \caption{Left, polaritonic absorption spectra for a \qty{5}{\angstrom} intermolecular separation H-aggregate $(\text{H}_2)_{N\ge6}$ (same as  \autoref{fig:cavity spectra SI}).
    Right, spectra for a system with double dimension compared to left, $(\text{H}_2)_{N\ge12}$, and coupling strength rescaled by $1/\sqrt{2}$.
    The LP and MP excitation energies are the same in these two realizations, consistent with the TC predictions in the thermodynamic limit.
    }
    \label{fig:cavity thermo spectra SI}
\end{figure*}

Finally, we investigate the spectral effect of decreasing the coupling strength while increasing the number of dimers to keep the ratio $N/V_{el}$ fixed, where $V_{el}$ is the quantization volume (connected to the light-matter couplings strength $\lambda = \sqrt{\frac{1}{\varepsilon_0 V}}$).
In  \autoref{fig:cavity thermo spectra SI}, we show the absorption spectra (LP and MP branches) for the \qty{5}{\angstrom} intermolecular separation H-aggregate.
The left panel is the same as \autoref{fig:cavity spectra SI}, while the right panel is for a system with twice as much dimers and coupling strength rescaled by $1/\sqrt{2}$ to maintain the $N/V_{el}$ ratio.
The LP and MP excitation energies are the same in the two realizations, which is consistent with the TC predictions in the thermodynamic limit. 
At the same time, the peaks are more intense due to the larger number of oscillators contributing to the excitations.

\clearpage
\subsection{TC-Kasha model for the molecular polaritons}\label{TC-Kasha polaritons}
\begin{figure*}[!ht]
    \centering
    \includegraphics[width=1\textwidth]{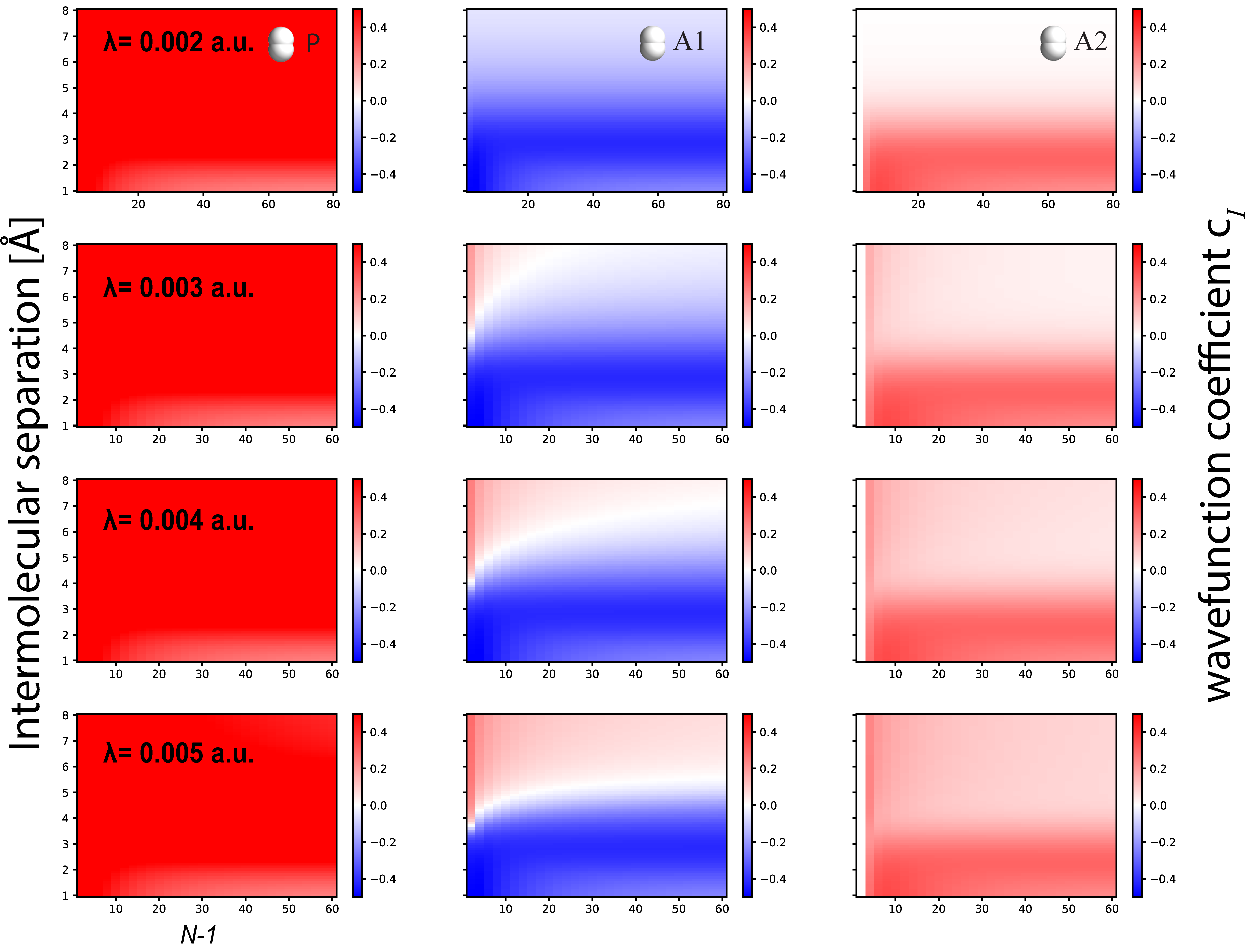}
    \caption{Lp wave function coefficients $c$ of P, A1, and A2 (left to right) with $\lambda=$ \qtylist{0.002;0.003;0.004;0.005}{\atomicunit} from top to bottom.}
    \label{fig:SI model_c1 different g noscaling}
\end{figure*}

\begin{figure*}[!ht]
    \centering
    \includegraphics[width=1\textwidth]{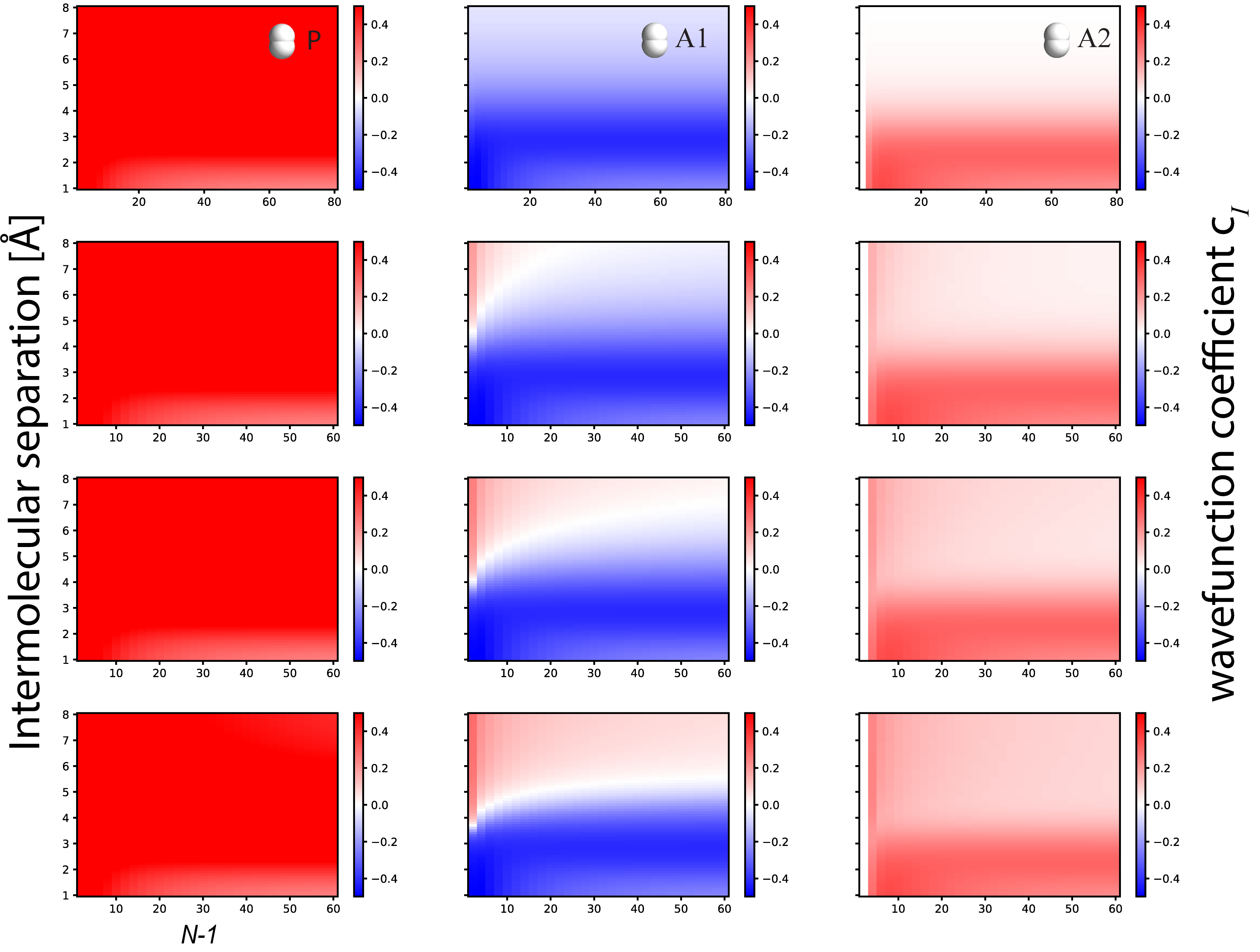}
    \caption{Lp wave function coefficients $c$ of P, A1, and A2 (left to right) without (top) cavity, where the cavity strength is scaled as $\lambda=\lambda_0/\sqrt{N}$ with $\lambda_0=$  \qtylist{0.0;0.01;0.0125;0.015}{\atomicunit}}
    \label{fig:SI model_c1}
\end{figure*}

The sign flip in the local transition density of A1 in  \autoref{fig: SI LP 5} can be modeled using a simplified picture that merges the TC and the Frenkel-Kasha models.
Following the same procedure as the TC model, we consider the first excited state manifold of the $N$ hydrogen dimers, each with the same excitation energy as the isolated molecules.
Each H$_2$ is then coupled to the photon field via its isolated transition dipole moment as in \autoref{Eq: H_TC matrix}.
Following the same assumptions as in the Kasha model in \autoref{Eq: Kasha hamiltonian}, we introduce the intermolecular interactions via the transition dipole-dipole coupling of \autoref{Eq: dipole-dipole coupling}.
For simplicity, we couple only the nearest dimers, and the Hamiltonian matrix for the TC-Kasha model then reads
\begin{equation}\label{Eq: H_TC-Kasha matrix}
    \begin{pmatrix}
        \omega & \tilde {g_0}' & \tilde g_0 &\tilde g_0 &\tilde g_0 & \cdots\\
        \tilde {g_0}' & \omega' & V' & 0 &0& \cdots\\
        \tilde g_0 & V' & \omega& V & 0&\cdots\\
        \tilde g_0 &0& V & \omega& \ddots & \ddots\\
        \vdots & \vdots & \ddots& \ddots & \ddots& \ddots
    \end{pmatrix}.
\end{equation}
Diagonalizing the Hamiltonian in \autoref{Eq: H_TC-Kasha matrix}, we obtain the eigenvalues and eigenvectors corresponding to 3 polaritonic branches and $N$-2 (quasi) dark states.
Compared to the \textit{ab initio} simulations, which are limited by the steep computational costs of the CC methods, it is possible to perform simulations with large $N$.
In  \autoref{fig:SI model_c1 different g noscaling}, we show the coefficients of P (left), A1 (center), and A2 (right) for the LP from \autoref{Eq: H_TC-Kasha matrix} for coupling strengths $\lambda=$ \qtylist{0.002;0.003;0.004;0.005}{\atomicunit} as a function of the intermolecular separation and the chain length $N$.
The qualitative behavior and the order of magnitude of the coefficients are the same in all cases and agree with the \textit{ab initio} results in  \autoref{fig: SI LP 5}.
The effects of rescaling the coupling strength $\lambda$ for a fixed intermolecular separation by a factor $\alpha$ shifts the avoided crossing from length $N$ roughly to $N'\sim N/\alpha^2$, which is consistent to the thermodynamic limit requirement $N_p/V_{el}  = \text{const.}$ for $N_p, V_{el} \rightarrow \infty$, where $V_{el}$ is the quantization volume $\lambda\propto \sqrt{V_{el}^{-1}}$.
Therefore, this local sign flip involving A1, which stems from the competition between longitudinal and transverse fields, is persistent even in the thermodynamic limit.
In  \autoref{fig:SI model_c1}, we show the coefficients of P (left), A1 (center), and A2 (right) for the LP obtained from the matrix \autoref{Eq: H_TC-Kasha matrix} where the bare coupling strength is $\lambda_0=$ \qtylist{0.0;0.01;0.0125;0.015}{\atomicunit} further rescaled by $1/\sqrt{N}$, that is, $\lambda = \lambda_0/\sqrt{N}$.
In agreement with the \textit{ab initio} simulations, outside the cavity (first row of  \autoref{fig:SI model_c1}), the coefficients of P, A1, and A2 display alternate signs due to the intermolecular forces, and the contribution of A1 and A2 to the lowest excitation, mainly associated with P, increases for shorter distances.
When the photon coupling is switched on, the coefficient of A1 still displays a sign flip, and the order of magnitude of the coefficients is the same in the three simulations.
The position at which such a sign flip occurs depends on $\lambda_0$, the intermolecular separation, the chain length $N$, and the excitation energy of the impurity.
Therefore, the described effects are resistant to the thermodynamic limit for a single impurity P.

The simplified TC-Kasha model successfully reproduces the spectral behavior of the QED-CC results.
However, in this framework, the transition and excited state densities can be reconstructed from \autoref{Eq: H_TC-Kasha matrix} only from the \textit{isolated} H$_2$ densities, which are available as an input for the model.
Moreover, {the ground state is assumed to be unchanged}.
For the H$_2$ system under consideration, the modifications of the ground and excited state densities are non-negligible for intermolecular distances below \qty{5}{\angstrom}, as can be seen from \autoref{fig: SI GS densities}, \ref{fig: SI ES1GS densities}, and \ref{fig: SI ES2GS densities}.
Therefore, the mixing and rescaling of a few isolated densities coming from the eigenvectors of \autoref{Eq: H_TC-Kasha matrix} could be unsuited for describing such local modifications, which can be responsible for alterations of chemical processes such as molecular reactivity.
On the other hand, QED-CC includes all the quantum chemistry effects due to wavefunction overlap and Coulomb forces.
Therefore, when the effect of the solvent is non-perturbative, although a simplified optical model can be used to reproduce the spectral polaritonic properties, an \textit{ab initio} description of the solute-solvent will be necessary for a detailed investigation of the single-molecule (local) chemistry changes observed in several polaritonic chemistry experiments.
While the current discussion has focused on electronic strong coupling, the same considerations hold for vibrational excitations. 
Quantum chemistry is then necessary to model the potential energy surfaces on which the nuclei move.
An analogous interplay between longitudinal and transverse field effects is also expected in the framework of vibrational strong coupling.
In particular, solute-solvent interactions, such as hydrogen bonds, will significantly modify the molecular vibrations.
Moreover, the light-matter strong coupling can modify intermolecular forces, and it is suggested in particular that van der Waals forces can be particularly affected \cite{haugland2020intermolecular, cao2023cavity, philbin2022molecular, haugland2023understanding}.
While a few-level CI model might not capture such electron-photon correlation effects, they are naturally included in \textit{ab initio} QED.

\begin{figure*}[!ht]
    \centering
    \includegraphics[width=1\textwidth]{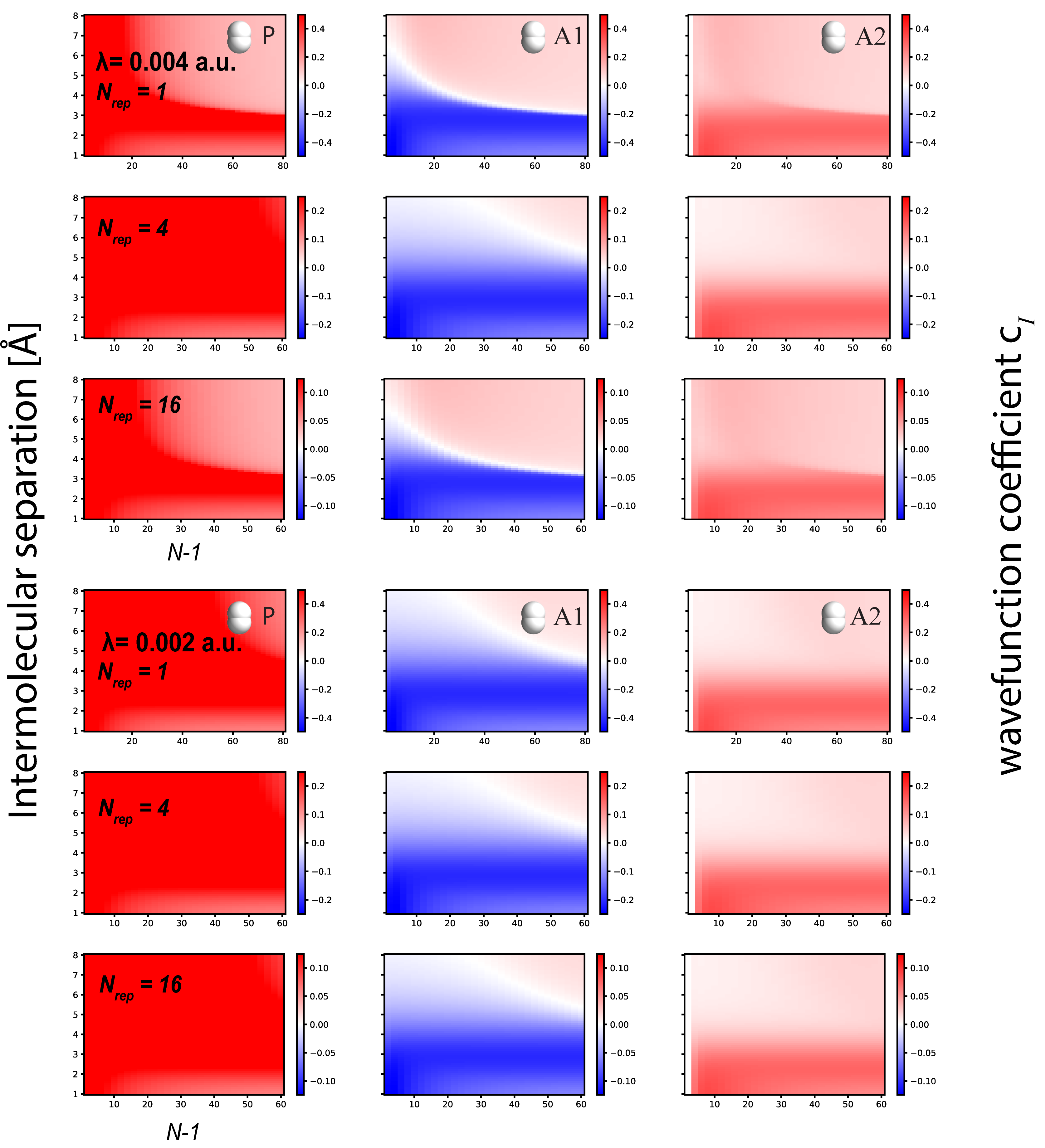}
    \caption{LP wave function coefficients $c$ of P, A1, and A2 (left to right).
    The first three panels refer to a bare coupling strength of $\lambda_0=$ \qty{0.004}{\atomicunit}, while for the last three $\lambda_0$ = \qty{0.002}{\atomicunit}.
    For each $\lambda_0$, the calculations are repeated for 1, 4, and 16 duplicates of the chain (from top to bottom), and the coupling strength is rescaled by $1/\sqrt{N_{rep}}$ i.e. $\lambda=\frac{\lambda_0}{\sqrt{N_{rep}}}$.}
    \label{fig:SI model_c2 rescale Nreplicas}
\end{figure*}

\begin{figure*}[!ht]
    \centering
    \includegraphics[width=1\textwidth]{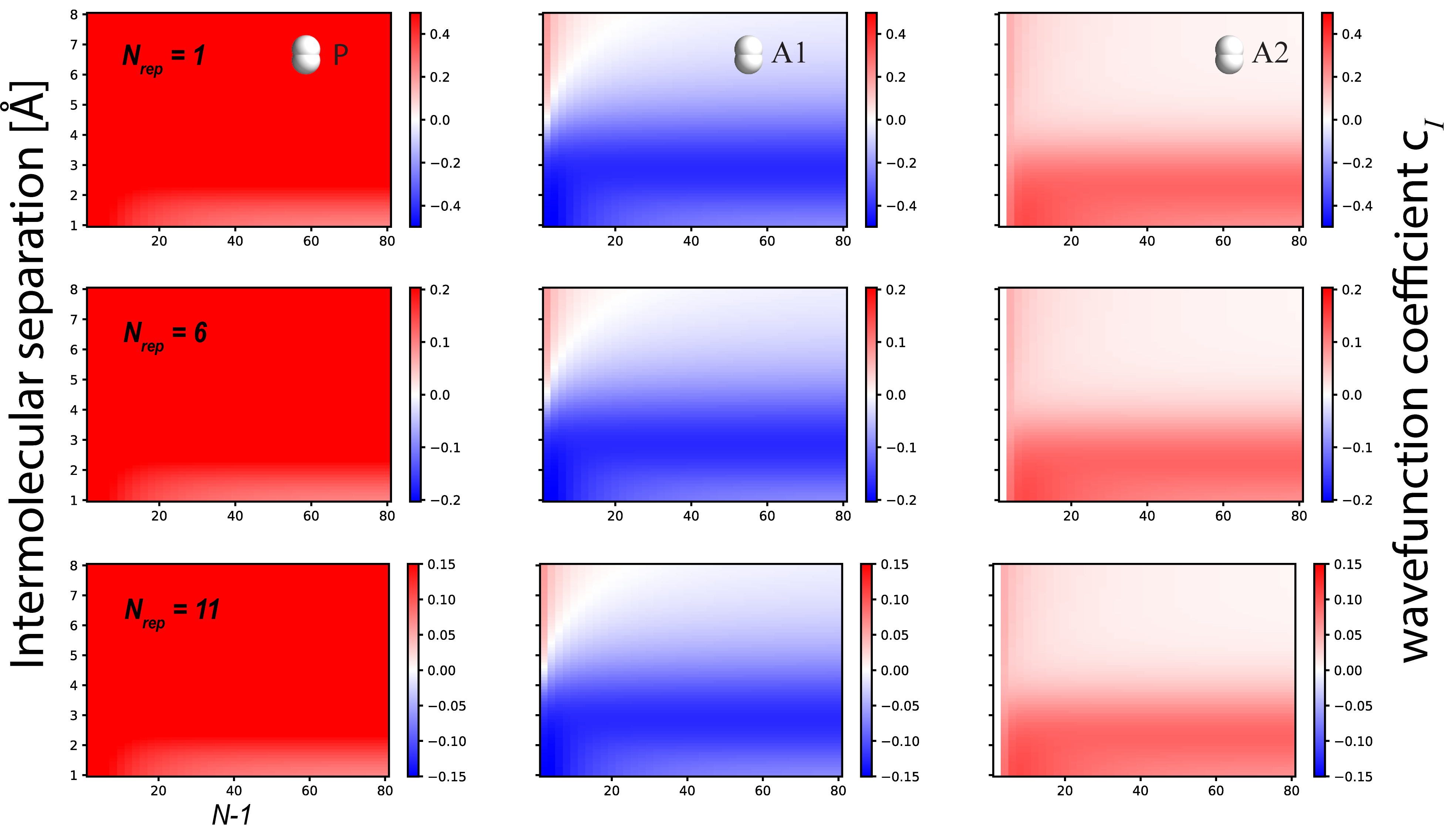}
    \caption{Lp wave function coefficients $c$ of P, A1, and A2 (left to right) for 1, 6, and 11 duplicates (from the top down to the bottom) of the H-aggregate system with up to 40 bare molecules. The cavity strength is scaled as $\lambda=0.01/\sqrt{N_{tot}}$ to investigate the thermodynamic limit $N, V_{el} \rightarrow \infty,~N/V_{el} = c$. }
    \label{fig:SI model_c2}
\end{figure*}

So far, we have focused on a single perturbed dimer P.
However, the TC-Kasha model also allows us to study $N_{rep}$ replicas of the H$_2$ chain.
In  \autoref{fig:SI model_c2 rescale Nreplicas}, we show the TC-Kasha eigenvector coefficients for P, A1, and A2 (from left to right) for coupling strengths $\lambda_0$ = \qty{0.004}{\atomicunit} (first three panes) and $\lambda_0$ = \qty{0.002}{\atomicunit} (last three panels) for number of chain replicas $N_{rep}$ = \numlist{1;4;16} (from top to bottom for each $\lambda_0$).
The light-matter coupling strength is then rescaled by $1/\sqrt{N_{rep}}$, that is, $\lambda = \lambda_0/\sqrt{N_{rep}}$.
For each $\lambda_0$, the results show the same qualitative behavior for every $N_{rep}$. 
We still predict a sign flip in the A1 coefficients, which means the qualitative behavior resists the thermodynamic limit.
Nevertheless, notice that the coefficients in this case decrease of a factor $(\sqrt{N_{rep}})^{-1}$ and the coefficients will asymptotically approach zero.
Similar results are obtained from the \textit{ab initio} simulations, although the computational costs limit the number of replicas we can study using QED-CC.
\begin{figure*}[!ht]
    \centering
    \includegraphics[width=\textwidth]{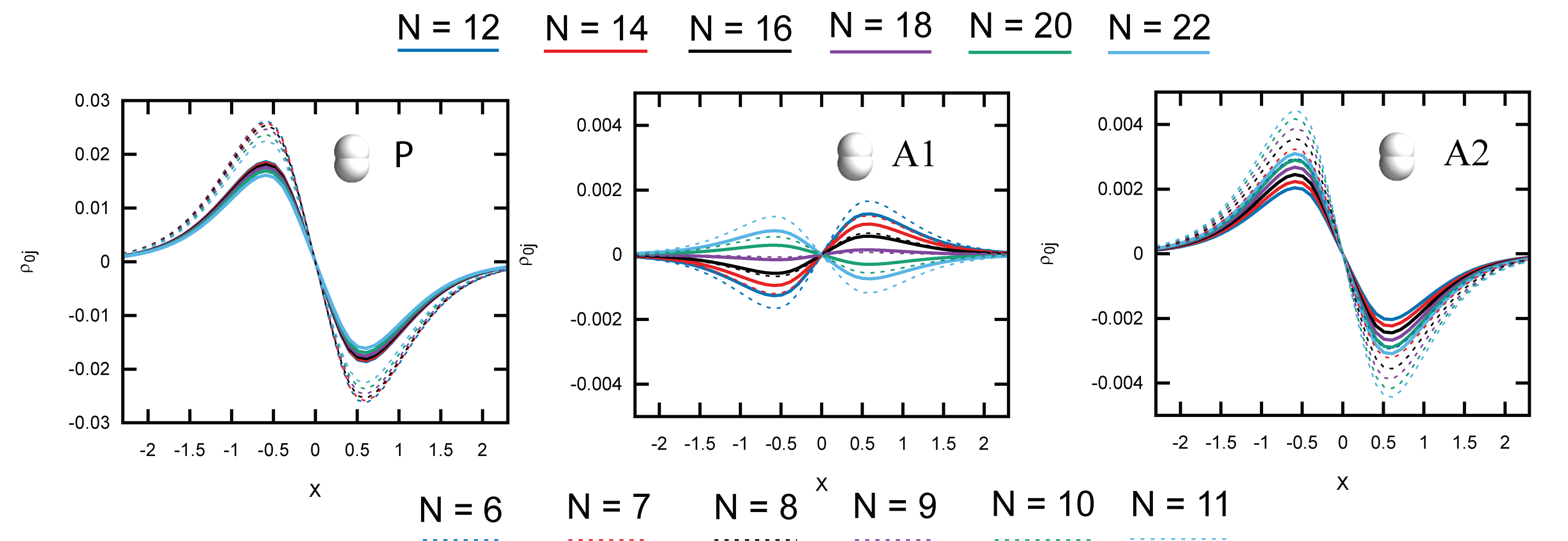}
    \caption{The solid lines show the local transition densities for P, A1, and A2 for a chain containing two perturbed dimers and intermolecular separation of 5\AA.
    Compared to Fig.\autoref{fig: SI LP 5}, the coupling strength has been rescaled by $(\sqrt{2})^{-1}$, and the two perturbed H$_2$ are separated by 6 dimers so that their mutual influence is negligible.
    The dotted lines show for reference the same results as in  \autoref{fig: SI LP 5} for a chain with a single perturbed dimer.
    The results for these two realizations are qualitatively the same and display the same order of magnitude.
    The differences are likely due to different intermolecular forces arising from the different number of dimers, dipole self-energy, and electron-photon correlation.}
    \label{fig: SI image therm}
\end{figure*}
In  \autoref{fig: SI image therm}, we show the LP integrated local transition density matrix for a chain containing 2 perturbed dimers for 5{\AA} intermolecular distance, with coupling strength rescaled by the factor $1/\sqrt{2}$.
The two perturbed dimers are separated by 6 unperturbed H$_2$ so that their direct influence is negligible.
For comparison, the transition densities for a single perturbed dimer and unscaled light-matter coupling strength are shown in dotted lines.
In agreement with the TC-Kasha model, the qualitative sign-inversion behavior of A1 is reproduced, the position $N$ at which the sign-flip occurs is also the same, and the contribution of each dimer decreases by roughly the same scaling factor as the coupling strength.
Nevertheless, we point out that the number of replicas is smaller than the number of coupled dimers, that is, $N_{tot}=N\times N_{rep}$.
Therefore, the presented results show a slower decay in the thermodynamic limit as $1/N_{rep}$ instead of $1/N_{tot}$.\\
In  \autoref{fig:SI model_c2}, we plot the coefficients for P, A1, and A2 (left, center, and right panels, respectively) for up to 80 unperturbed dimers (upper panels) for a bare light-matter coupling strength $\lambda$ of 0.01 a.u. and different intermolecular distances.
In the second and third rows of  \autoref{fig:SI model_c2}, we then duplicate the system, adding 5 (second row) and 10 (last row) identical replicas and consequently rescaling the fundamental coupling strength $\lambda / \sqrt{N\times N_{rep}}$.
The figures show the same qualitative sign flip behavior as a function of $N$ and the intermolecular separation.
This confirms again that such a local change, stemming from the interplay between longitudinal (intermolecular forces) and transverse (photon) fields, is resistant to the thermodynamic limit.\\

\clearpage

Using a simplified TC-Kasha model also allows us to study the effect of disorder in the system.
To this end, we consider P and its first solvation shell (A1) to be fixed while introducing disorder in the remaining dimers.
Therefore, we model only long-range disorder while short-range disorder (in the vicinity of P) is suppressed for simplicity.
A Gaussian distribution of the excitation energies, centered at the unperturbed dimers' excitation energies, introduces an energy broadening.
Such an energy broadening can be thought of as a distribution of bond lengths in the dimers caused by thermal fluctuations. 
Moreover, the orientational disorder is introduced by assuming the dimers can form an angle $\theta_k\in[0,\pi/2]$ with the X axis ($k$ labels the different disordered dimers). 
At the same time, for simplicity, we keep them orthogonal to the Y axis.
The restriction $\theta_k\in[0,\pi/2]$ (instead of $\theta_k\in[0,\pi]$) ensures that the local transition dipole moments are still aligned in the same direction to avoid fictitious flipping of the wavefunction coefficients (see \autoref{eq: disordered polaritons}).
We also assume all the dimers to be kept at a fixed distance.
Therefore, the Hamiltonian to be diagonalized here reads
\begin{equation}\label{Eq: H_TC-Kasha disorder matrix}
    \begin{pmatrix}
        \omega & \tilde {g_0}' & \tilde g_0 &\tilde g_0(3) &\tilde g_0(4) & \cdots\\
        \tilde {g_0}' & \omega' & V' & 0 &0& \cdots\\
        \tilde g_0 & V' & \omega& V(2,3) & 0&\cdots\\
        \tilde g_0(3) &0& V(2,3) & \omega(3)& V(3,4) & \ddots\\
        \tilde g_0(4) &0& 0&V(3,4) & \omega(4)& \ddots & \\
        \vdots & \vdots & \ddots& \ddots & \ddots& \ddots
    \end{pmatrix}.
\end{equation}
In \autoref{Eq: H_TC-Kasha disorder matrix}, the dimer labeled 1 is P, A1 is labeled 2, and the remaining N-2 dimers have energy disorder embedded in the excitation energy on the diagonal $\omega(k)$, and orientational disorder which causes a decrease of the light-matter coupling strength compared to the ordered case $\tilde g_0(k)=\tilde g_0 \cos(\theta_k)$.
\begin{figure*}[!ht]
    \centering
    \includegraphics[width=1\textwidth]{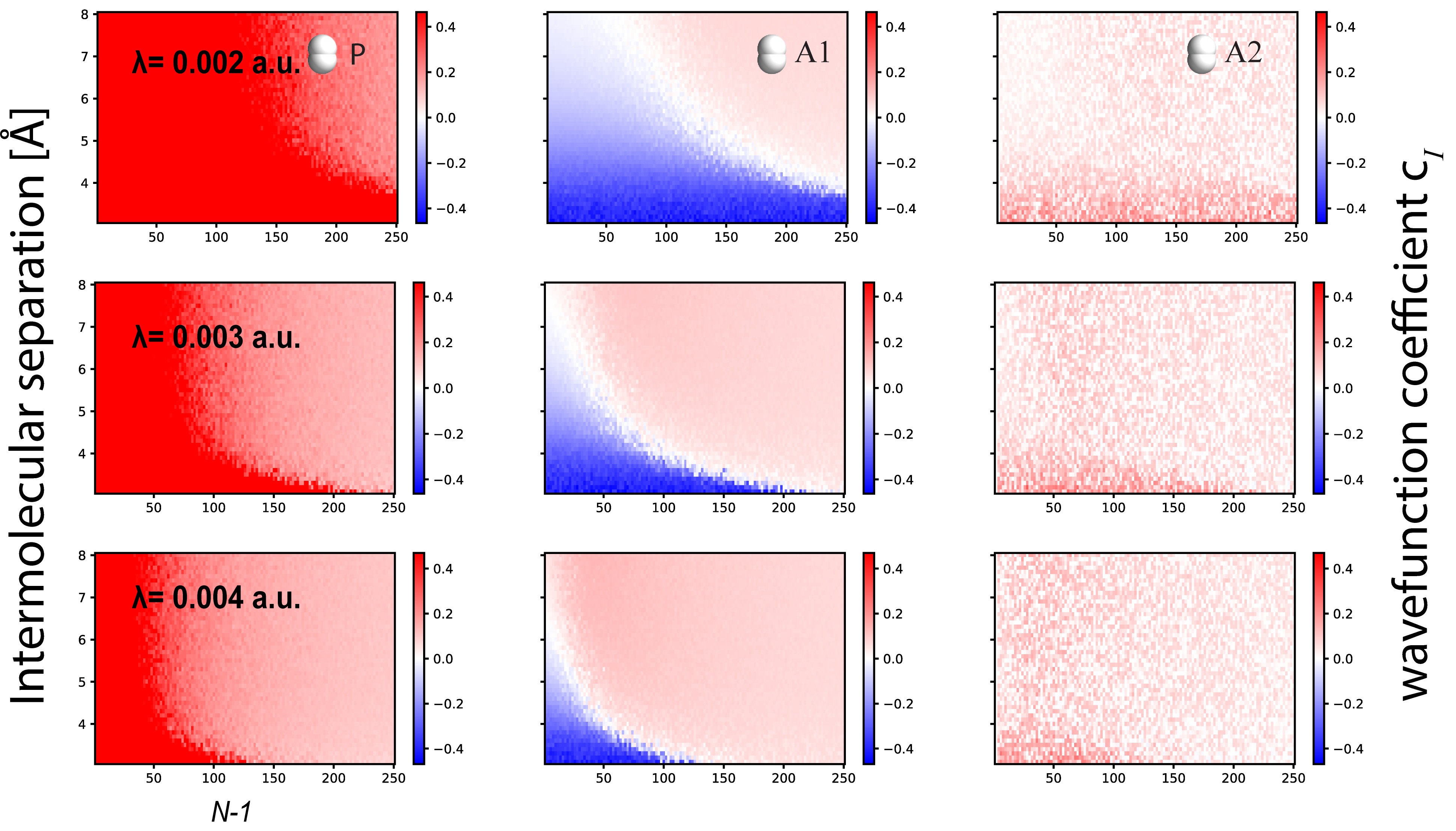}
    \caption{LP wave function coefficients $c$ of P, A1, and A2 (left to right) with coupling strengths $\lambda$ = \qtylist{0.002;0.003;0.004}{\atomicunit} (from the top down to the bottom) of the H-aggregate system with up to 250 bare molecules. 
    Disorder is introduced in the system via a Gaussian energy broadening of the dimers (except for P and A1), with Gaussian width \qty{0.02}{\eV}.
    Moreover, the same dimers are given a random angle with the X axis (while keeping them perpendicular to Y) $\theta_k\in[0,\pi/2]$.}
    \label{fig:SI disorder model scaled 0}
\end{figure*}
Moreover, the orientational disorder also affects the dipole-dipole couplings $V(m,n)$, as seen in \autoref{Eq: dipole-dipole coupling}.
Refinement of such simplified disorder can be obtained by relaxing the rotation constraints and permitting a displacement of the H$_2$.
In  \autoref{fig:SI disorder model scaled 0}, we show the LP coefficients obtained from the matrix in \autoref{Eq: H_TC-Kasha disorder matrix} for P, A1, and A2 (from left to right) for coupling strengths $\lambda$ = \qtylist{0.002;0.003;0.004}{\atomicunit} (from the top down to the bottom).
These results are analogous to  \autoref{fig:SI model_c1 different g noscaling}.
The order of magnitude of the wave function coefficients and the qualitative sign flip of the A1 coefficients are again obtained, although for a larger number of dimers $N$.
Indeed, the collective coupling of the system is decreased because of disorder since the coupling to the cavity, compared to the perfectly ordered case, is scaled by $\cos{\theta_k}$ for each dimer.
Therefore, a larger number of coupled molecules is generally necessary to reach the avoided crossing between the LP and the MP.
Nevertheless, the fact that this effect resists the introduction of long-range disorder proves that it stems from the interplay between two distinct molecular environments: the local chemical environment (solvation shell) and the collective optical environment (polaritonic dressing).
It also confirms the local nature of such changes.
Therefore, we expect the presented results always to be relevant for systems characterized by short-range order, such as solute-solvent systems or aggregates.

\clearpage

\section{J-aggregate polaritonic excitations}
\begin{figure}[!ht]
    \centering
    \includegraphics[width=.45\textwidth]{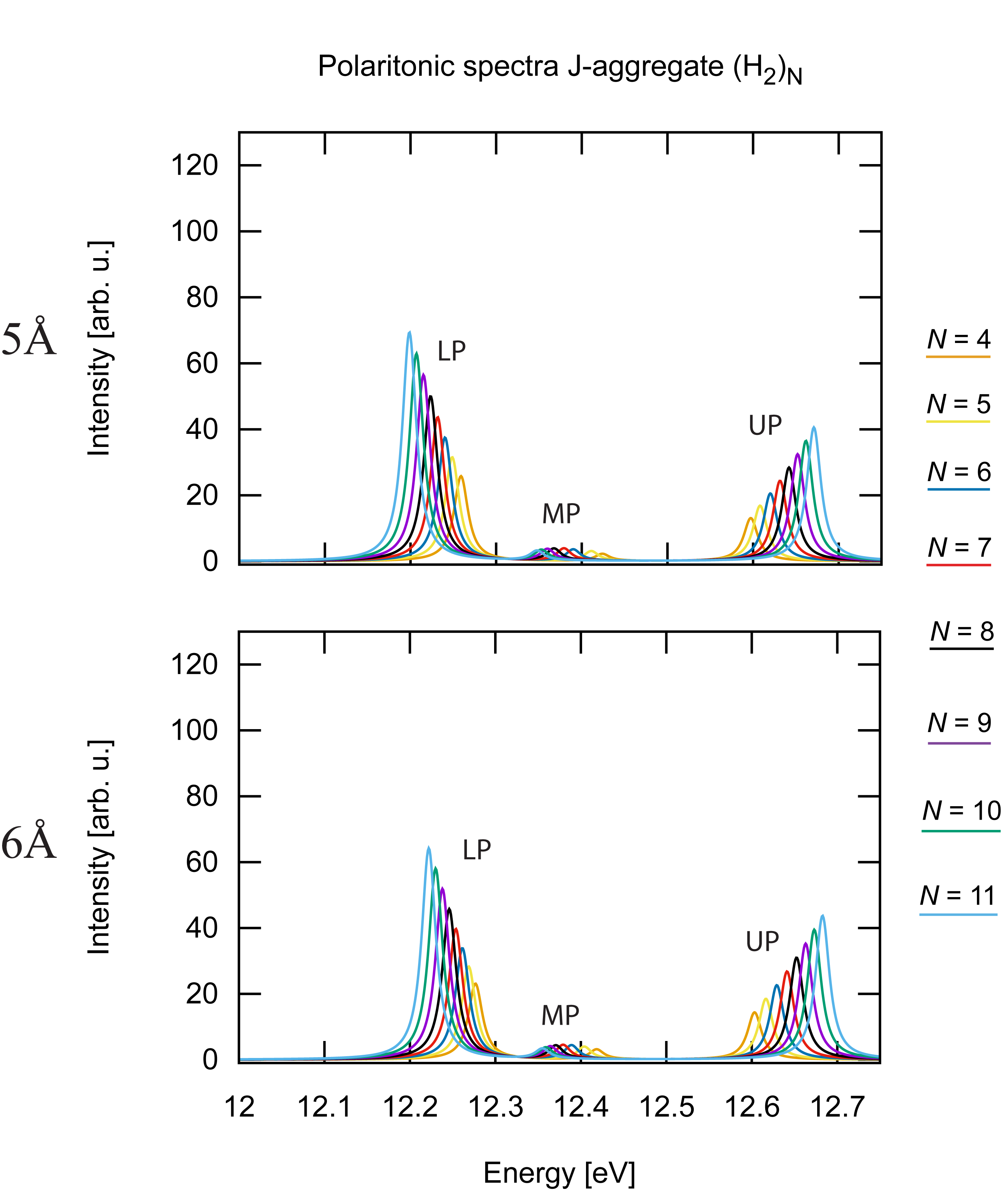}
    \caption{Polaritonic absorption spectra for a J-aggregate $(\text{H}_2)_{N\ge4}$ for \qty{5}{\angstrom} (upper panel) and \qty{6}{\angstrom} (lower panel) intermolecular separation.
    From the interaction between the photon field and the H-aggregate electronic excitations E1 and E2, three polaritonic branches, the lower, middle, and upper polariton, emerge.
    }
    \label{fig:Jaggr cavity spectra SI}
\end{figure}
\begin{figure*}[!ht]
    \centering
    \includegraphics[width=\textwidth]{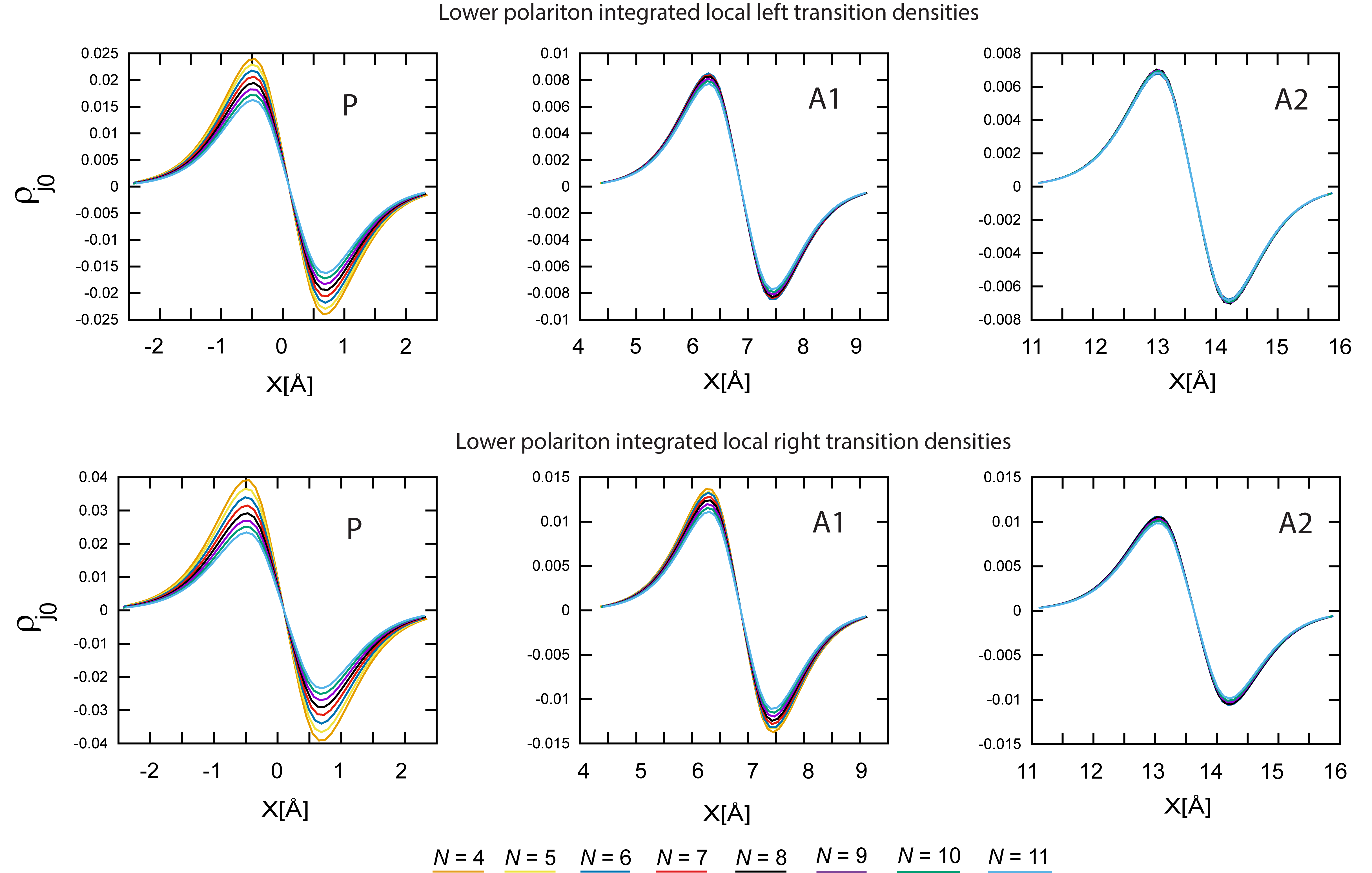}
    \caption{Local left and right transition densities for the lower polariton for an intermolecular separation of \qty{6}{\angstrom} in a J-aggregate configuration head-tail arrangement, see  \autoref{fig:Si structures}).
    The local transition dipole moments are all aligned for all $N$.
    }
    \label{fig: SI LP Jaggr 6}
\end{figure*}
\begin{figure*}[!ht]
    \centering
    \includegraphics[width=\textwidth]{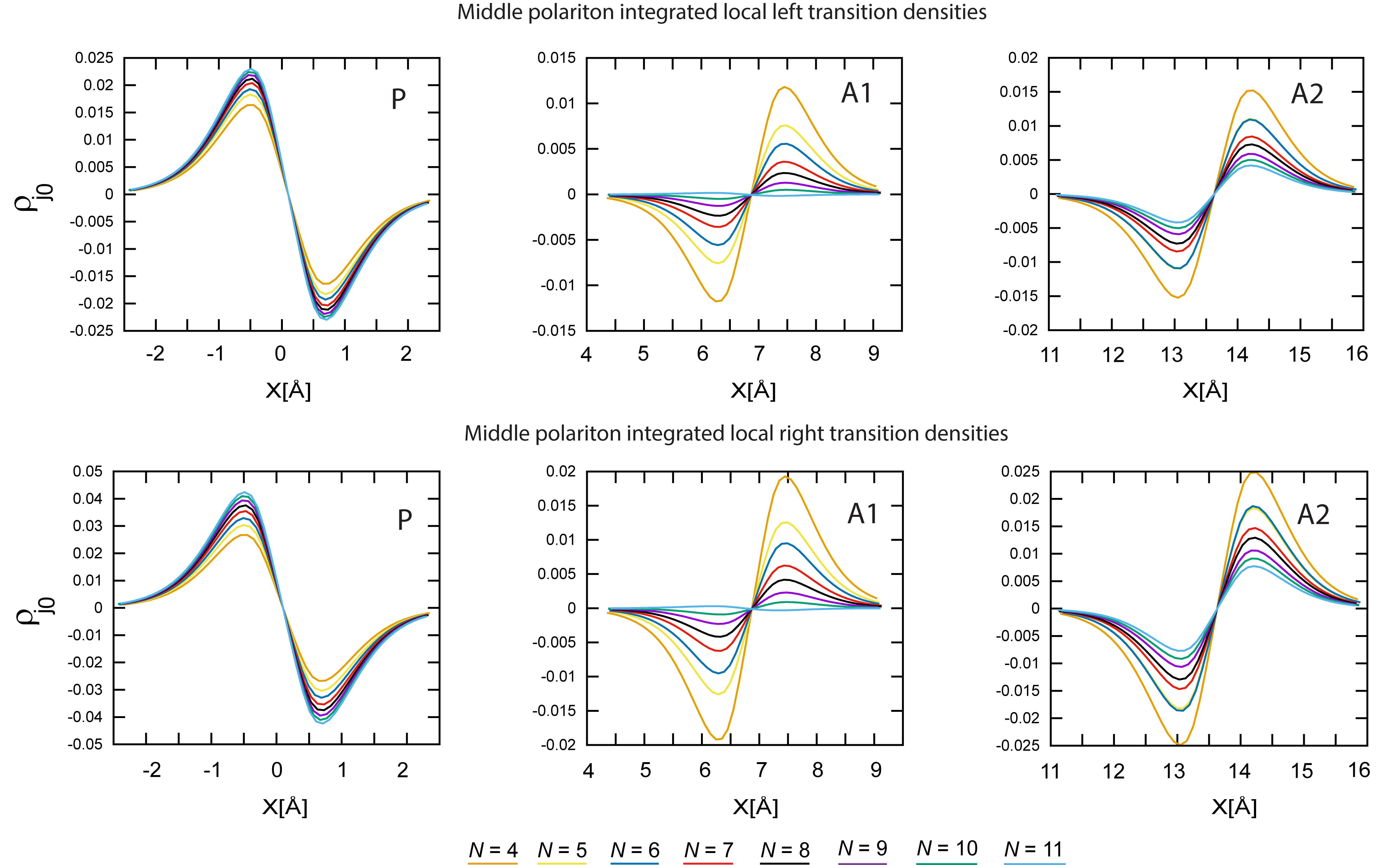}
    \caption{Local left and right transition densities for the middle polariton for an intermolecular separation of \qty{6}{\angstrom} in a J-aggregate configuration head-tail arrangement, see  \autoref{fig:Si structures}).
    The local transition dipole moment of A1 changes sign with $N$.
    }
    \label{fig: SI MP Jaggr 6}
\end{figure*}

In  \autoref{fig:Jaggr cavity spectra SI}, we show the dressed absorption spectra for  J-aggregate (see  \autoref{fig:Si structures}) for \qtylist{5;6}{\angstrom} intermolecular separation.
The cavity frequency is tuned to the undressed excitation of the isolated unperturbed dimers.
Compared to the H-aggregates (see  \autoref{fig:cavity spectra SI}), the middle polariton branch is less intense and loses oscillator strength with shorter intermolecular separations.
This reflects the different effects the local arrangement can have on a solute due to the strong directionality of intermolecular forces.
As the intermolecular forces become more relevant (shorter distances), the excitations redshift, in agreement with the predictions from the Kasha model (See \autoref{TC-Kasha polaritons}).
In  \autoref{fig: SI LP Jaggr 6}, we plot the LP left and right transition densities for the J-aggregate with an intermolecular distance of \qty{6}{\angstrom}.
Contrary to the H-aggregates, the local transition dipole moments are aligned for every $N$ and show a smaller dependence on the chain length.
In  \autoref{fig: SI MP Jaggr 6}, we show the corresponding middle polariton.
As $N$ increases, the transition dipole moment of A1, initially antialigned with P as in E2, changes sign.
As for the H-aggregate, this proves again a local (single-molecule) change stemming from longitudinal and transverse fields.
Moreover, the local changes here are more relevant for the middle polaritonic branches (instead of the lower branch as the H-aggregate), which provides a simple proof-of-concept example of how different chemical environments can influence chemical properties due to the directionality of the intermolecular forces.

\clearpage
\bibliography{lit}